\documentclass[useAMS,usenatbib]{mn2e}

\usepackage{graphicx,color}
\usepackage{latexsym,amsfonts,amssymb}
\usepackage{url}
\usepackage[authoryear]{natbib}
\usepackage{rotating}
\usepackage{lscape}
\usepackage{xspace}
\usepackage{txfonts}
\usepackage{graphicx}
\usepackage{epstopdf}
\usepackage{footmisc}
\usepackage{placeins}
\usepackage{lscape}
\usepackage{mathrsfs}
\usepackage{pstricks}

\newcommand{\WFI}{\texttt{WFI}\xspace}

\newcommand{\FORS}{\texttt{FORS}\xspace}
\newcommand{\MegaPrime}{\texttt{MegaPrime}\xspace}
\newcommand{\LBT}{\texttt{LBT}\xspace}
\newcommand{\LBC}{\texttt{LBC}\xspace}

\newcommand{\SExtractor}{\texttt{SExtractor}\xspace}
\newcommand{\Swarp}{\texttt{Swarp}\xspace}
\newcommand{\Weightwatcher}{\texttt{Weightwatcher}\xspace}

\newcommand{\eqref}[1]{eq.~(\ref{#1})}

\title[The Massive Galaxy Cluster XMMU J1230.3+1339 at z $\sim$ 1]{The Massive Galaxy Cluster XMMU J1230.3+1339 at z $\sim$ 1:\\ \vspace{0.25cm} \Large{Colour-magnitude relation, Butcher-Oemler effect, X-ray and weak lensing mass estimates.}
}

\author[Lerchster et al.]{M. Lerchster$^{1,2,3}$\thanks{E-mail: mike@usm.lmu.de, mlerchst@mpe.mpg.de}
, S. Seitz$^{1,2}$, F. Brimioulle$^{1}$, R. Fassbender$^{2}$, M. Rovilos$^{2}$,  H. B\"ohringer$^{2}$,
\newauthor 
D. Pierini$^{2}$, M. Kilbinger$^{1,3}$, A. Finoguenov$^{2,4}$, H. Quintana$^{5}$, and R. Bender$^{1,2}$.
\\                      
$^{1}$University Observatory Munich, Ludwigs-Maximilians University Munich, Scheinerstr. 1, 81679 Munich, Germany\\
$^{2}$Max-Planck-Institute for extraterrestrial Physics, Giessenbachstra{\ss}e, 85748 Garching, Germany\\
$^{3}$Excellence Cluster Universe, Technical University Munich, Boltzmannstr. 2, 85748 Garching, Germany\\
$^{4}$University of Maryland, Baltimore County, 1000 Hilltop Circle,  Baltimore, MD 21250, USA\\
$^{5}$Pontificia Universidad Catolica de Chile, Department of Astronomy and Astrophysics, Casilla 306, Santiago, Chile
}

\begin{document}

\date{Accepted ... Received ... ; in original form 2010 August 06}

\pagerange{\pageref{firstpage}--\pageref{lastpage}} \pubyear{2010}

\label{firstpage}

\maketitle

\begin{abstract}
We present results from the multi-wavelength study of XMMU J1230.3+1339 at z $\sim$ 1. We analyze deep multi-band wide-field images from the Large Binocular Telescope, multi-object spectroscopy observations from VLT, as well as space-based serendipitous observations, from the GALEX and Chandra X-ray observatories. We apply a Bayesian photometric redshift code to derive the redshifts using the FUV, NUV and the deep U, B, V, r, i, z data. We make further use of spectroscopic data from \texttt{FORS2} to calibrate our photometric redshifts, and investigate the photometric and spectral properties of the early-type galaxies. We achieve an accuracy of $\triangle z/(1+z)$ = 0.07 (0.04) and the fraction of catastrophic outliers is $\eta$ = 13 (0) \%, when using all (secure) spectroscopic data, respectively. 
The $i - z$ against $z$ colour-magnitude relation of the photo-z members shows a tight red-sequence with a zero point of 0.935 mag, and slope equal to -0.027. We observe evidence for a truncation at the faint end of the red-cluster-sequence and the Butcher-Oemler effect, finding a fraction of blue galaxies $f_b \approx$  0.5.
Further we conduct a weak lensing analysis of the deep 26\arcmin $\times$ 26\arcmin r-band \texttt{LBC} image. The observed shear is fitted with a Single-Isothermal-Sphere and a Navarro-Frenk-White model to obtain the velocity dispersion and the model parameters, respectively. Our best fit values are, for the velocity dispersion $\sigma_{\rm SIS}$ = 1308 $\pm$ 284 km/s, concentration parameter $c$ = 4.0$^{+14}_{-2}$ and scale radius $r$$_{\rm s}$ = 345$^{+50}_{-57}$ kpc.
From a 38 ks Chandra X-ray observation we obtain an independent estimate of the cluster mass. In addition we create a S/N map for the detection of the matter mass distribution of the cluster using the mass-aperture technique. We find excellent agreement of the mass concentration identified with weak lensing and the X-ray surface brightness. 
Combining our mass estimates from the kinematic, X-ray and weak lensing analyses we obtain a total cluster mass of $M^{\rm tot}_{\rm 200}$ = (4.56 $\pm$ 2.3)  $\times$ 10$^{14}$ M$_{\sun}$.
This study demonstrates the feasibility of ground based weak lensing measurements of galaxy clusters up to $z \sim 1$.\\

\end{abstract}


\section{Introduction}
\label{sec:intro}
Clusters of galaxies are receiving an increasing interest in astronomical studies in the last decade for a variety of reasons. They are the largest gravitationally bound objects in the universe and provide ideal organized laboratories for the cosmic evolution of baryons, in the hot Intra Cluster Medium (ICM) phase and galaxies. They are also ideal tracers of the formation of the large-scale structure.
 
The combination of optical, spectroscopic, lensing and X-ray studies allows us to get important insights into physical properties of the cluster members, the ICM and the determination of the gravitational mass and its distribution with independent methods.
The dynamical state of the ICM can also reveal information about the recent merger history, thus providing an interesting way to test the concept of hierarchical cluster formation, as predicted in the cold dark matter (CDM) paradigm.

In the past, clusters of galaxies have been many time the subject of combined weak lensing
and dynamical studies (e.g. Radovich et al.\citealt{radovich08}), weak and strong lensing plus X-ray studies (e.g. Kausch et al.\citealt{kausch07}), and weak lensing and X-ray studies (e.g. Hoekstra 
et al.\citealt{hoekstra07}).
In addition, Taylor et al. \citet{taylor04} for ground based data and Heymans et al. \citet{heymans08} for 
space based data showed for the $z = 0.16$ supercluster A901/2 (COMBO-17 survey) that a three-dimensional (3D) lensing analysis combined with photometric redshifts can reveal the 3D dark matter distribution to high precision.

Wide-field instruments, such as the ESO Wide-Field Imager \WFI at the ESO2.2m with a field of view (FOV) of 34\arcmin$\times$33\arcmin, the \MegaPrime instrument mounted at the 3.6m CFHT with a FOV of $\approx$1 square degree, the Suprime-Cam with a FOV of 34\arcmin$\times$27\arcmin at the 8.2m Subaru telescope and more recently the Large Binocular Camera (\LBC) instruments mounted at the 8.4m LBT (Hill \& Salinari 1998) with a FOV of 26\arcmin$\times$26\arcmin, are particularly well suited for the multi-band and weak lensing studies of clusters. One can obtain multi-band photometry for evolution studies based on spectral energy distribution (SED) fitting techniques and one can also apply weak lensing studies to the clusters, because the FOV is capable of sampling the signal well beyond the cluster size.

In this paper we present an combined optical, kinematic, weak lensing, and X-ray analysis of  XMMU J1230.3+1339. This cluster of galaxies is located at $\alpha$ = 12h30m18s, $\delta$ = +13$^{\circ}$39\arcmin00\arcsec (J2000), redshift z$_{\rm cl}$ = 0.975 and Abell richness\footnote{See richness definition in Abell et al. \cite{abell89}.} class 2 (see Fassbender et al. 2010, Paper I hereafter, for more details).
The cluster was detected within the framework of the XMM-Newton Distant Cluster Project (XDCP) described in Fassbender et al. \cite{fassbender08}.

We combine optical 6-band data, multiobject spectroscopy data, space based UV data (GALEX) and X-ray (Chandra) data.

In this work we present a detailed analysis of the optical data, estimating photometric redshifts and performing a SED classification, exploring the colour-magnitude and colour-colour relations. Based on this we conduct a weak lensing analysis and estimate independently and self consistently the cluster mass from X-ray, kinematics and weak lensing, complementing the analysis presented in Paper I. 
In section \ref{sec:data} we give an overview of the data acquisition and reduction.
We briefly outline the reduction of the \LBC data and describe the acquisition of the spectroscopic redshifts in Sec. \ref{sec:data}. In Sec. \ref{sec:optanal} we explain the method used to derive the photometric catalogues and redshifts, we discuss the accuracy of the derived redshifts for the cluster, and we show the distribution of its galaxies.
In Sec. \ref{sec:wlanal} we present the weak lensing analysis (mass distribution) and constraints for the clusters, whereas the estimates for the kinematic analysis (velocity dispersion) are shown in section \ref{sec:kinanal}.
The X-ray analysis (surface-brightness profile, global temperature) is presented in section \ref{sec:xrayanal}, and the mass estimates are shown in section \ref{sec:mass-est}. We discuss and summarise our findings in section \ref{sec:discon} and \ref{sec:conclu}.

Throughout the paper we assume a $\Lambda$CDM cosmology with $\Omega_{\rm M} = 0.27$, $\Omega_{\rm \Lambda}$ = 0.73, and H$_{\rm 0}$ = 72 km s$^{-1}$ Mpc$^{-1}$. Therefore one arcsecond angular distance corresponds to a physical scale of 7.878 kpc at the cluster redshift.

\section{Data Acquisition and Reduction}
\label{sec:data}

\begin{figure*}
\centering
\includegraphics[width=15.5cm]{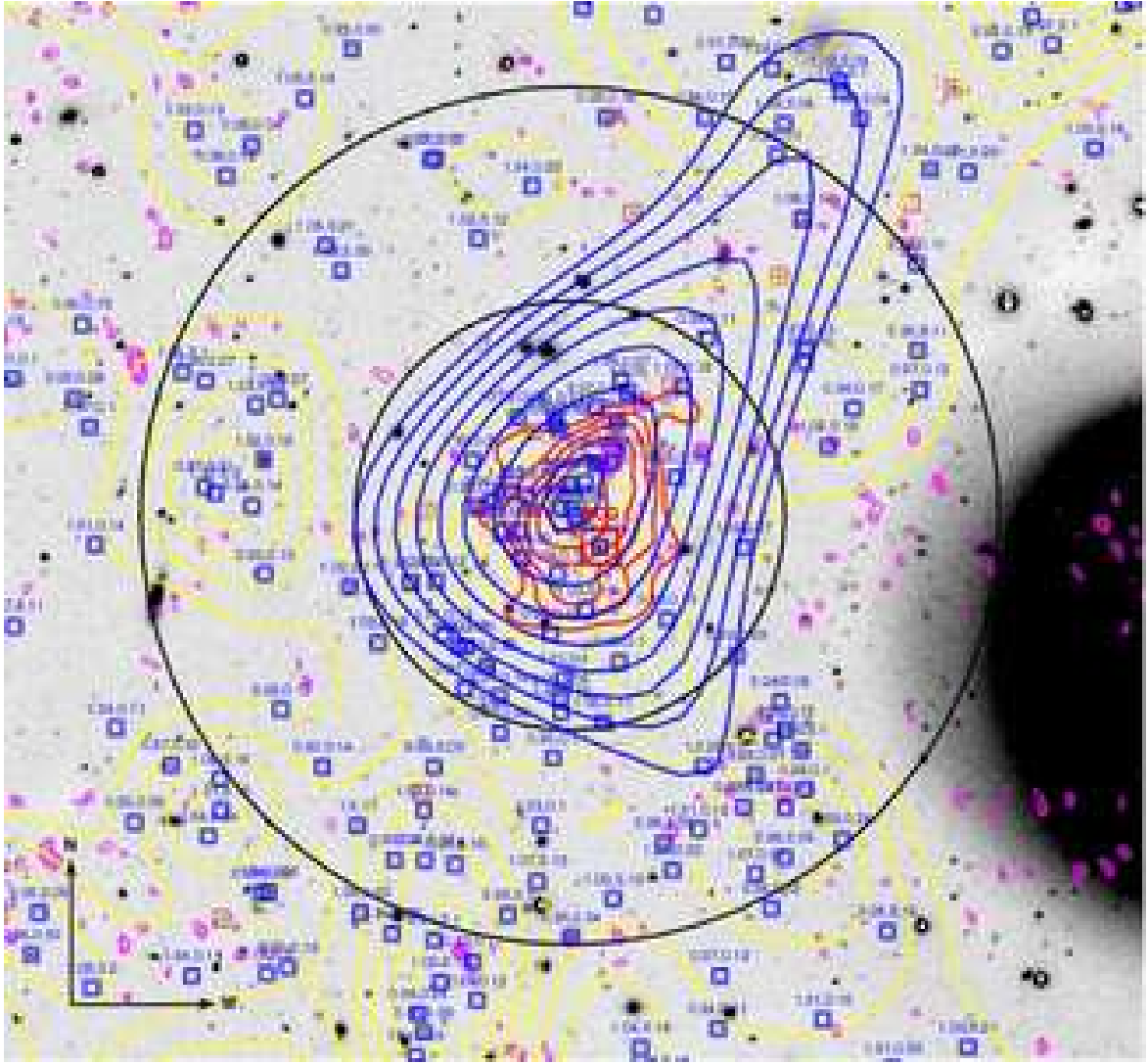}
\caption{XMMU J1230.3+1339 at z = 0.975: 6\arcmin$\times$6\arcmin LBT-i band image (5.94 ks), overlaid with colour coded contours. In yellow: the surface brightness contours of the smoothed light of cluster members, selected to lie at z$_{cluster}$  - 0.05 $\le$ z$_{phot}$ $\le$ z$_{cluster}$ + 0.05; in blue: weak lensing M$_{ap}$ significance contours (1 - 5); in red: X-ray contours (0.3 - 2.4 keV); in magenta: VLA-FIRST (0.3 - 0.9 mJy) contours. The red big box marks the location of the BCG, the red small boxes mark the high confidence spectroscopic confirmed members (13), the blue small boxes (numbers are redshift and error) mark the photo-z selected cluster members, and the black circles mark a radius of 0.5 and 1 Mpc from the cluster centre, respectively (note: r$_{200}$$\approx$ 1 Mpc). The large object to the west is the foreground galaxy NGC 4477.
} 
\label{Fig1aXMMUJ1230}
\end{figure*}       
\begin{figure*}
\centering
\includegraphics[width=15.5cm]{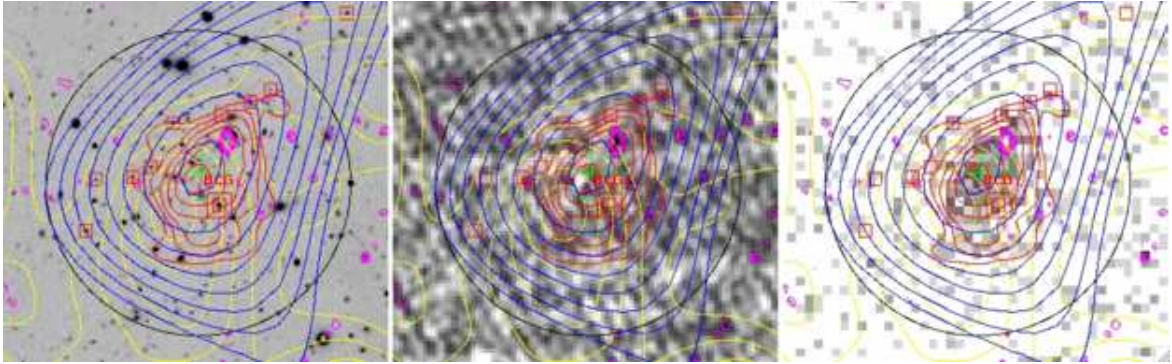}
\caption{Left panel: same as Figure \ref{Fig1aXMMUJ1230} for the inner 0.5 Mpc (black circle) of the cluster, FOV 3\arcmin$\times$3\arcmin. Middle panel: VLA-First 1.4 GHz-band image overlaid with colour coded contours. Right panel: X-ray (0.6 - 2.4 keV) image overlaid with colour coded contours. In yellow the luminosity contours of the cluster members, selected to lie at z$_{\rm cluster}$  - 0.05 $\le$ z$_{phot}$ $\le$ z$_{\rm cluster}$ + 0.05, in blue weak lensing M$_{ap}$ significance contours (1 - 5), and in red X-ray contours (0.6 - 2.4 keV).} 
\label{Fig1bXMMUJ1230}
\end{figure*}       
\begin{figure*}
\centering
\includegraphics[width=15.5cm]{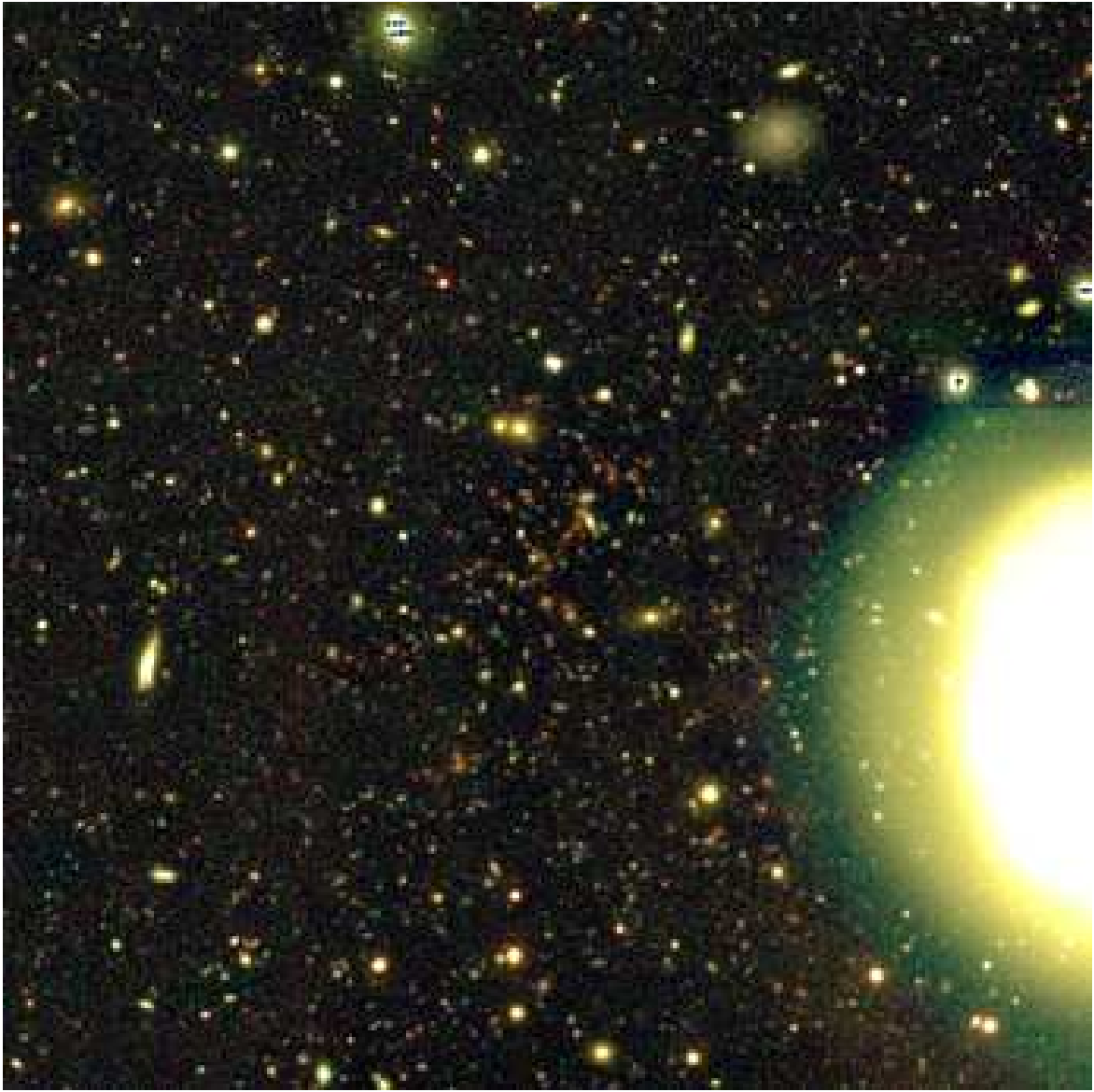}   
\caption{The 6\arcmin$\times$6\arcmin, three-colour image of the cluster XMMU J1230.3+1339, with r-channel (5.94 ks i-band), g-channel (3.24 ks r-band) and b-channel (5.94ks B-band) colours. The cluster members appear as reddish objects. The large, yellowish object on the right side is the foreground galaxy NGC 4477.} 
\label{Fig2XMMUJ1230}
\end{figure*}       
%

\subsection{Optical ground based data}
\label{sec:optdat}
\begin{table*}
\caption{Cluster optical data overview.}
\centering
\renewcommand{\footnoterule}{}  
\resizebox{17.5cm}{!} {
\begin{tabular}{lcccccccccl}
\hline \hline
Field/Area & Telescope/Instrument & FOV & Filter & N$^a$ & expos. time & zero-point & $m_{\rm lim}$ $^b$ & seeing & $\sigma_{\rm sky}$  \\
{[sq. deg.]}  &         ~           &  [arcmin*arcmin]   ~    &    ~    & ~ &   [s]       &    [AB mag]       &    [AB mag]       &   [$\arcsec$]     &      [ADU]             \\
\hline
XMMU J1230.3+1339 & LBT/\LBC\texttt{B} & 26*26 & U & 24 & 4320 & 32.22 & 26.97 & 0.85 & 3.2 \\
(0.19) & LBT/ \LBC\texttt{B} & 26*26 & B & 17 & 3060 & 32.98 & 26.90 & 0.87 & 6.9 \\
~ & LBT/\LBC\texttt{B} & 26*26 & V& 13 & 2340 & 33.27 & 26.32 & 0.85 & 15.3 \\
~ & LBT/\LBC\texttt{R} & 26*26 & r$^c$ & 12 & 2160 & 33.44 & 26.18 & 0.95 & 20.3\\
~ & LBT/\LBC\texttt{R} & 26*26 & i & 27 & 4860 & 33.27 & 25.80 & 0.87 & 24.5 \\
~ & LBT/\LBC\texttt{R} & 26*26 & z & 44 & 7960 & 32.79 & 25.37 & 0.78 & 23.5 \\
\hline
\end{tabular}
}
Note: $^a$ number of single exposures (t$_{\rm exp}=180$ s) used for the final stack; $^b$ the limiting magnitude is defined in Eq. \ref{eq:limMag}; $^c$ lensing band.
\label{tab:dataprop}
\end{table*}
In Table 1 we summarise the properties (e.g., instrument, filter, exposure time, seeing) of all optical data used in this analysis. The limiting magnitude quoted in column 8 is defined as the 5-$\sigma$ detection limit in a 1.0\arcsec aperture via:
\begin{equation}
m_{lim}=ZP-2.5 * log(5\sqrt{N_{pix}} \sigma_{sky}),
\label{eq:limMag} 
\end{equation}
where ZP is the magnitude zeropoint, $N_{pix}$ is the number of pixels in a circle with radius $1.0\arcsec$ and $\sigma_{sky}$ the sky background noise variation.
Since this equation assumes Poissonian noise whereas the noise of reduced, stacked data is typically correlated, this estimate represents an upper limit.

\subsubsection{\LBC-observations and Data Reduction}
We observed XMMU J1230.3+1339 in 6 optical broad bands (U, B, V, r, i, z), with the 8.4m Large Binocular Telescope (LBT) and the \LBC instruments  in ``binocular" mode in an observing run from 28.02.2009 until 02.03.2009.
The \LBC cameras (see e.g. Giallongo et al.\citealt{giallongo08}, Speziali et al.\citealt{speziali08}) have  a 4 CCD array ($2048\times 4096$ pixel in each CCD; 0.224\arcsec pixel scale; 26\arcmin$\times$26\arcmin total field-of-view). Three chips are aligned parallel to each other while the fourth is rotated by 90 degrees and located above them.
The U,B,V images were obtained with the blue sensitive instrument (3500 - 9000 $\AA$) and in parallel to the r,i,z images with the red sensitive instrument (5000 - 10000 $\AA$). The filter transmission curves are shown in Figure \ref{FigFilter}.
For the single frame reduction and the creation of image stacks for the scientific exploitation, we follow the method described in detail in Rovilos et al. \cite{rovilos09}. Here we quickly summarise the most important reduction steps:

\begin{enumerate}

\item [1.] Single frame standard reduction (e.g. bias correction, flat-fielding, defringing).

\item [2.] Quality control of all 4 chips of each exposure to identify bad ones (e.g. chips with a large fraction of saturated pixels, etc.).

\item [3.] Identification of satellite tracks by visual inspection and creation of masks.

\item [4.] Creation of weight images with \Weightwatcher (Bertin \& Marmo \citealt{bertin07}) for each chip, accounting for bad pixels and the satellite track masks.

\item [5.] Extraction of catalogues with \SExtractor (Bertin \& Arnouts \citealt{bertin96}) for astrometric calibration.

\item[6.] Astrometric calibration with the USNO-A2 astrometric catalogue (Monet \citealt{monet98}) using the \texttt{IRAF} {\it ccmap} package. This yields an external accuracy with an $rms$ of 0.25 arcsec and an internal accuracy with an $rms$ of 0.066 arcsec (Rovilos et al. \citealt{rovilos09}).

\item [7.] Flux calibration with standard star observations in each of our 6 bands using Landolt standards or SDSS-DR7. 

\item [8.] Background subtraction and measure of the variation of $\sigma_{\rm sky}$ of each exposure to identify bad images (e.g. those with a value 3 $\sigma$ above the median value).

\item [9.] PSF control of all exposures to identify bad patterns (e.g. strong variation of the direction and amplitude of the measured star ellipticity on each chip of each exposure, etc.)

\item [10.] Co-addition with the \Swarp software (Bertin \citealt{bertin08}).

\item [11.] Masking of the image stacks to leave out low-SNR regions and regions affected by stellar diffraction spikes and stellar light halos as well as asteroid tracks.

\end{enumerate}

\begin{figure}
\centering
\includegraphics[width=4.15cm,angle=0]{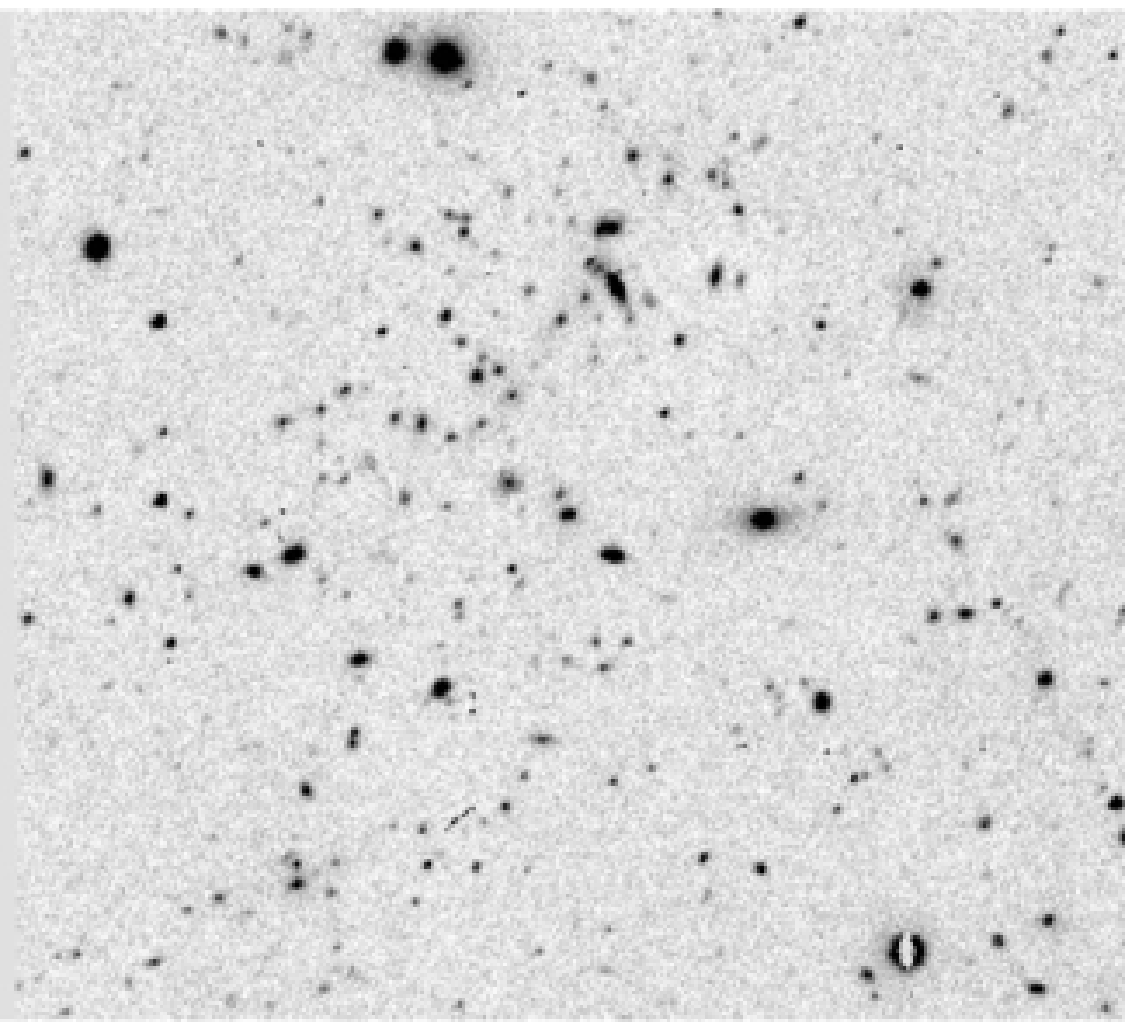}
\includegraphics[width=4.15cm,angle=0]{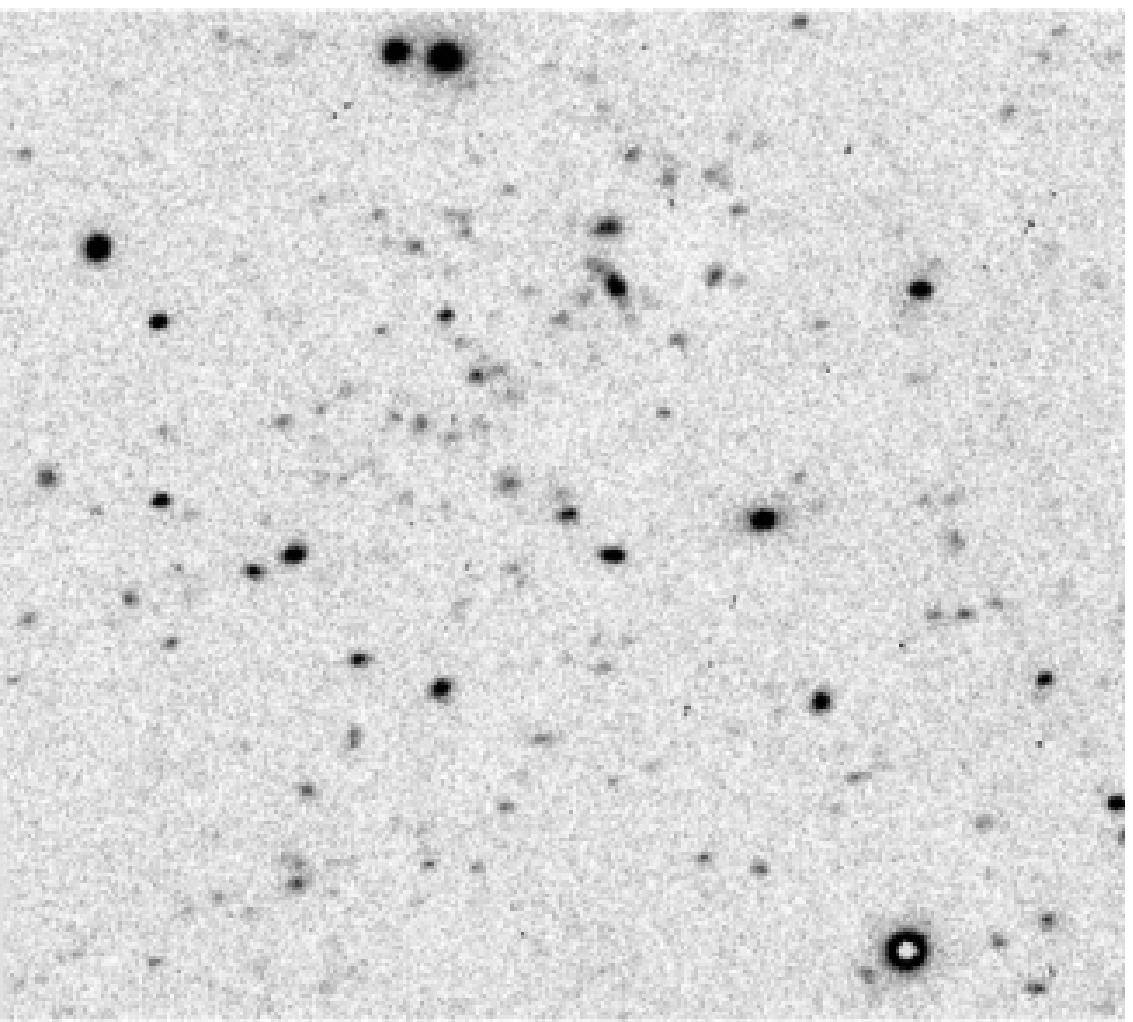}   
\caption{\LBT i-band single frames: the left panel shows a ``good'' frame (FWHM 0.64\arcsec) whereas the right panel shows a ``bad'' one (FWHM 1.20\arcsec).} 
\label{FigXMMUJ1230_LBT_frames}
\end{figure}

The number of frames which remain in this procedure is different from band to band, see column 5 in Table 1. Starting from 50,33,20,20,33,50 images in the U,B,V,r,i,z filters, we at the end only stack 24,17,13,12,27,44 frames for the photometric and weak lensing analyses. In Figure \ref{FigXMMUJ1230_LBT_frames} we show as an example one ``good" and one ``bad" single frame exposure.

\begin{figure*}
\centering
\includegraphics[width=5.5cm]{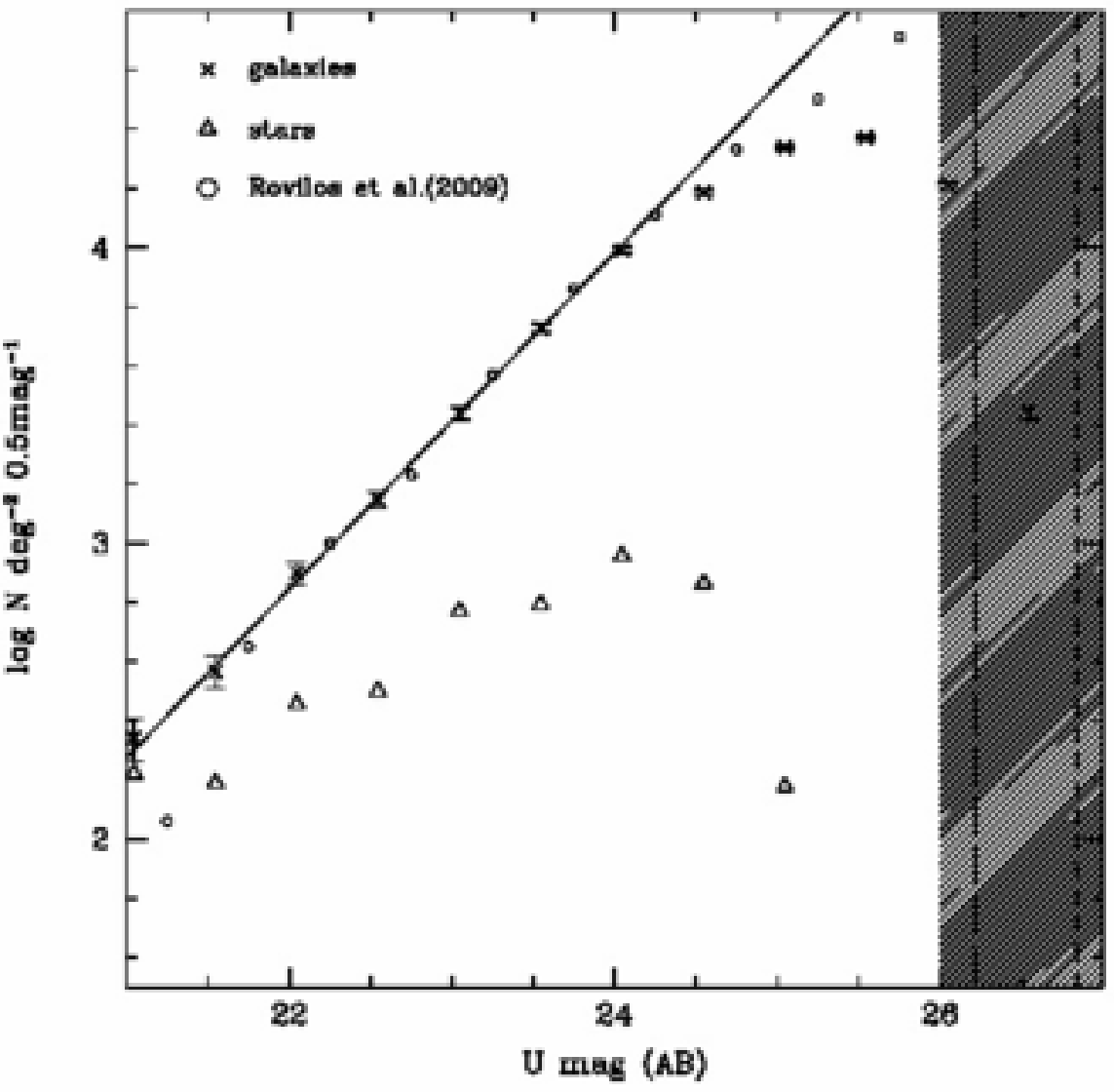}
\includegraphics[width=5.5cm]{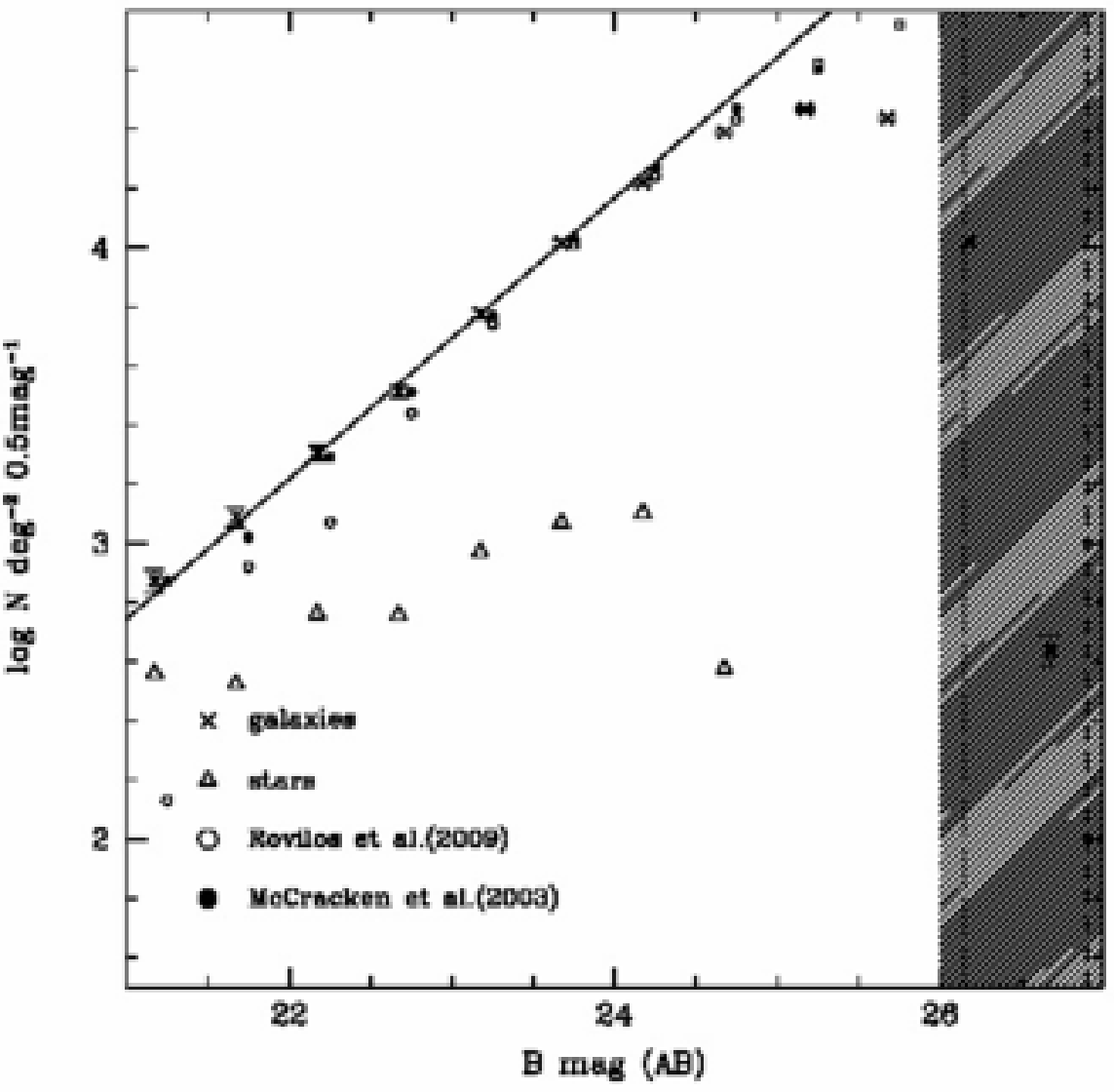}
\includegraphics[width=5.5cm]{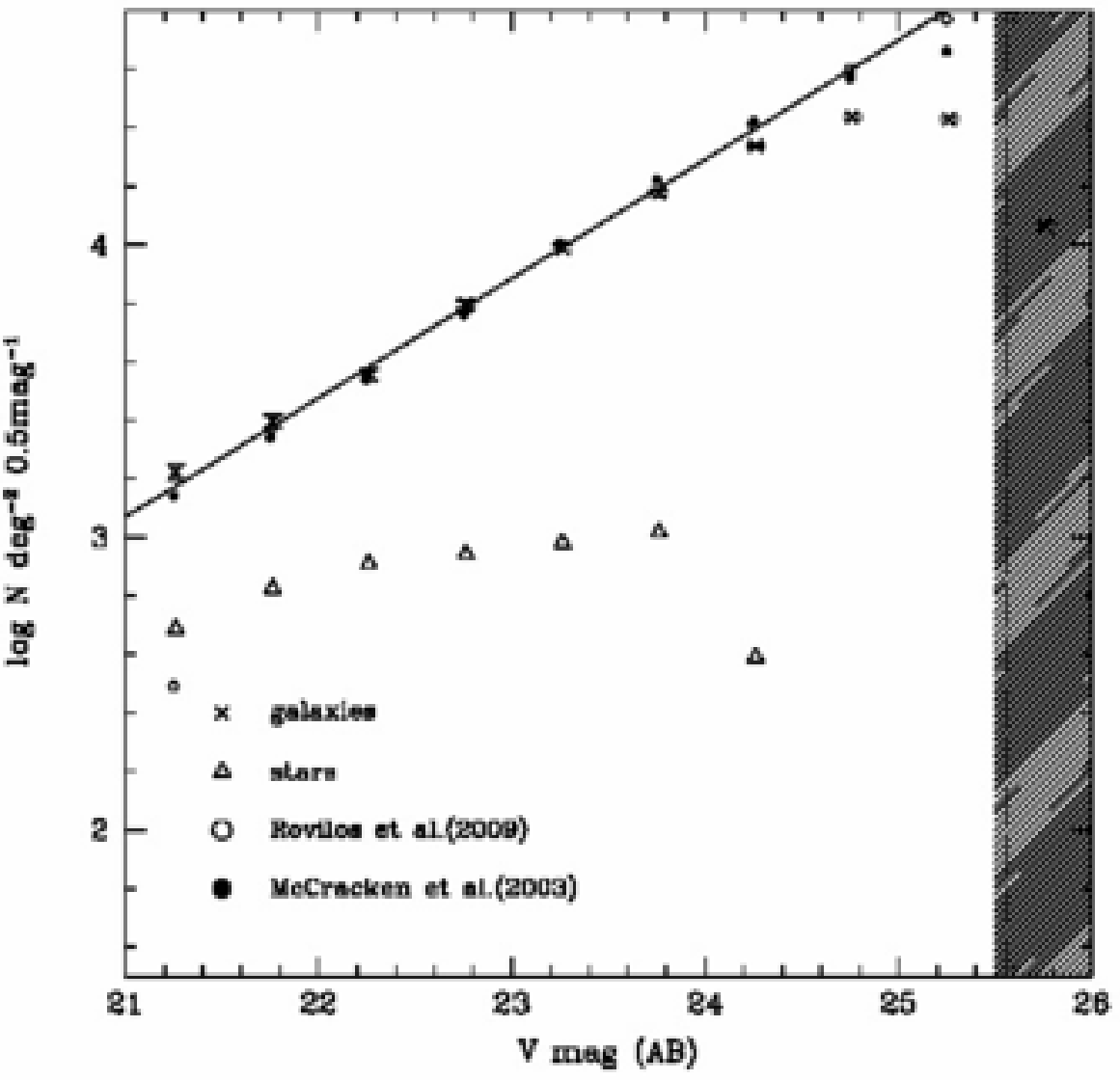}
\includegraphics[width=5.5cm]{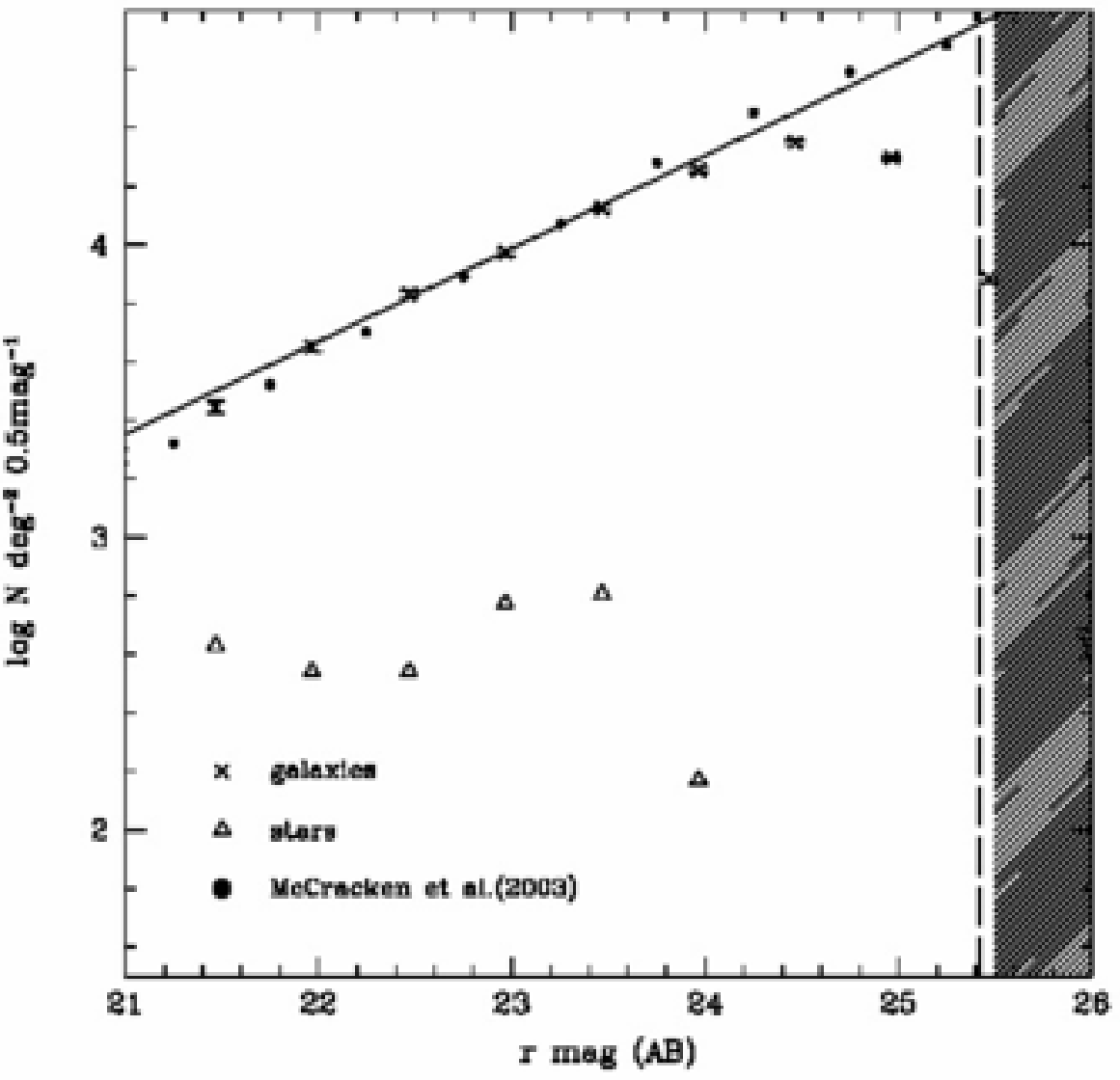}
\includegraphics[width=5.5cm]{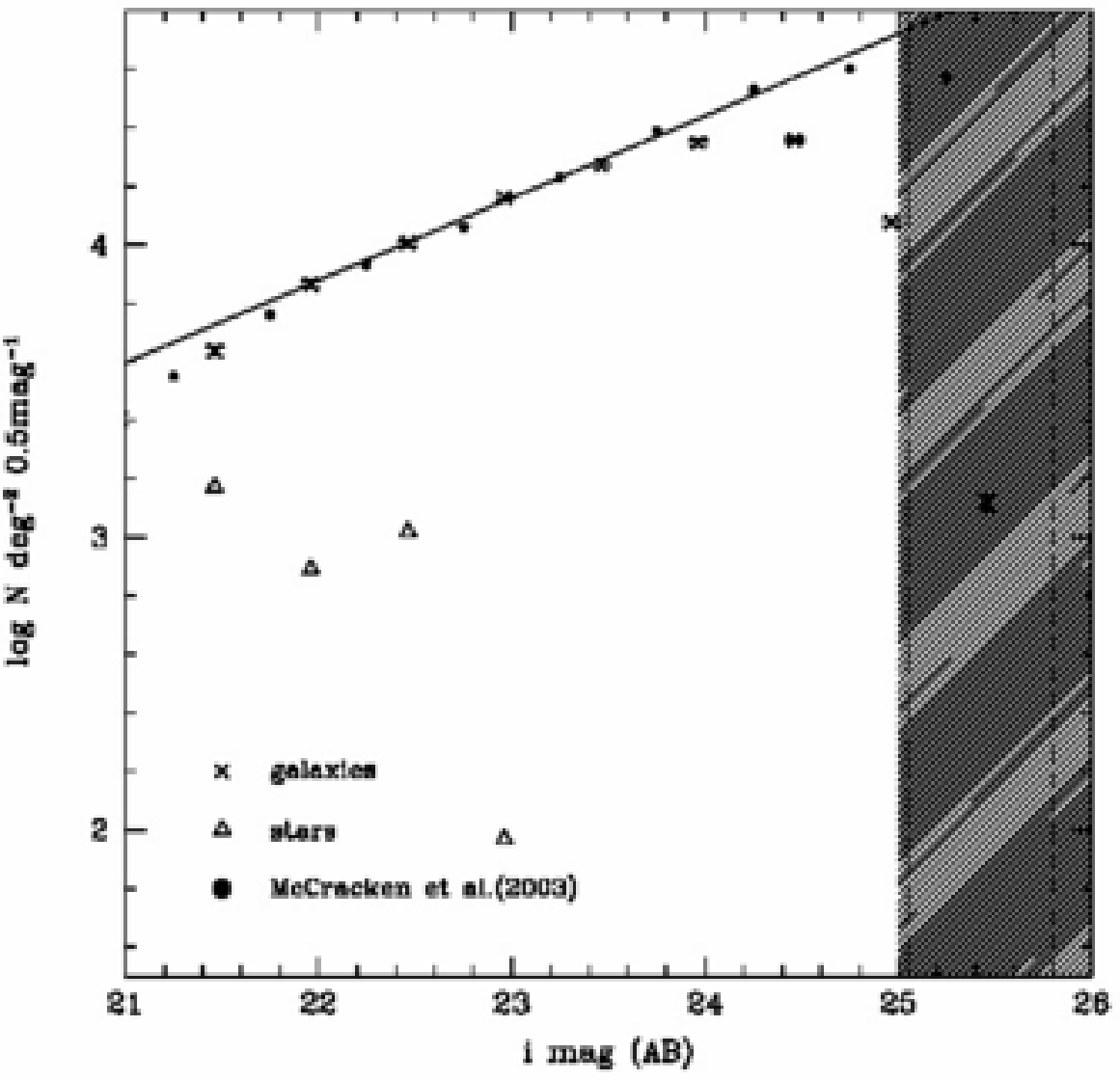}
\includegraphics[width=5.5cm]{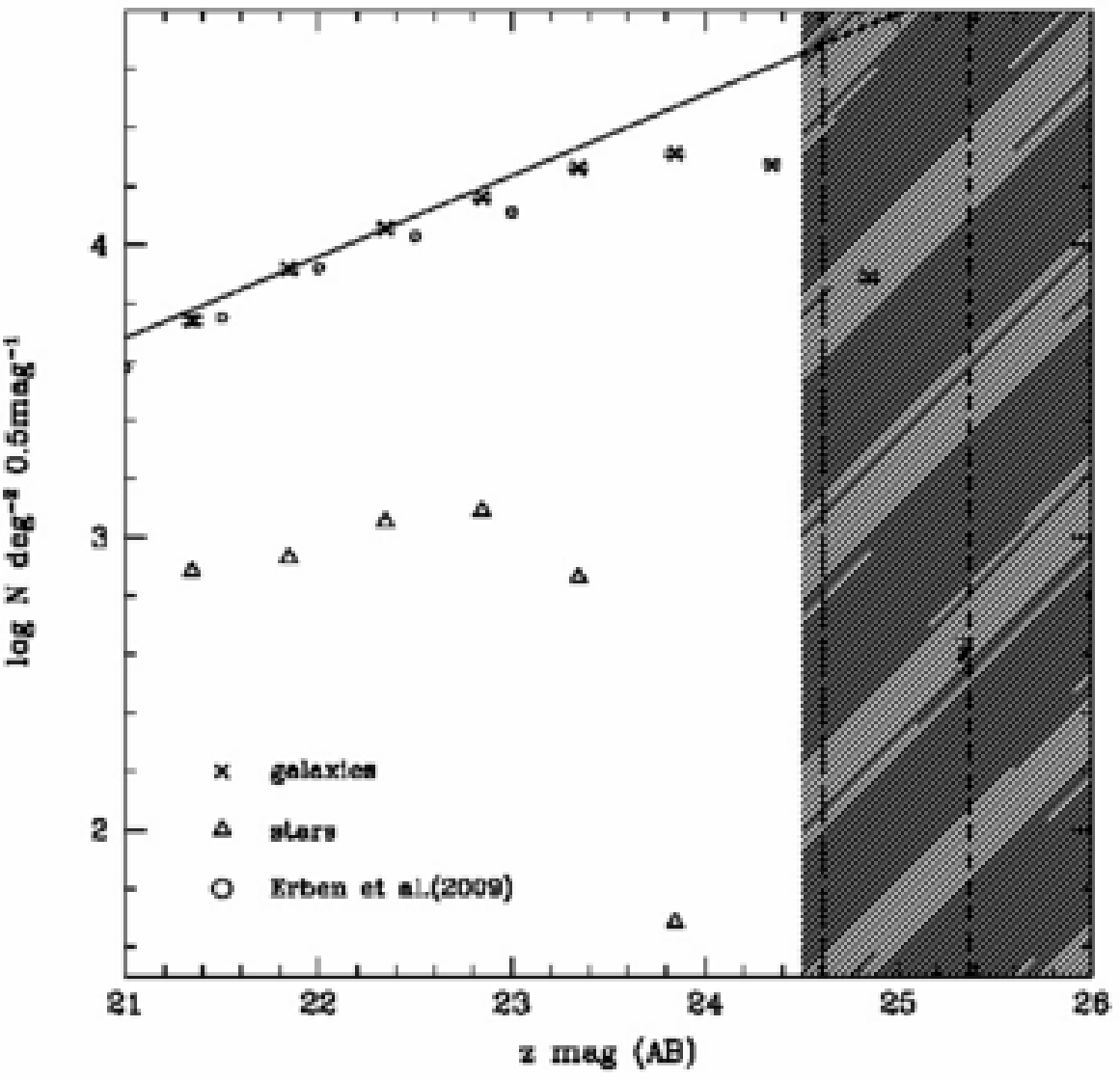}
\caption{Number counts per 0.5 mag and sq. degree of galaxies (crosses) and stars (open triangles) for the image stacks in the 6 (U, B, V, r, i, z) filters. The black line shows a linear fit to the galaxies number counts. For comparison we plot the number counts (open circles) derived from Rovilos et al. (2009) and (filled circles) McCracken et al. (2003), see text for details. The short/long dashed lines indicate the estimated limiting magnitudes respectively, defined as a point source  5$\sigma$ detection limit in a 1.0/2.0\arcsec aperture.
We used Eq. \ref{eq:limMag} for the noise levels of the image stacks listed in the Table 1. By comparing our numbercounts to deeper literature data and the extrapolation of the linear fit we can estimate the 50\% completeness for the detection. We have shaded the region where the completeness drops below 50\% in each filter.}
\label{Figmagdist}
\end{figure*}

We derive our number counts by running \SExtractor on the final image stacks with the corresponding weight map. We convolve the data with a Gaussian kernel and require for a detection to have 4 contiguous pixels exceeding a S/N of 1.25 on the convolved frame. Stars are then identified in the stellar locus, plotting the magnitude versus the half-light radius in Figure \ref{FigStars} and identify high signal to noise but unsaturated stars.
In Figure \ref{Figmagdist} we show the achieved number counts for the final image stacks. Table \ref{tab:dataprop} summarises the properties of each image stack. For comparison we plot in Figure \ref{Figmagdist}, for the U, B, V band, the number counts derived from Rovilos et al. \cite{rovilos09} for the deep Lockman Hole observation with the LBT. We do not show the uncertainties for these literature results, which have large errors at the bright end (U, B, V mag $<$ 22.5). In addition we compare our B, V, r and i band number counts to those from McCracken et al. \cite{mccracken03} for the VVDS-Deep Field. For the z band the comparison is with the number counts from Erben et al. \cite{erben09} for the CFHTLS D1 Field. In terms of amplitude and slope, our agreement with the literature measurements is very good.
The comparison of our data to deeper literature data and the linear number-count fits added with a solid line shows that the 50$\%$ completeness for an extented galaxy roughly falls together with the prediction for a point source within a 2\arcsec diameter aperture (long dashed line) according to Equation 1.
%

\subsection{Spectroscopy}
\label{sec:spec}
In Table 2 we give an overview of all spectroscopic data used in this analysis.
\begin{table*}
\begin{minipage}[t][]{\columnwidth}
\caption{Spectroscopic data overview.}
\centering
\renewcommand{\footnoterule}{}  
\begin{tabular}{lccccccl}
\hline \hline
ID & RA & DEC & Mag\_Auto  & $z_{spec}$\footnote{Measurement error: $\sigma_z\sim 0.0002$.} & Flag\footnote{Flags and confidence classes - 0: no redshift, 1: 50 \% confidence, 2: 75 \% confidence, 3: 95 \% confidence, 4: 100 \% confidence} & ref.\footnote{1: Adelman-McCarthy et al. \cite{adelman07}; 2: Paper I } & comments  \\
~   &  [deg]   &  [deg]    &  $i_{AB}$ &  ~ & ~ & ~      &     \\
\hline
34  &  187.5598939  &  13.7739387  &  17.1514  &  0.1300  &  4        & 1\\ 
35  &  187.6228679  &  13.7854572  &  16.8121  &  0.1573  &  4	 & 1\\ 
44  &  187.5528796  &  13.4550020  &  16.5786  &  0.0830  &  4	 & 1\\ 
46  &  187.5733437  &  13.4931093  &  16.3311  &  0.0832  &  4	 & 1\\ 
47  &  187.5878466  &  13.4371208  &  17.0668  &  0.0838  &  4	 & 1\\ 
50  &  187.6041016  &  13.4623009  &  16.2996  &  0.0829  &  4	 & 1\\ 
57  &  187.3367446  &  13.6621547  &  20.2306  &  0.3918  &  4	 & 1\\ 
58  &  187.4747995  &  13.6681400  &  17.4096  &  0.1447  &  4	 & 1\\ 
59  &  187.4367670  &  13.5817725  &  18.3299  &  0.4377  &  4	 & 1\\ 
61  &  187.6479182  &  13.6215654  &  18.3519  &  0.2439  &  4	 & 1\\ 
\hline
67  &  187.5400466  &  13.6773148  &  24.3561  &  0.9642  &  4	 & 2 & [OII] emitter\\  
69  &  187.5498491  &  13.6759957  &  22.0181  &  0.9691  &  4	 & 2\\  
80  &  187.5740730  &  13.6573756  &  21.9994  &  0.9758  &  4	 & 2\\  
82  &  187.5681276  &  13.6472714  &  20.8572  &  0.9786  &  3	 & 2 & BCG\\  
83  &  187.5767988  &  13.6519471  &  22.2796  &  0.9793  &  3	 & 2\\  
84  &  187.5791935  &  13.6507665  &  22.2532  &  0.8616  &  3	 & 2\\  
85  &  187.5838292  &  13.6503877  &  22.8933  &  0.9690  &  4	 & 2\\  
86  &  187.5851446  &  13.6441070  &  21.9350  &  0.9767  &  4	 & 2\\   
\hline
\end{tabular}
\end{minipage}
\label{tab:specprop}
\end{table*}
%

\subsubsection{VLT Spectroscopy and Data Reduction}
\label{sec:specvlt}
For the kinematic study of the cluster and the calibration of the photometric redshifts we use the data taken in April 2008 and June 2008 with the FOcal Reducer and low dispersion Spectrograph (\FORS) mounted at the UT1 of the Very Large Telescope (VLT), which were taken as part of the VLT-program 081.A-0312 (PI: H. Quintana).
The MOS/MXU spectroscopy was performed with FORS2 (see e.g. Appenzeller et al.\citealt{appenzeller98}) in medium resolution (R = 660) mode (grism GRIS\_300I+21). 
The data were reduced using the \texttt{IRAF}\footnote{\url{http://iraf.noao.edu}} software package. For the redshift determination (redshift, error, confidence level) the spectra were cross-correlated with template spectra using the \texttt{IRAF} package $RVSAO$. More details can be found in Paper I.

The spectroscopic observations provide 38 spectra, where 17 of them matched in position with objects from our photometric catalogue. For the comparison to photometric redshifts we consider galaxies with trustworthy $\geq$ 95\% confidence spectroscopic redshifts only. Due to low S/N (at high redshift)  and the limited wavelength ranges of the spectra only 8 objects could be considered.

We have been awarded additional FORS2 spectroscopy for $\sim$ 100 photo-z selected cluster members; the analysis of this data will be described in an upcoming paper.

\subsubsection{SDSS Spectroscopy}
\label{sec:specsdss}

The field of XMMU J1230.3+1339 has complete overlap with the Sloan Digital Sky Survey\footnote{Funding for the SDSS and SDSS-II has been provided by the Alfred P. Sloan Fundation, the Participating Institutions, the National Science Foundation, the U.S. Department of Energy, the National Aeronautics and Space Administration, the Japanese Monbukagakusho, the Max Planck Society, and the Higher Education Funding Council for England. The SDSS Web Site is 
\url{http://www.sdss.org/}.}
(SDSS). Via the flexible web-interface \texttt{SkyServer\footnote{\url{http://cas.sdss.org/astro/en/tools/search/SQS.asp}}} of the Catalog Archive Server (CAS), we get access to the redshift catalogue. For our purpose we only consider objects clearly classified as a galaxy with a redshift $\geq$ 95\% confidence. 
By matching the SDSS catalogue with our catalogue we find 13 objects with spectroscopic redshifts between 0 and 0.6 in the SDSS-DR7 (Adelman-McCarthy et al. 2007). For the comparison to photometric redshifts we consider 10 of the 13 objects for which the confidence level is larger than 3.

\subsection{GALEX UV data}
We include near-UV (NUV, 1770-2730 \AA) and far-UV (FUV, 1350-1780 \AA) data from the Galaxy Evolution Explorer\footnote{\url{http://galex.caltech.edu}} (GALEX). The instrument provides simultaneous co-aligned FUV and NUV images with spatial resolution 4.3 and 5.3 
arcseconds respectively, with FOV of $\approx$ 1.2$^{\circ}$. The zero-points have been taken from GALEX Observers's Guide\footnote{\url{http://galexgi.gsfc.nasa.gov/docs/galex/Documents/ERO_data_description_2.htm}}, ZP$_{FUV}$ = 18.82 mag$_{AB}$ and ZP$_{NUV}$ = 20.08 mag$_{AB}$.

We use data taken during the nearby galaxy survey (NGS) and the guest investigator program (GII)  and publicly released in the GR4. In total we make use of three pointings, GI1\_109010\_NGC4459, NGA\_Virgo\_MOS09, and NGA\_Virgo\_MOS11, with an total exposure time of $\approx$ 8.4 ks in NUV and $\approx$ 3.0 FUV.
After retrieving the data from the web-interface \texttt{MAST\footnote{\url{http://galex.stsci.edu/GR4/}}}, we further process them with public available tools\footnote{\url{http://astromatic.iap.fr/}} (SCAMP, SWarp) and obtain resampled and stacked images.

\begin{figure*}
\centering
\includegraphics[width=15.5cm]{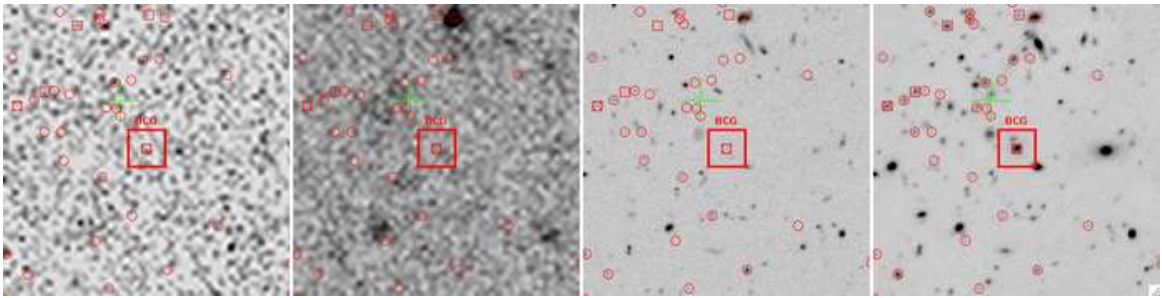}
\caption{1.5\arcmin$\times$1.5\arcmin zoom at the BCG (large red box) in FUV-, NUV-GALEX, LBT-U and LBT-i band, overlaid (red circle) galaxies used in our photometric analysis. Small red boxes mark objects with spectroscopic redshifts.} 
\label{FigXMMUJ1230UV}
\end{figure*}

The GALEX data are rather shallow. Nevertheless they are useful to improve the quality of the photometric redshifts, in particular to disentangle SED degeneracies in the blue part of the spectra as detailed in Gabasch et al. \cite{gabasch08}. Figure \ref{FigXMMUJ1230UV} shows that some galaxies can clearly be detected in the GALEX filters (FUV,NUV), e.g. foreground spiral galaxy above the brightest cluster galaxy (BCG) shows an NUV-excess.

%

\subsection{X-ray data}
\begin{table*}
\begin{center}
\caption{Cluster X-ray data overview.}
\label{tab:fieldpos}
\begin{tabular}{lcccccccl}
\hline \hline
Object & Telescope & Instrument & Filter/Mode & total exposure & clean exposure &  Net counts$^a$ & Obs-Id & comments  \\
~                 &         ~          &        ~           &        &   ~     &  [s]                  & ~     & ~ & ~\\
\hline
\texttt{XMMU J1230.3+1339}  & Chandra & ACIS-S & VFAINT & 37681 & 36145 & 639 & 9527 & chip S3\\
\hline
\end{tabular}
\end{center}
Notes: $^a$ number of net detected counts in the 0.5 - 2 keV band.
\end{table*}
%

\subsubsection{Chandra observations}
\label{sec:xrayobs}
We use the archival observation by Chandra (PI: Sarazin, Obs-Id 9527), as listed in Table 3.
The cluster centre lies on the S3 chip of the Advanced CCD Imaging Spectrometer (ACIS) array.  The observation was conducted in VFAINT mode, and reprocessed accordingly using the Chandra software\footnote{\url{http://cxc.harvard.edu/ciao}} CIAO 4.1 (Friday, December 5, 2008) and related calibration files, starting from the level 1 event file. We used the CIAO 4.1 tools ACIS\_PROCESS\_EVENTS, ACIS\_RUN\_HOTPIX to flag and remove bad X-ray events, which are mostly due to cosmic rays. A charge transfer inefficiency (CTI) correction is applied during the creation of the new level 2 event file\footnote{\url{http://cxc.harvard.edu/ciao/threads/createL2/}}.
This kind of procedure reduces the ACIS particle background significantly compared to the standard grade selection\footnote{\url{http://asc.harvard.edu/cal/Links/Acis/acis/Cal_prods/vfbkgrnd}}, whereas source X-ray photons are practically unaffected (only about 2\% of them are rejected, independent of the energy band, provided there is no pileup\footnote{\url{http://cxc.harvard.edu/ciao/ahelp/acis_pileup.html}}).
Upon obtaining the data, aspect solutions were examined for irregularities and none were detected.  Charged particle background contamination was reduced by excising time intervals during which the background count rate exceeded the average background rate by more than 20\%, using the CIAO 4.1 LC\_CLEAN routine, which is also used on the ACIS background files\footnote{\url{http://cxc.harvard.edu/ciao/threads/acisbackground/}}. Bad pixels were removed from the resulting event files, and they were also filtered to include only standard grades 0, 2, 3, 4, and 6.  After cleaning, the available exposure time was 36145 seconds.

Images, instrument maps and exposure maps were created in the 0.5 - 7.0 keV band using the CIAO 4.1 tools DMCOPY, MKINSTMAP and MKEXPMAP, binning the data by a factor of 2.  Data with energies above 7.0 keV and below 0.5 keV were excluded due to background contamination and uncertainties in the ACIS calibration, respectively.
Flux images were created by dividing the image by the exposure map, and initial point source detection was performed by running the CIAO 4.1 tools WTRANSFORM and WRECON on the 0.4-5.0 keV image.  These tools identify point sources using ``Mexican Hat'' wavelet functions of varying scales.  An adaptively smoothed flux image was created with CSMOOTH.  
This smoothed image was used to determine the X-ray centroid of the cluster, which lies at an RA, Dec of 12:30:17.02, +13:39:00.94 (J2000).  The uncertainty associated with this position can be approximated by the smoothing scale applied to this central emission (2.5$\arcsec$).  Determined in this manner, the position of the X-ray centroid is 15$\pm$ 2.5 $\arcsec$ (110.85$\pm$ 18.5 h$_{72}^{-1}$ kpc) from the BCG of XMMU J1230.3+1339.  Figure~\ref{Fig1bXMMUJ1230} shows an LBT image of XMMU J1230.3+1339 with adaptively smoothed X-ray flux contours overlayed.

\subsection{Radio data}
For the identification of X-ray point sources we make use of Faint Images of the Radio Sky at 20-cm from the FIRST survey (see Becker et al.\citealt{becker95}). 
Using the NRAO Very Large Array (VLA) and an automated mapping pipeline, images with 1.8\arcsec pixels, a typical rms of 0.15 mJy, and a resolution of 5\arcsec are produced, and we retrieve this data through the web-interface\footnote{\url{http://third.ucllnl.org/cgi-bin/firstcutout}}. 

Figures~\ref{Fig1aXMMUJ1230} and \ref{Fig1bXMMUJ1230} show LBT images of XMMU J1230.3+1339 with smoothed radio contours overlaid, see Paper I for more details.

\section{Optical analysis}
\label{sec:optanal}

In this section we give an overview of the catalogue creation, the photometric redshift estimation, 
and the galaxy classification. Furthermore we describe in detail the construction of luminosity maps and the colour-magnitude relation. 
\subsection{Photometric Catalogues}
\begin{figure}
\centering
\includegraphics[width=8.cm]{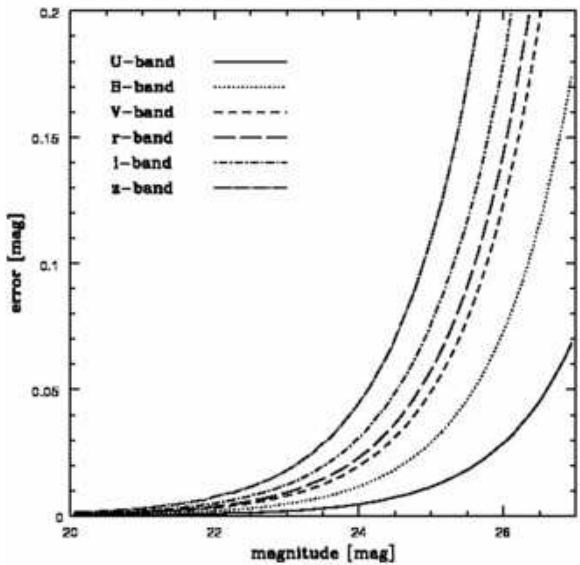}
\caption{Photometric error as a function of apparent magnitude in the 6 (U, B, V, r, i, z) filters.
}
\label{FigCatErr}
\end{figure}

For the creation of the multi-colour catalogues we first cut all images in the filter U, B, V, r, i, z to the same size. After this we measure the seeing in each band to find the band with the worst seeing. We then degrade the other bands with a Gaussian filter to the seeing according to the worst band. For the object detection we use \SExtractor in dual-image mode, using the unconvolved $i$-band image as the detection image, and measure the fluxes for the colour estimations and the photometric redshifts on the convolved images. We use a Kron-like aperture (\texttt{MAG\_AUTO}) for the total magnitudes of the objects, and a fixed aperture (\texttt{MAG\_APER}) of $\approx$ 2\arcsec diameter (corresponding to $\approx$15 kpc at the cluster redshift) for the colours of the objects. Figure \ref{FigCatErr} shows the median photometric error as a function of apparent magnitude for the different filters.
The magnitude error is from \SExtractor which does not consider correlated noise. Comparing the formal magnitude error with the 50\% detection limits suggests a scaling of a factor of two.
To account for Galactic extinction we apply the correction factor  $E(B-V) = 0.022$ retrieved from the NASA Extragalactic Database\footnote{\url{http://nedwww.ipac.caltech.edu/}}, from the dust extinction maps of Schlegel et al. \cite{schlegel98}, to our photometric catalogue.

\begin{figure}
\centering
\includegraphics[width=8.cm]{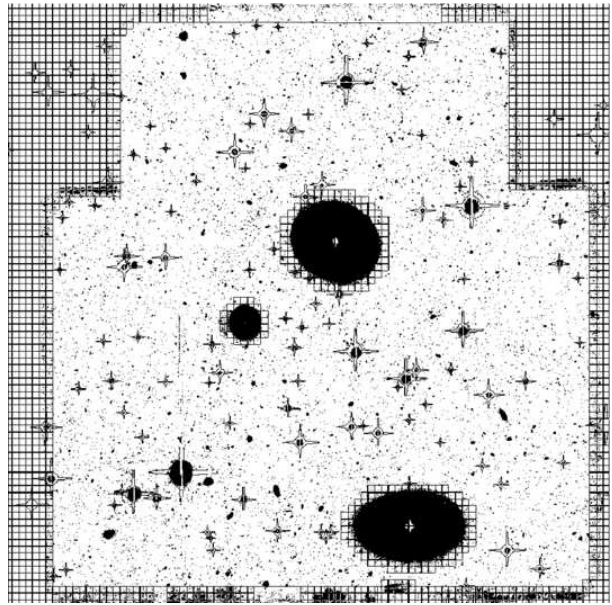}
\caption{Semi-automatic image masking for the 26\arcmin$\times$26\arcmin LBT field. The polygon squares mark objects from the density analysis and the stars cover sources identified in the GSC-1, GSC-2 and USNO-A2 Standard Star Catalogues, see Erben et al. (2009) for further details.}
\label{FigImaMask}
\end{figure}

To account for extended halos of very bright stars, diffraction spikes of stars, areas around large and extended galaxies and various kinds of image reflections we generate a mask, shown in Figure \ref{FigImaMask}. We do not consider objects within these masks in the photometric and shape catalogue later on.

\begin{figure*}
\includegraphics[width=4cm]{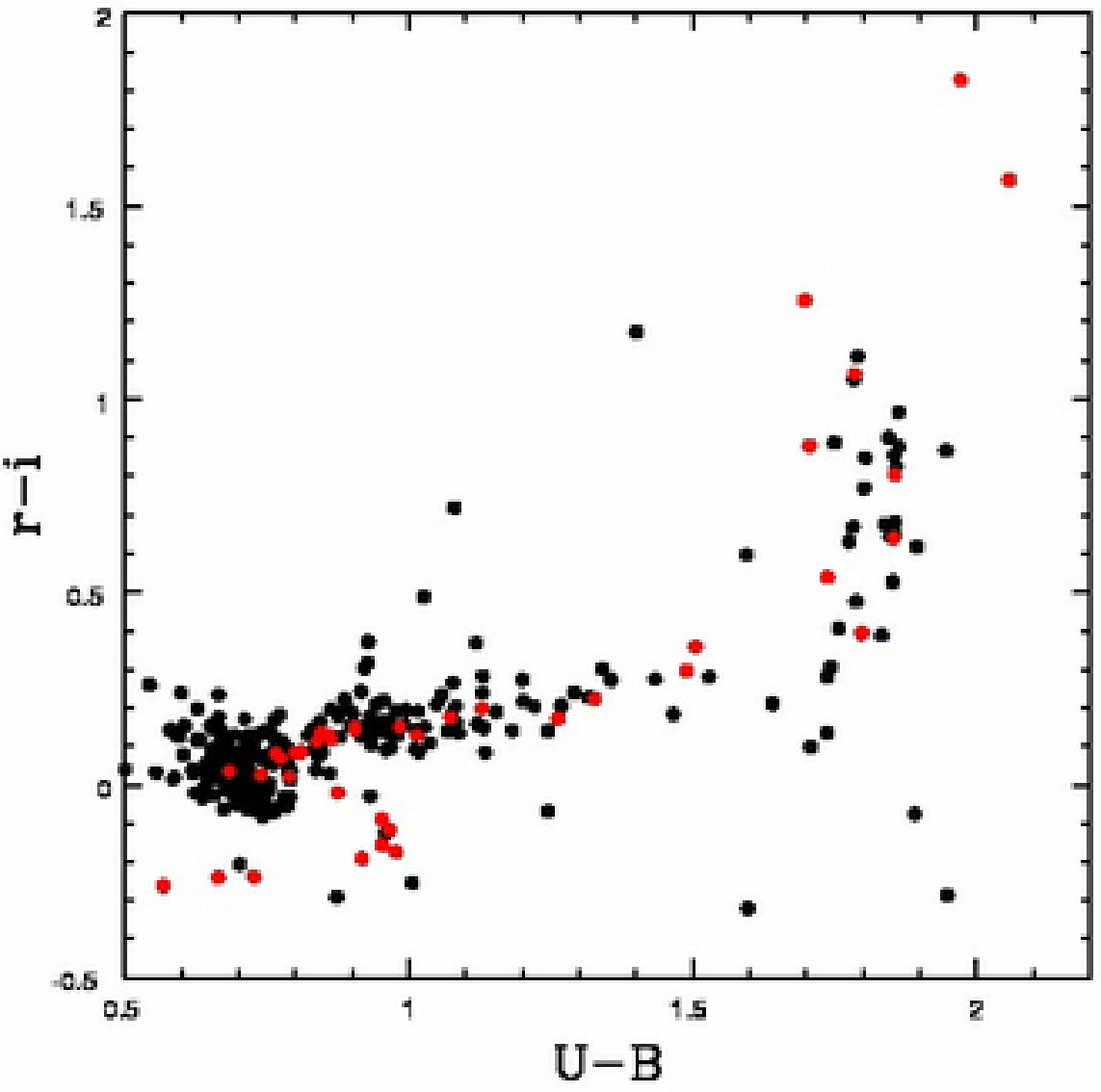}
\includegraphics[width=4cm]{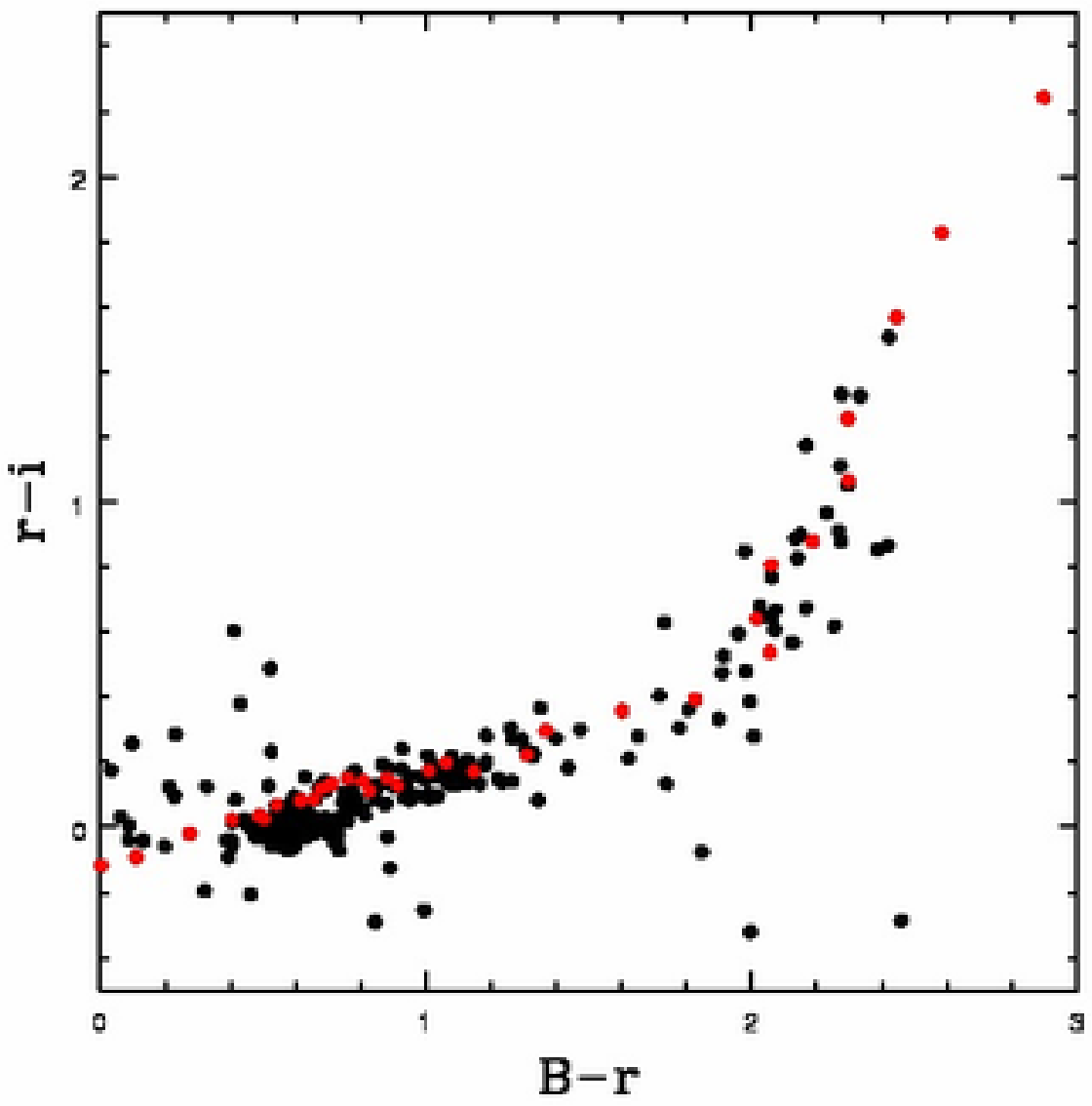}
\includegraphics[width=4cm]{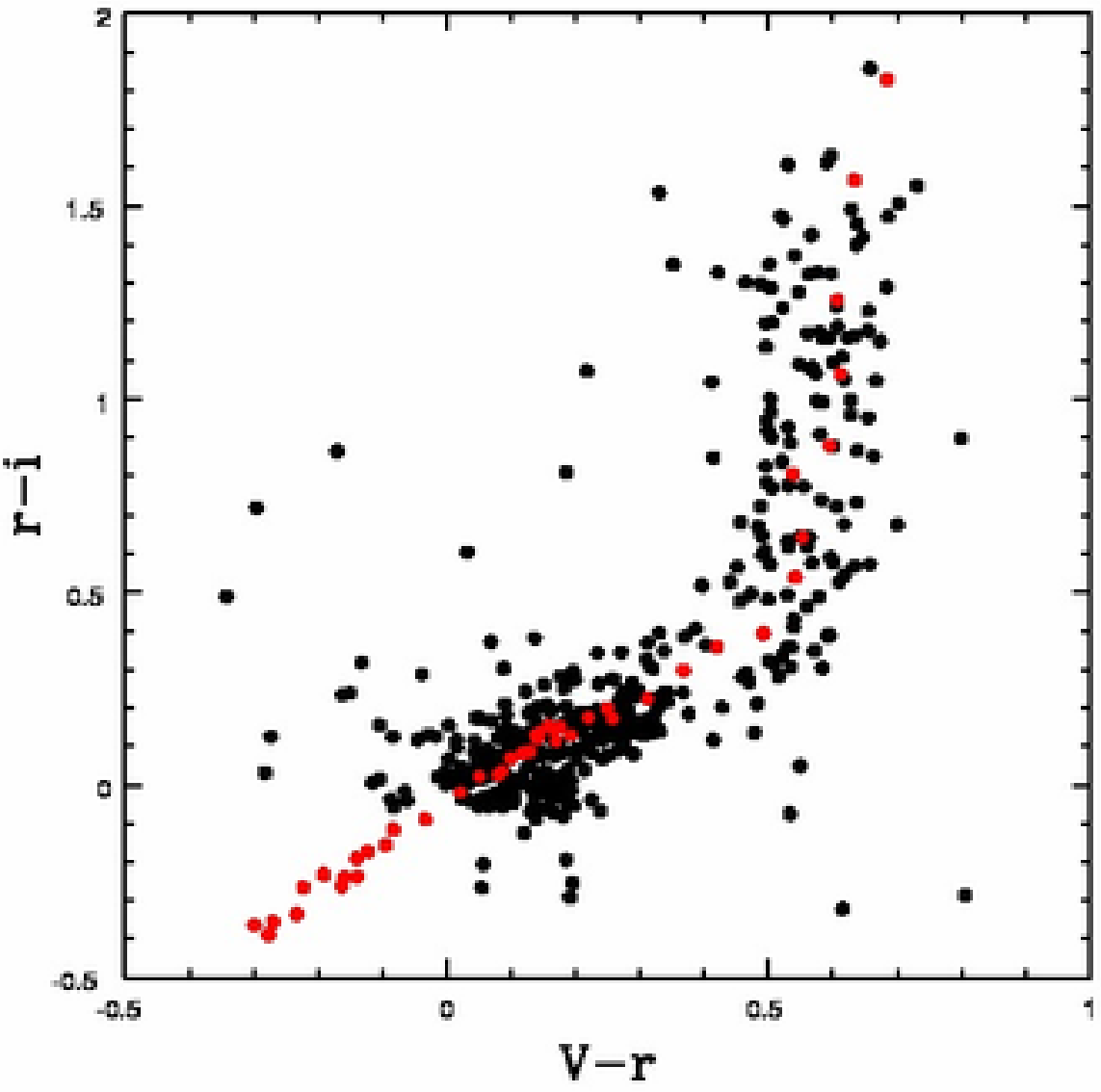}
\includegraphics[width=4cm]{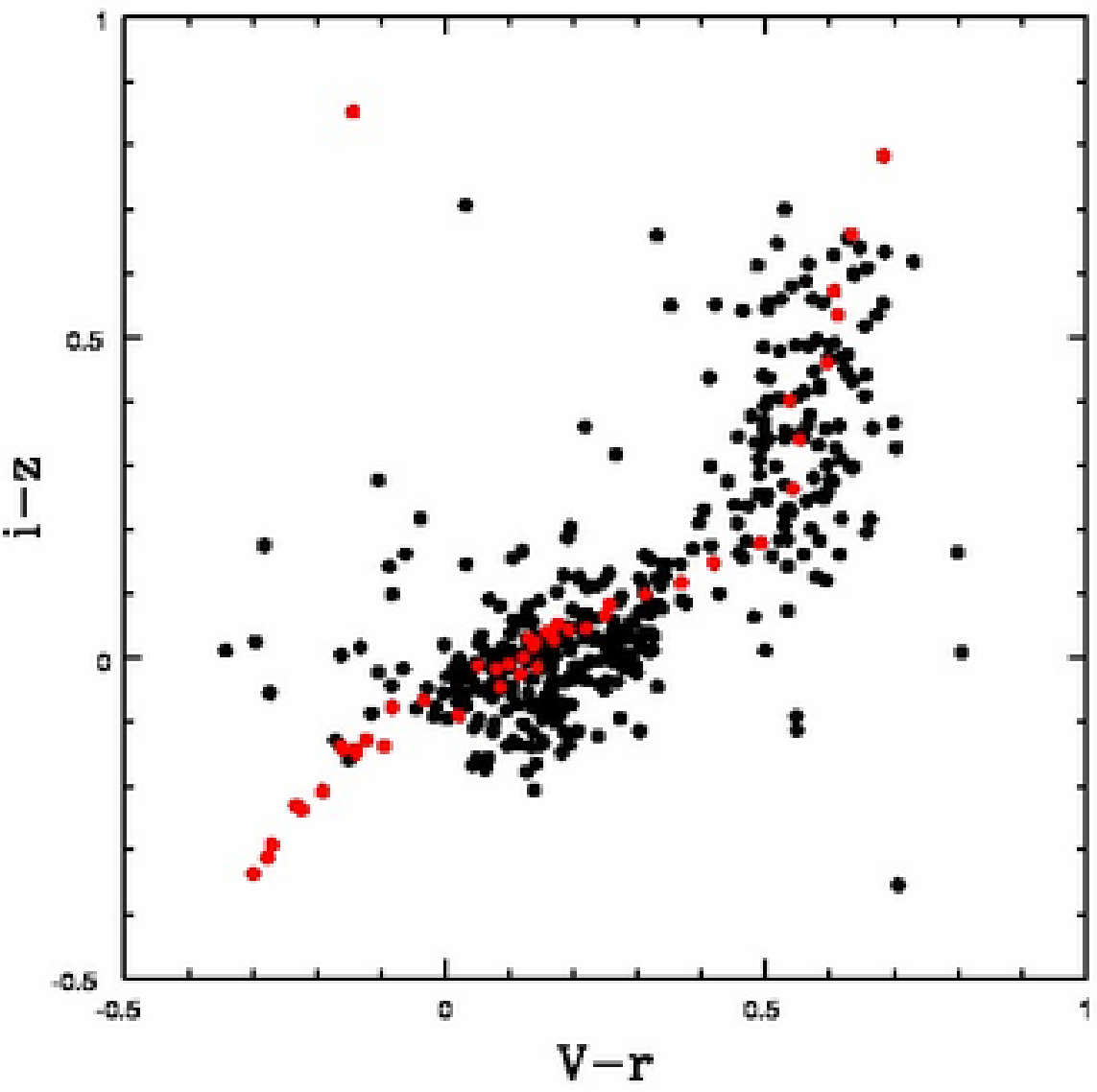}
\centering
\caption{ Colour-colour diagrams of stars (black dots), plotted against the Pickles (1998) stellar library (red dots). The stars used in this plot are selected with 17.0 $<$ mag $<$ 23.0 in all bands. The photometric errors (as shown in Figure 7) are of the order of 0.02 mag. 
All colours perform well.
To obtain these colour-colour diagrams we have applied zeropoint corrections to all the bands, in detail -0.12, 0.02, 0.01, -0.15, 0.00 and 0.10 magnitudes for U,B,V, r, i, and z bands. 
 }
\label{FigCC}
\end{figure*}
%

\subsection{Photometric Redshifts}
We use the Bayesian \texttt{PHOTO-z} code (Bender et al.\citealt{bender01}) to estimate the photometric redshifts. 
The method is similar to that described by Ben\'itez et al.\cite{benitez00}. The redshift probability for each SED type is obtained from the likelihood according to the $\chi^2$ of the photometry of the object relative to the SED model, and from a prior describing the probability for the existence of an object with the given SED type and the derived absolute magnitude as a function of redshift.
The code has been successfully applied on the MUNICS survey (Drory et al.\citealt{drory04}), the FORS Deep Field (Gabasch et al.\citealt{gabasch04a} \&\citealt{gabasch06}), the COSMOS (Gabasch et al.\citealt{gabasch08}), the Chandra Deep Field (Gabasch et al.\citealt{gabasch04b}), and the CFHTLS (Brimioulle et al.\citealt{brimioulle08}).

\begin{figure}
\centering
\includegraphics[width=8.cm]{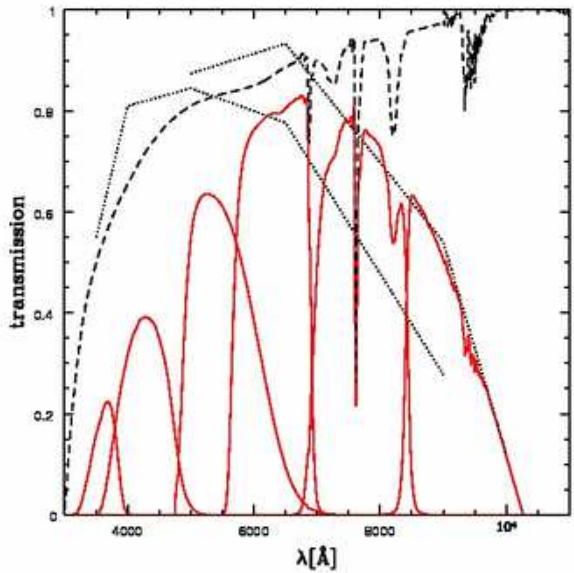}
\caption{Transmission curves of the 6 (Bessel-like U,B,V from LBCB; Sloan-like r,i,z from LBCR) filters. The dotted lines show the CCD efficiency of the LBCB (left) and LBCR (right). LBCB has the more blue sensitive curve. The dashed line shows the atmospheric absorption at the observatory site. The red lines correspond to the effective filter transmission curves, corrected for CCD efficiency and atmospheric absorption.}
\label{FigFilter}
\end{figure}

Small uncertainties in the photometric zero-point and an imperfect knowledge of the filter transmission curve (see Figure \ref{FigFilter}) have to be corrected. We use the colours of stars from the Pickles (1998) library for these adjustments. We compare the colours for this library of stars with the observed colours and apply zeropoint shifts to achieve a matching. We fix the i-band filter and shift the remaining ones by -0.12,0.02,0.01,-0.15 and 0.10 for the U,B,V,r and z band filters.
With additional spectroscopic information we can improve the calibration for the photometric redshift code by slightly adjusting the zero-point offsets for each band again. The match of our observed colours after zero-point shift to the Pickles (1998) stellar library colours is shown in Figure \ref{FigCC}.

Using all position matchable spectra (31), leads to an accuracy of $\sigma_{\triangle z/(1+z)}=0.074$ and the fraction of catastrophic outliers is $\sim 13\%$, as plotted in Figure \ref{FigPhotSpecComp}. The dotted lines are for $z_{phot}=z_{spec}\pm 0.15 ~(1+z_{spec})$. The fraction of catastrophic outliers is defined as $\eta=|z_{spec}-z_{phot}|/(1+z_{spec}) > 0.15$. $\sigma_{\triangle z/(1+z)}$ is the redshift accuracy measured with the normalised median absolute deviation, $1.48~ \times ~median (|\triangle z|/(1+z))$. Using only the secure spectra (17) which fulfill the quality criteria Flag $\ge$ 3 ($\ge$ 95 \% confidence, see Paper I) the accuracy becomes $\sigma_{\triangle z/(1+z)}=0.042$ with no catastrophic outliers. 

\begin{figure}
\centering
\includegraphics[width=9cm]{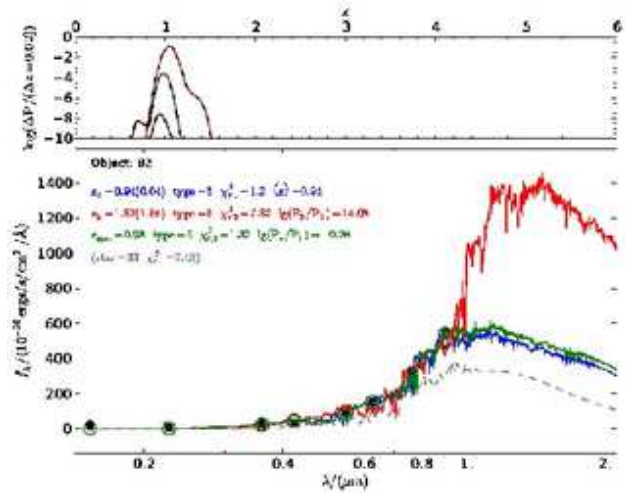} 
\caption{Redshift probability distribution (upper panel) and photometric redshift fit (lower panel) of the spectroscopic object 82, the BCG with $z_{\rm spec}=0.9786$. The different SED's are colour coded, the open circles represent the SED integrated within the various filter transmission curves, whereas the black dots represent the measured fluxes. The redshift for the best fitting SED is $z_{phot}=0.94\pm0.04$, which is in agreement with the spectroscopic one.} 
\label{FigPhotSpecLik}
\end{figure}   
\begin{figure*}
\centering
\includegraphics[width=8.cm]{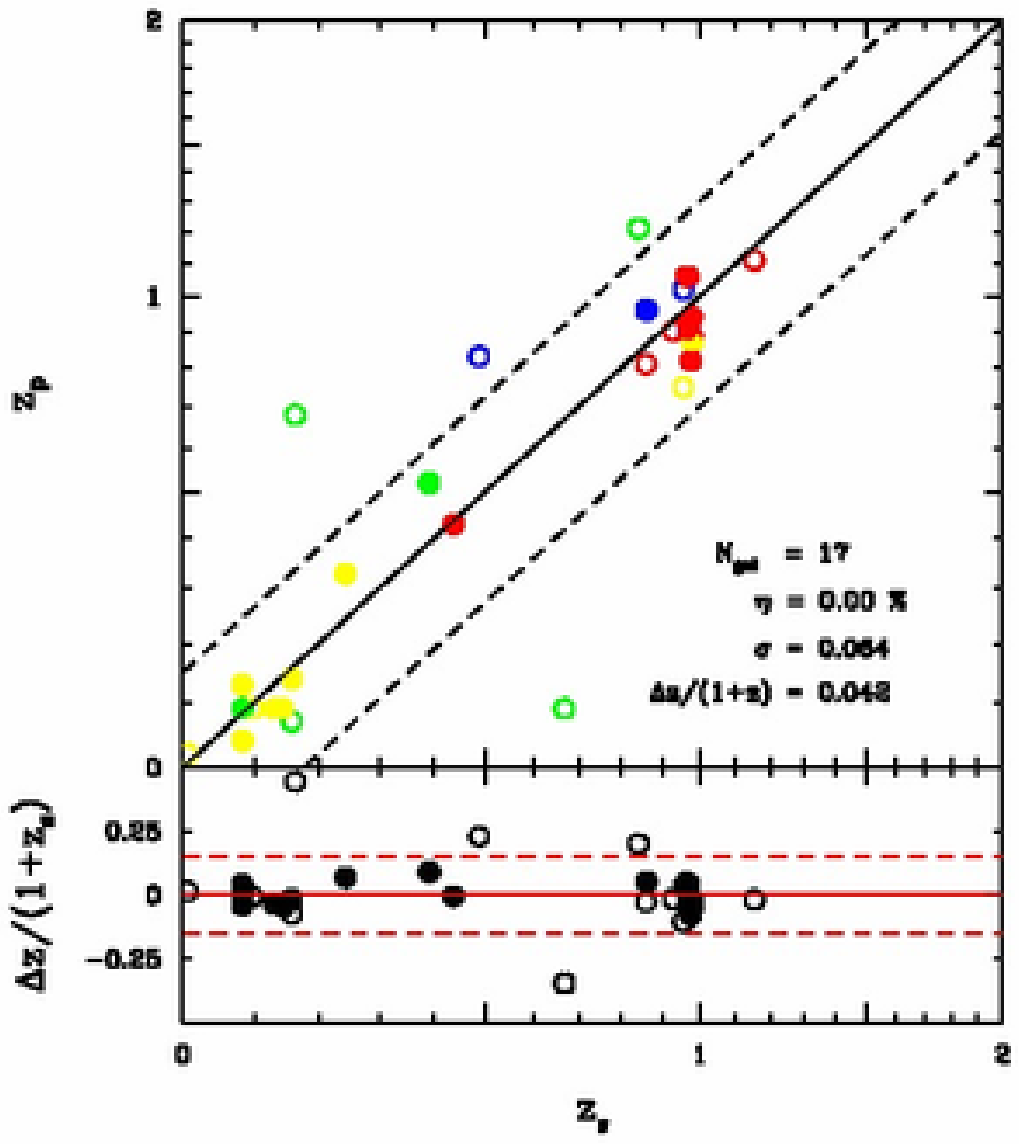}
\includegraphics[width=8.cm]{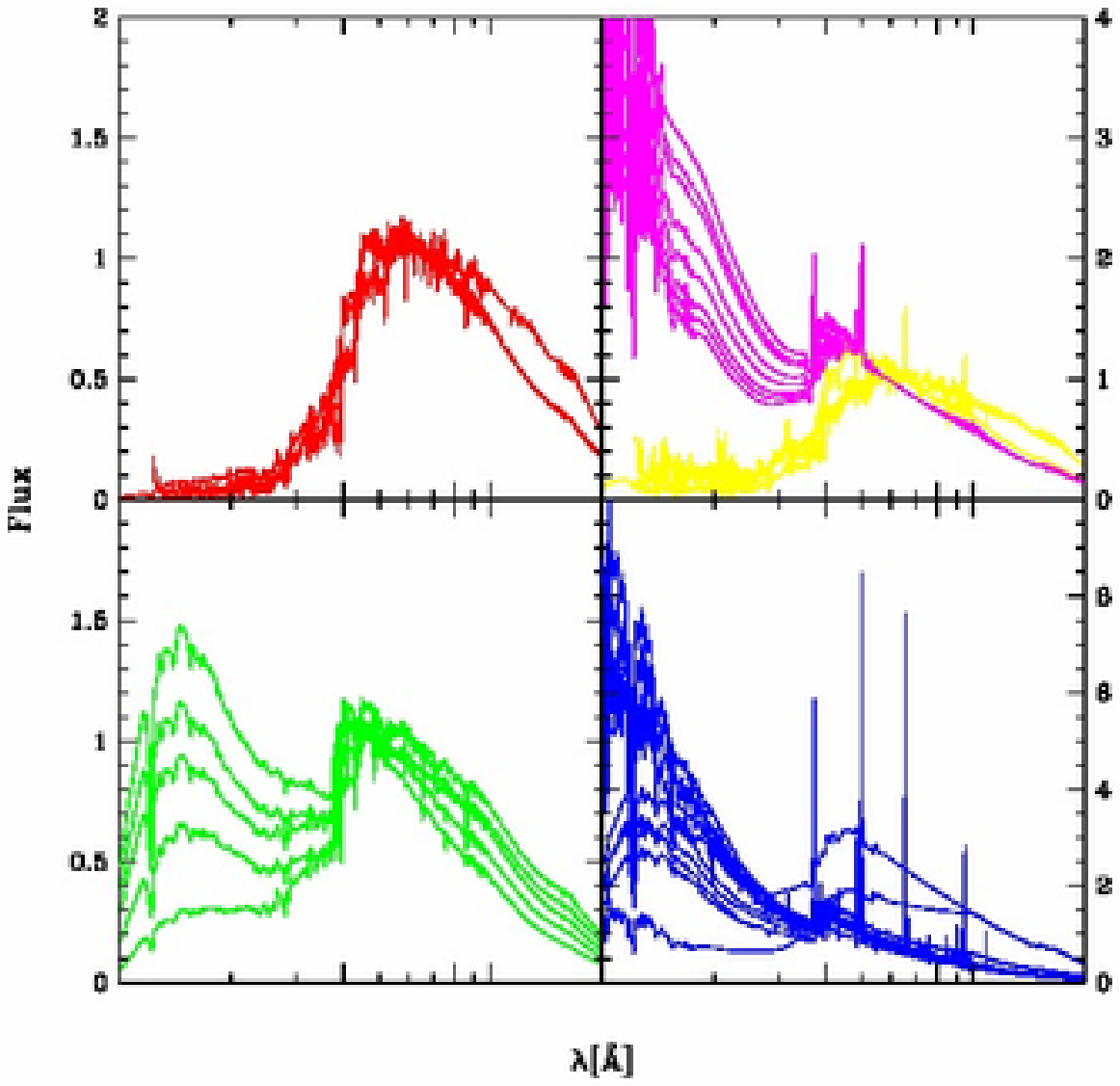}   
\caption{Photometric redshifts, plotted against the spectroscopic ones for the cluster field around XMMU J1230.3+1339. To distinguish the galaxy spectral types (see SED template plotted in the right panel) we use a colour-code: red dots represent early type galaxies (e.g., ellipticals, E/S0), yellow, magenta and green dots represent spiral like galaxy types (e.g., Sa/Sb,Sbc/Sc), and blue dots represent strong star-forming galaxies (e.g., Sm/Im \& SBm). The median residual is $(z_{phot}-z_{spec})/(1+z_{spec})=0.042$,  $\sigma_{\triangle z/(1+z)}=0.074$ and $\eta=0 \%$ for all secure matchable spectra of the cluster XMMU J1230.3+1339. Note that the filled circle mark secure spectroscopic redshifts.} 
\label{FigPhotSpecComp}
\end{figure*}   
\begin{figure}
\centering
\includegraphics[width=8.cm]{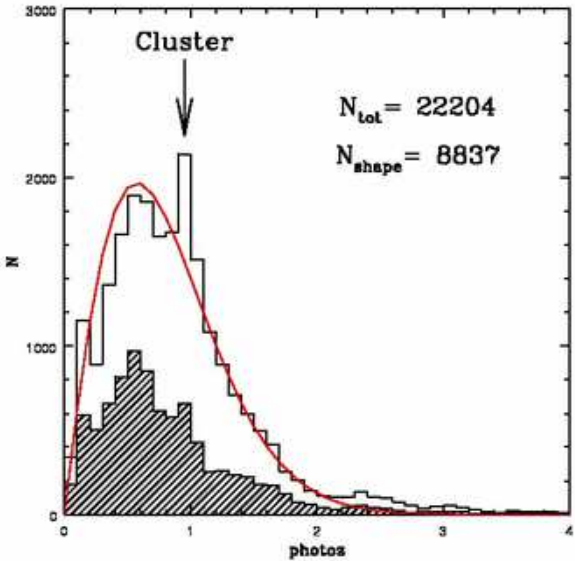}  
\caption{Redshift number distribution of all galaxies. The number distribution is shown as black histogram for the $26\arcmin\times26\arcmin$ field, with a median redshift $z_{median}=0.77$. The solid red curve represents a fit to the galaxy distribution, using the approach of Van Warbeke et al. (2001). The arrow indicates the galaxy cluster. The shaded distribution show the objects used for weak lensing analysis lateron. These are all objects for which photo-z and shape could be measured within a FOV 26\arcmin$\times$26\arcmin and outside the masked area of Figure 8.}
\label{FigHistoGal}
\end{figure} 

In Figure \ref{FigHistoGal} the galaxy redshift histogram of all objects in the field is shown. The galaxy redshift distribution can be parameterised, following Van Waerbeke et al.\citealt{waerbeke01}:
\begin{equation}
n(z_s)=\frac{\beta}{z_0 \Gamma\bigl(\frac{1+\alpha}{\beta}\bigr)}\bigl(\frac{z_s}{z_0}\bigr)^{\alpha} {\rm exp}\bigl[-\bigl(\frac{z_s}{z_0}\bigr)^{\beta}\bigr],
\end{equation}
where $(z_0, \alpha, \beta)$ are free parameters. The best fitting values for the cluster field are $z_0 = 0.77$, $\alpha = 1.0$, and $\beta = 1.6$. For the fit we have replaced the peak at $z=1$ by the average value of the lower and higher bin around the cluster. The fit is shown in Figure \ref{FigHistoGal}.

\begin{figure*}
\centering
\includegraphics[width=18cm]{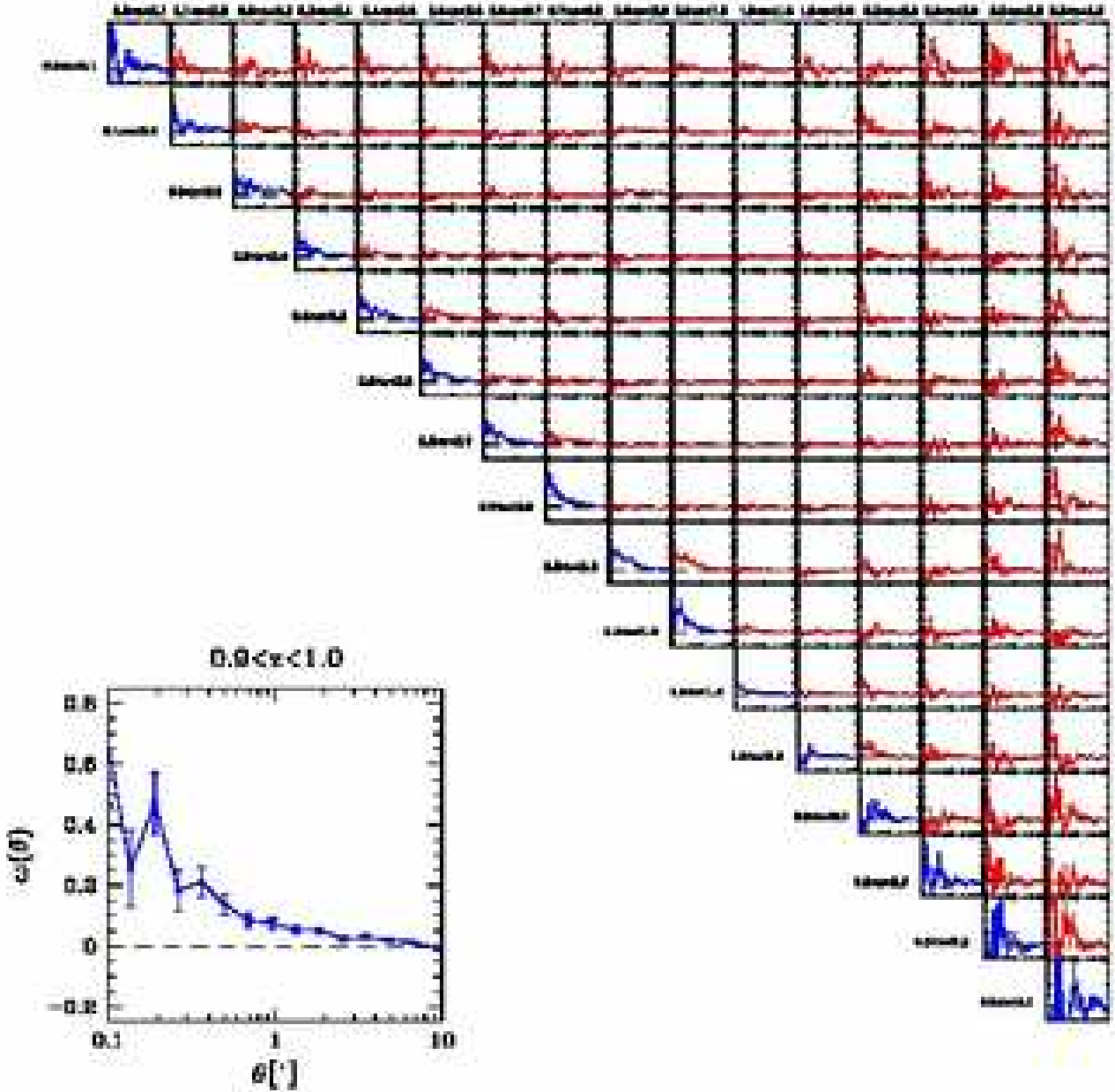}
\caption{Two-point angular-redshift correlation of galaxies in the field of the cluster XMMU J1230.3+1339 with $14.0<i_{AB}<26.0$ selected from the total sample as shown in Figure 13. The inlay shows as an example the auto-correlation for the redshift bin $0.9<z_{phot}<1.0$; the scale of the angular distance is logarithmic from 0.1 to 10 arcmin, whereas the correlation function scales linearly from -0.25 to 0.85 . The blue curves show the auto-correlation of each redshift-bin. The redshift bins start from 0 and increase by a stepsize of 0.1 until redshift 1.0 and stepsize of 0.5 from redshift 1.0 to redshift 4. The red curves show the cross-correlation of galaxy subsample pairs at different redshifts.}
\label{FigCorrelGal}
\end{figure*}

Since the number of spectroscopic data for this field is still limited we use another method to estimate the photometric redshift accuracy: we measure the two-point angular-redshift correlation of galaxies in photometric redshift slices with 14.0$<$ i$_{\rm AB}$ $<$26.0 selected from the total sample as shown in Figure \ref{FigHistoGal}, and show the results in Figure \ref{FigCorrelGal}. 
In this way the null hypothesis of non-overlapping redshift slices can be tested (see Erben et al.\citealt{erben09} and Benjamin et al.\citealt{benj10} for a detailed description of this technique).
For the calculation of the two-point correlation function we use the publically available code \texttt{ATHENA}\footnote{\url{www2.iap.fr/users/kilbinge/athena/}}.
The blue curves in Figure \ref{FigCorrelGal} show the autocorrelation of each redshift-bin. The red data points show the cross correlation of different redshift slices, which should be compatible with zero in absence of photometric redshift errors.
The redshift binning starts from 0 and increases with a stepsize of 0.1 up to redshift 1.0, and stepsize of 0.5 from redshift 1.0 to redshift 4. 
Nearly all cross-correlation functions of non-neighbouring photo-z slices show indeed an amplitude consistent with or very close to zero. This is a strong argument for the robustness of the photometric redshifts. No excessive overlap between low- and high-z slices is observed. The weak signal in the high redshift slice is due to the poor statistics in this and the neighbouring bins. Note that the last two bins contain less than 200 galaxies each. For the weak lensing analysis (Sect. \ref{sec:wlanal}) the vast majority of galaxies studied is limited to z = [1.0 - 2.5]. The correlation of all redshift intervals with galaxies in z $\epsilon$ [1.0 - 2.5] is very small.

\begin{figure*}
\centering
\includegraphics[width=15cm]{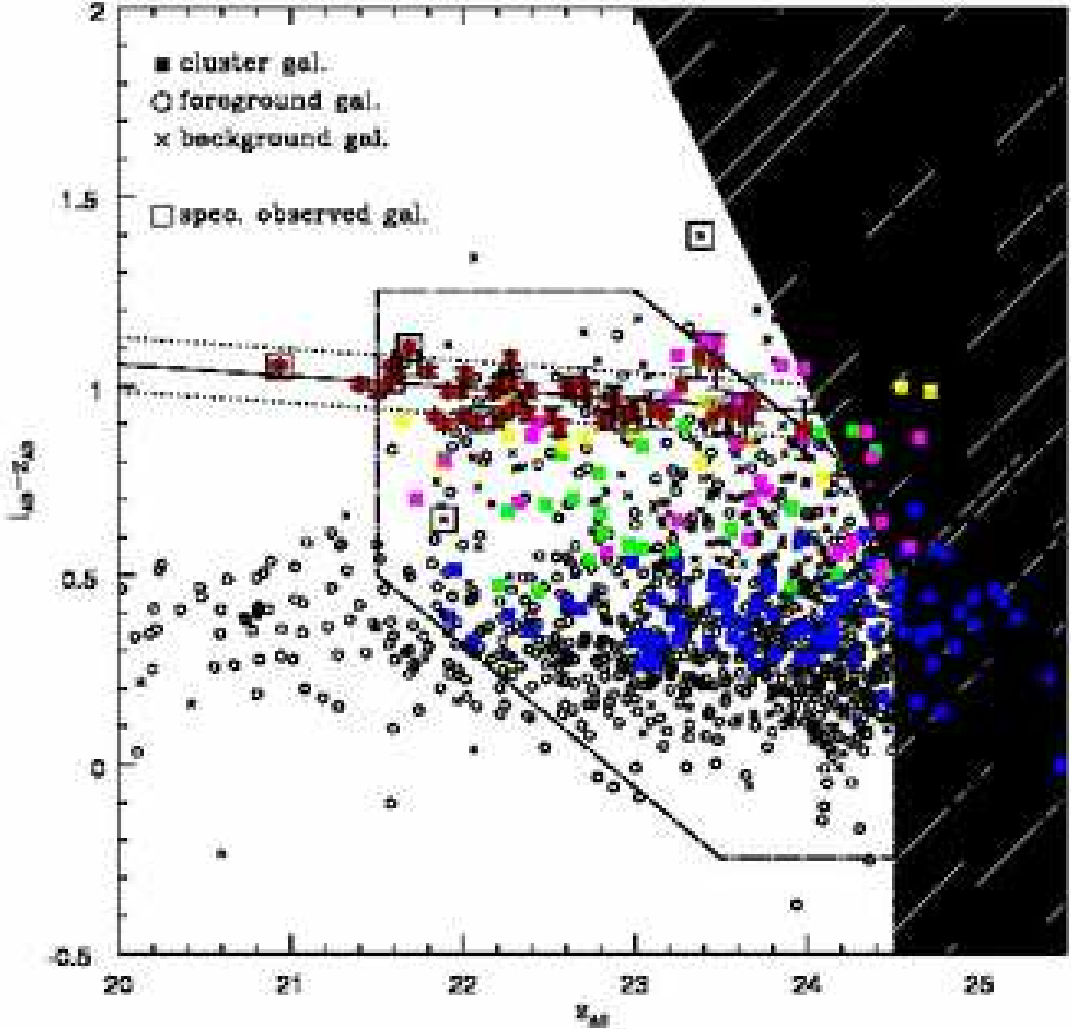}
\caption{Colour-magnitude relation, $i_{\rm AB}-z_{\rm AB}$ against $z_{\rm AB}$, FOV 6\arcmin$\times$6\arcmin, the dashed lines show the $5 \sigma$ limiting magnitudes and colours. The cluster members, are selected to lie at $z_{cluster}-0.05 \le z_{phot} \le z_{cluster}+ 0.05$. We use the same colour coding as shown in Figure 12. The black solid line refers to the linear fit to the passive cluster members, reproduced with red dots. The dotted lines correspond to the $3 \sigma$ region. The shaded region on the right indicates $>$ 50 \% incompleteness for a point source within a 2\arcsec aperture, which is roughly equal the 50\% completeness for an extended galaxy. Black boxes mark spectroscopically confirmed cluster galaxies. The dashed-dotted limited region marks the area where we select background galaxies for the WL analysis.} 
\label{FigXMMUJ1230_CMR}
\end{figure*}       
\begin{figure*}
\centering
\includegraphics[width=8cm]{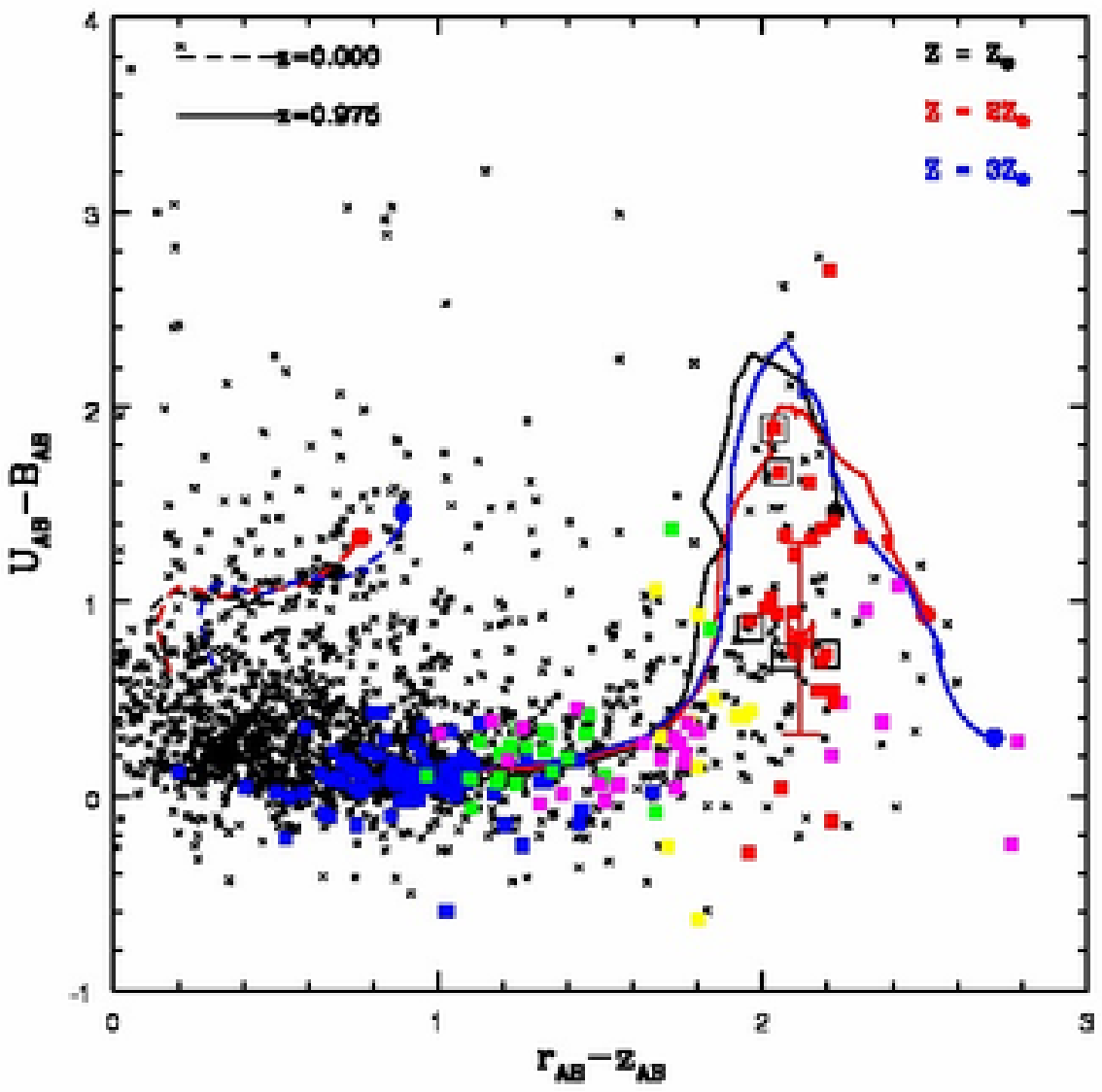}\hspace{1cm}
\includegraphics[width=8cm]{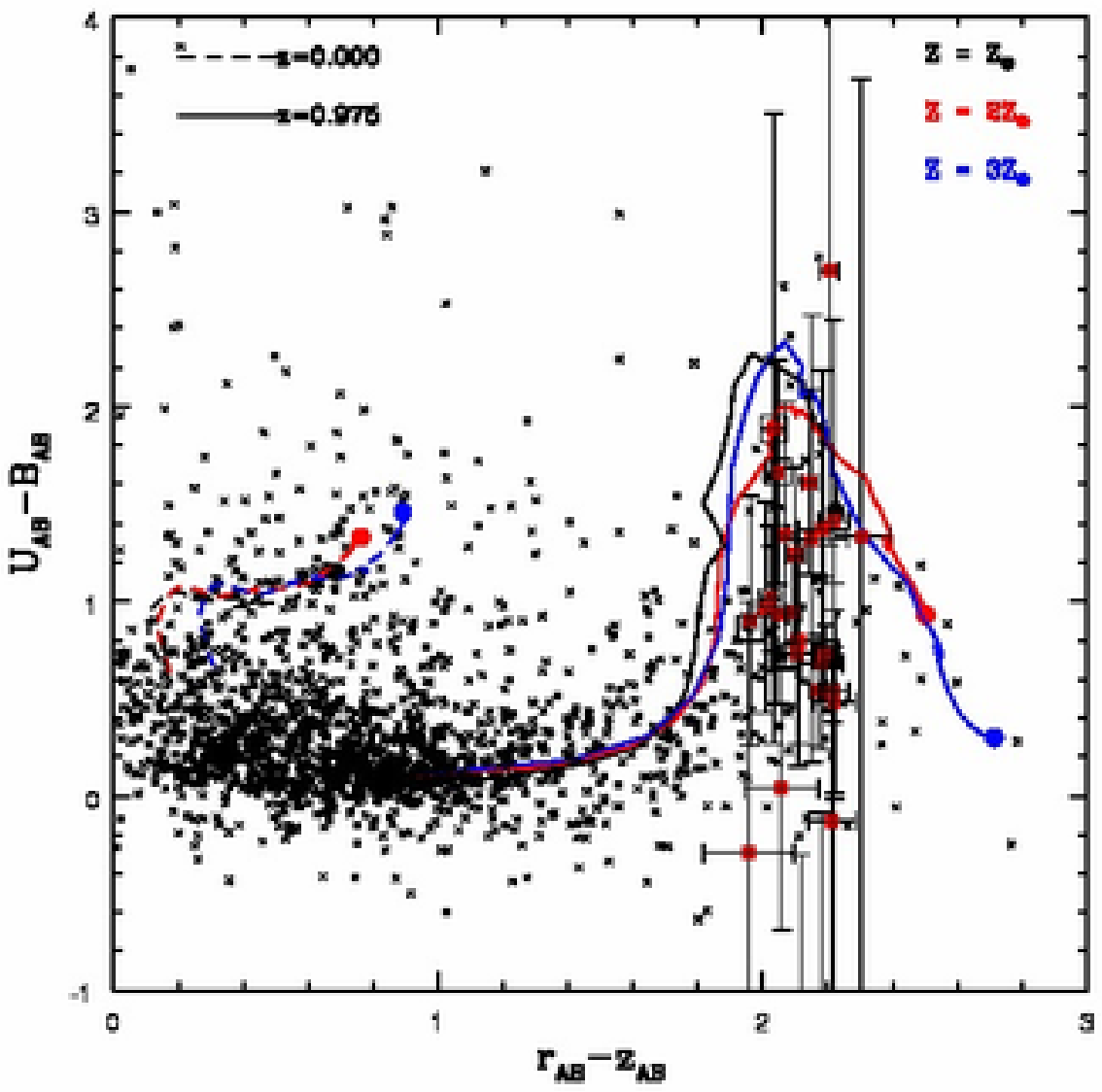}  
\caption{Colour-colour diagram with $U_{\rm AB}-B_{\rm AB}$ against $r_{\rm AB}-z_{\rm AB}$. The same FOV, photometric redshift selection and colour coding as in Figure 15 are used. On the right panel we show the colour-errors for the early type galaxies.
Overplotted we show single stellar population (SSP) models for passively evolving galaxies with Salpeter-like IMF and for three different metallicities (1, 2 and 3 times solar). The tracks show the time evolution from 100 Myr (filled circles) to 6 Gyr in steps of 100 Myr for two redshifts (0.0 and 0.975).
Black boxes in both figures mark spectroscopically confirmed cluster galaxies.} 
\label{FigXMMUJ1230_CCD}
\end{figure*}       
%

\subsection{Light distribution}
To quantify the light distribution we compute the absolute K-band luminosity for all the galaxies. Based on the SED fitting, performed during the photo-z estimation, we estimate for a given filter curve (K-band) the flux and the absolute magnitude. Next we map the flux of all galaxies of the cluster in a redshift interval $z_{cluster} \pm 0.05$ on a pixel grid, and smooth it with a circular $2\arcmin$ Gaussian filter.

\begin{figure}
\centering
\includegraphics[width=8.cm]{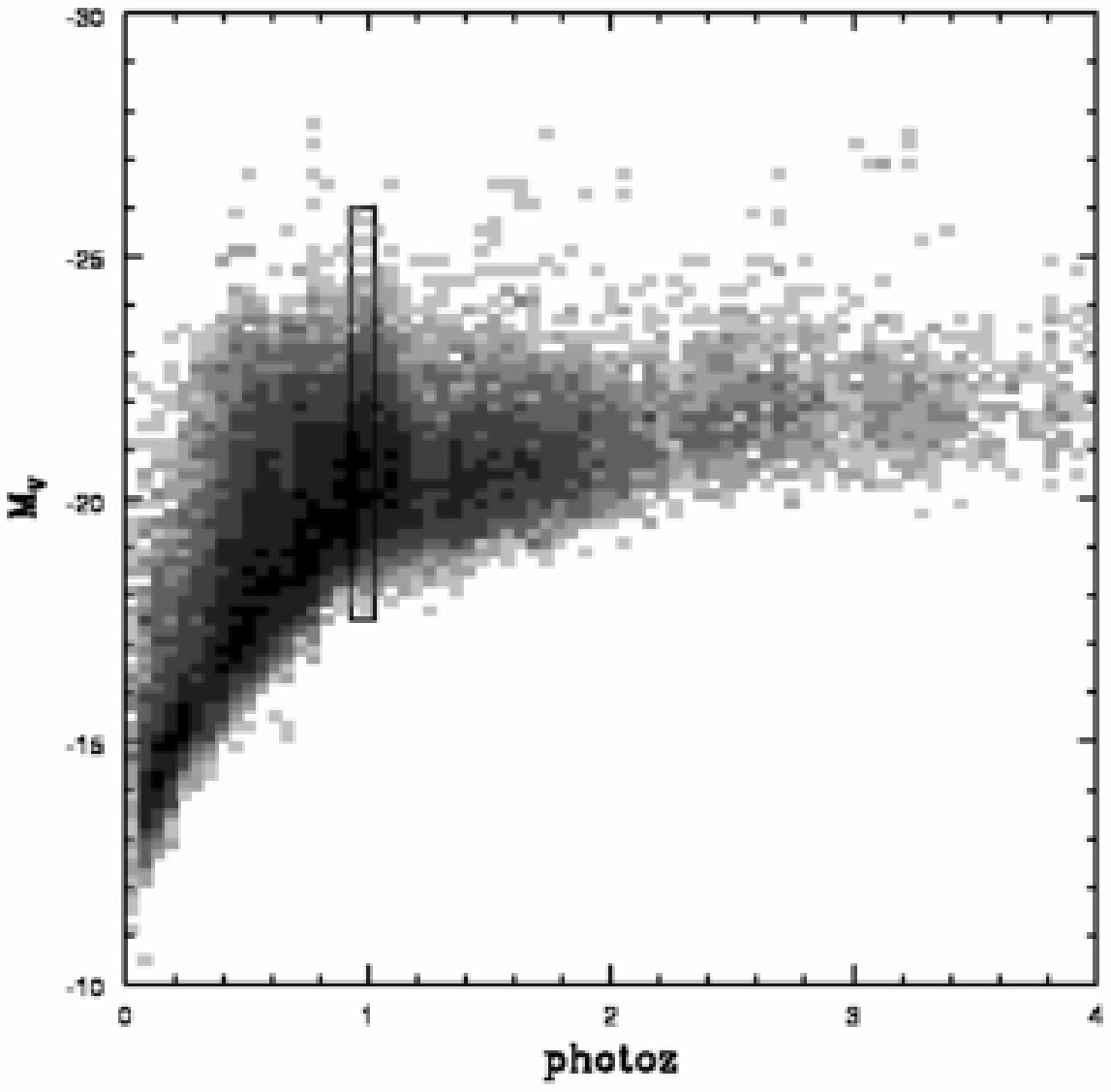}  
\caption{Absolute V-band magnitudes of galaxies in the observed 26\arcmin$\times$26\arcmin field as a function of redshift. The black box shows the region for the selection of cluster members  later on.} 
\label{FigPhotMagComp}
\end{figure}   

Figure~\ref{Fig1aXMMUJ1230} shows an LBT image of XMMU J1230.3+1339 with surface brightness contours overlayed. The cluster seems to be very massive, rich and has a lot of substructure, see also Paper I.

\subsection{Colour-Magnitude Relation}
The colour-magnitude relation (CMR, Visvanathan \& Sandage\citealt{visvanathan77}) is a fundamental scaling relation used to investigate the evolution of galaxy populations. The CMR of local galaxy clusters shows a red-sequence (RS, Gladders \& Yee\citealt{gladders00}) formed of massive red elliptical galaxies undergoing passive evolution.
For the following discussion we consider as cluster member those galaxies in a redshift interval $z_{cluster} \pm 0.05$, within a box of 6\arcmin$\times$6\arcmin centered on the X-ray centroid of the cluster. Our sample is neither spectroscopic nor does it use the statistical background subtraction as it is done when purely photometric sample are studied.
As shown in De Lucia et al. \cite{delucia07} for the ESO Distant Cluster Survey (EDisCS), results are not significantly affected whether one does use spectroscopic or photometric redshifts or does not apply photometric redshifts for member selection. Note that the accuracy of our photometric redshifts is better by a factor of 1.5 than those used for EDisCS (see details in Pello et al.\citealt{pello09} and White et al.\citealt{white05}). This is due to the fact that we can use deep 8 band data (FUV, NUV, U, B, V, r, i, z) in contrast to 5 band data (V, R, I, J, Ks) for EDisCS. 

The CMR for XMMU J1230.3+1339 is shown in Figure \ref{FigXMMUJ1230_CMR}. To distinguish the cluster galaxy spectral types we use a colour-code:  red dots represent early type galaxies (e.g., ellipticals, E/S0), yellow and green dots represent spiral like galaxy types (e.g., Sa/Sb,Sbc/Sc), and blue dots represent strongly star-forming galaxies (e.g., Sm/Im and SBm). 

\begin{figure*}
\centering
\includegraphics[width=8.cm]{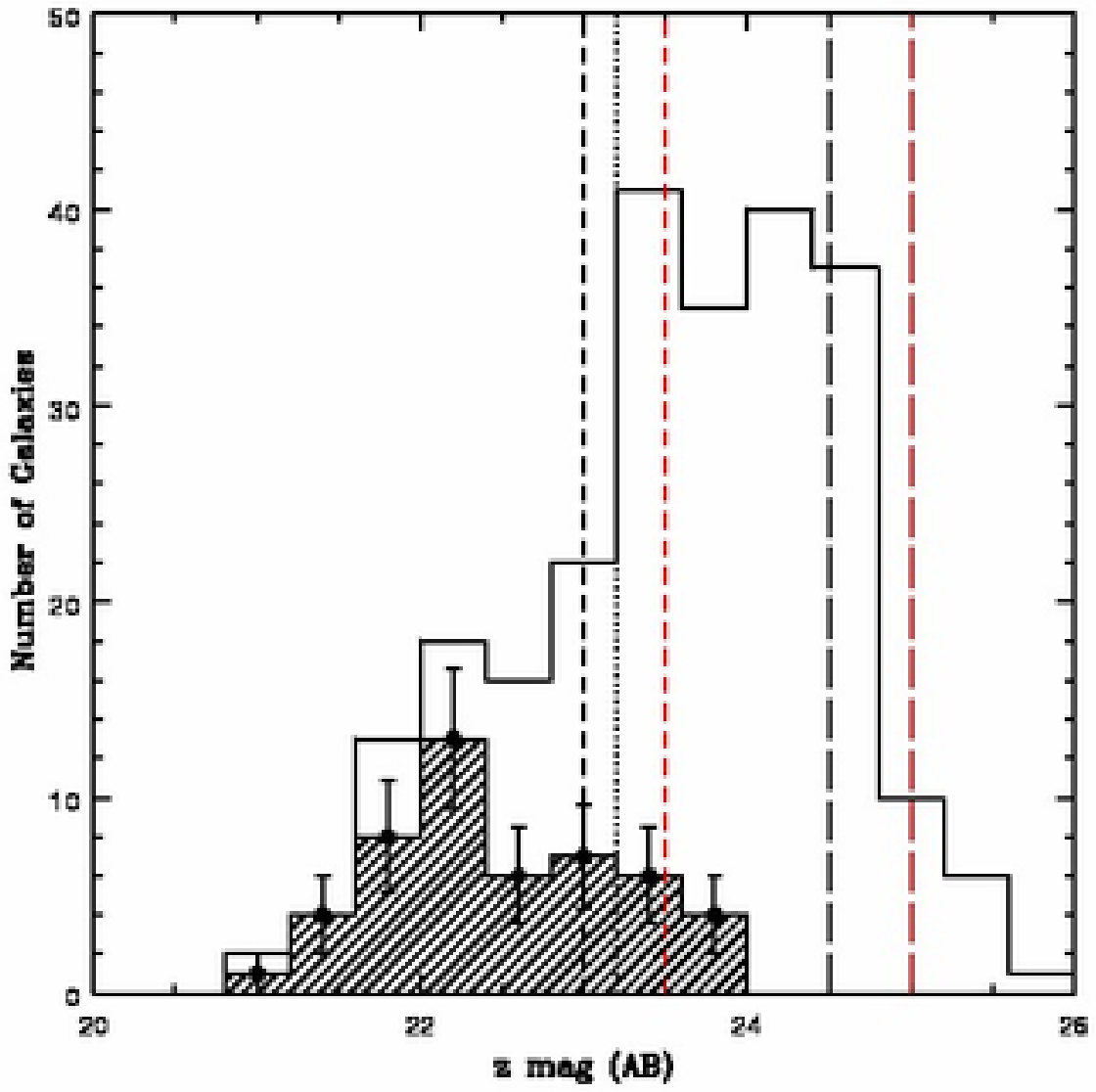}\hspace{1cm}
\includegraphics[width=8.cm]{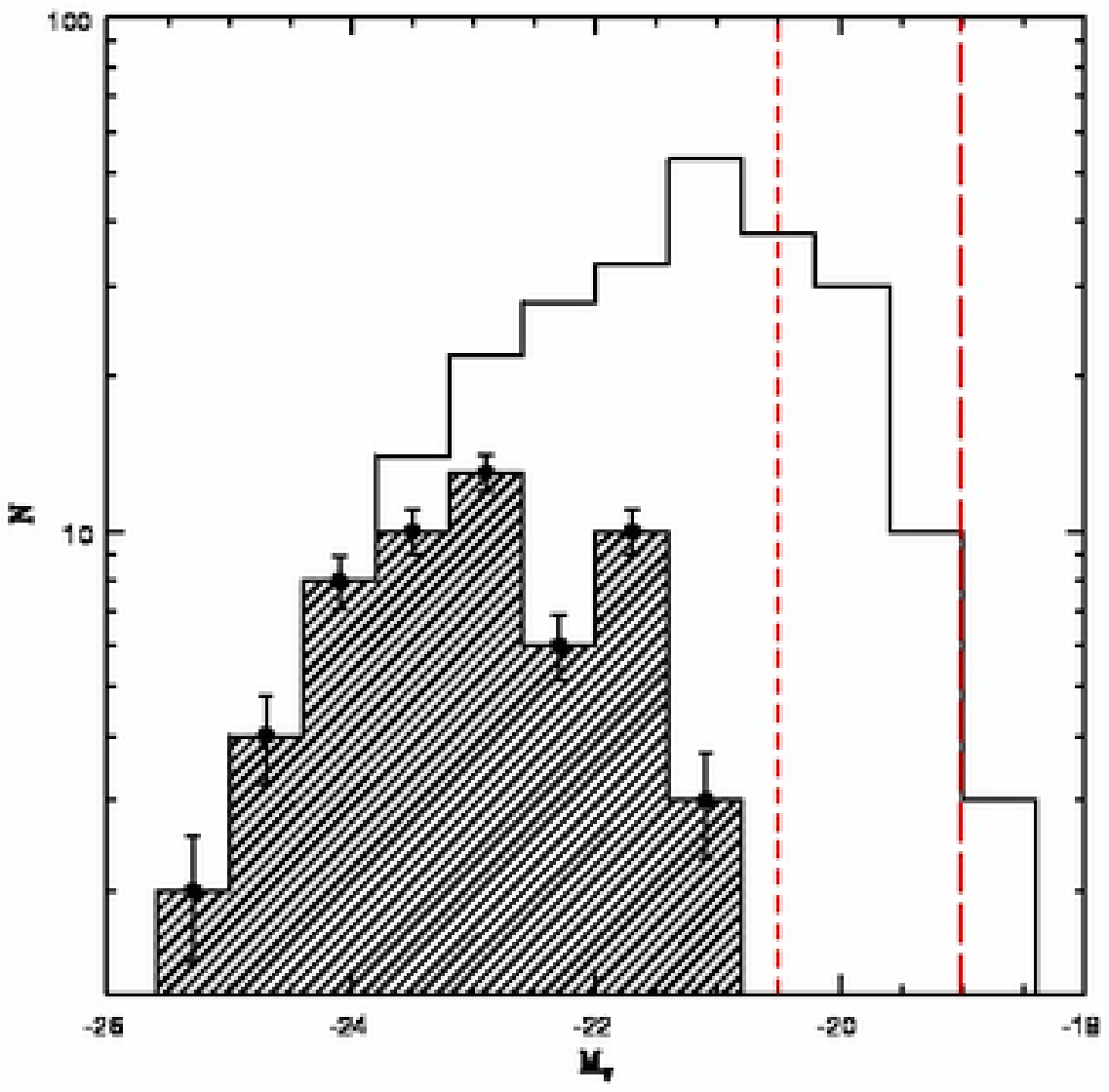}  
\caption{Number counts for the galaxies within a 6\arcmin$\times$6\arcmin region centered on XMMU J1230.3+1339 and a photometric redshift of $z_{cluster}-0.05 \le z_{phot} \le z_{cluster}+ 0.05$. The open and shaded histograms show all cluster galaxies and early type cluster galaxies, respectively. Error bars represent the Poisson errors. The red short (long-dashed) line represent the 100 (50)\% completeness in the i-filter, the detection band, and for comparison in black the z-filter. The dotted line represents $m(z)\approx23$, at which galaxies are classified into 'luminous' and 'faint'. There is a rarity of red galaxies with z $\ge$ 22.2 or i $\ge$ 23.2 or M$_V$ $\ge$ -24.2 where the detection efficiency is still 100\% in both bands.
} 
\label{FigPhotLumFct}
\end{figure*}   
\begin{figure}
\centering
\includegraphics[width=8cm]{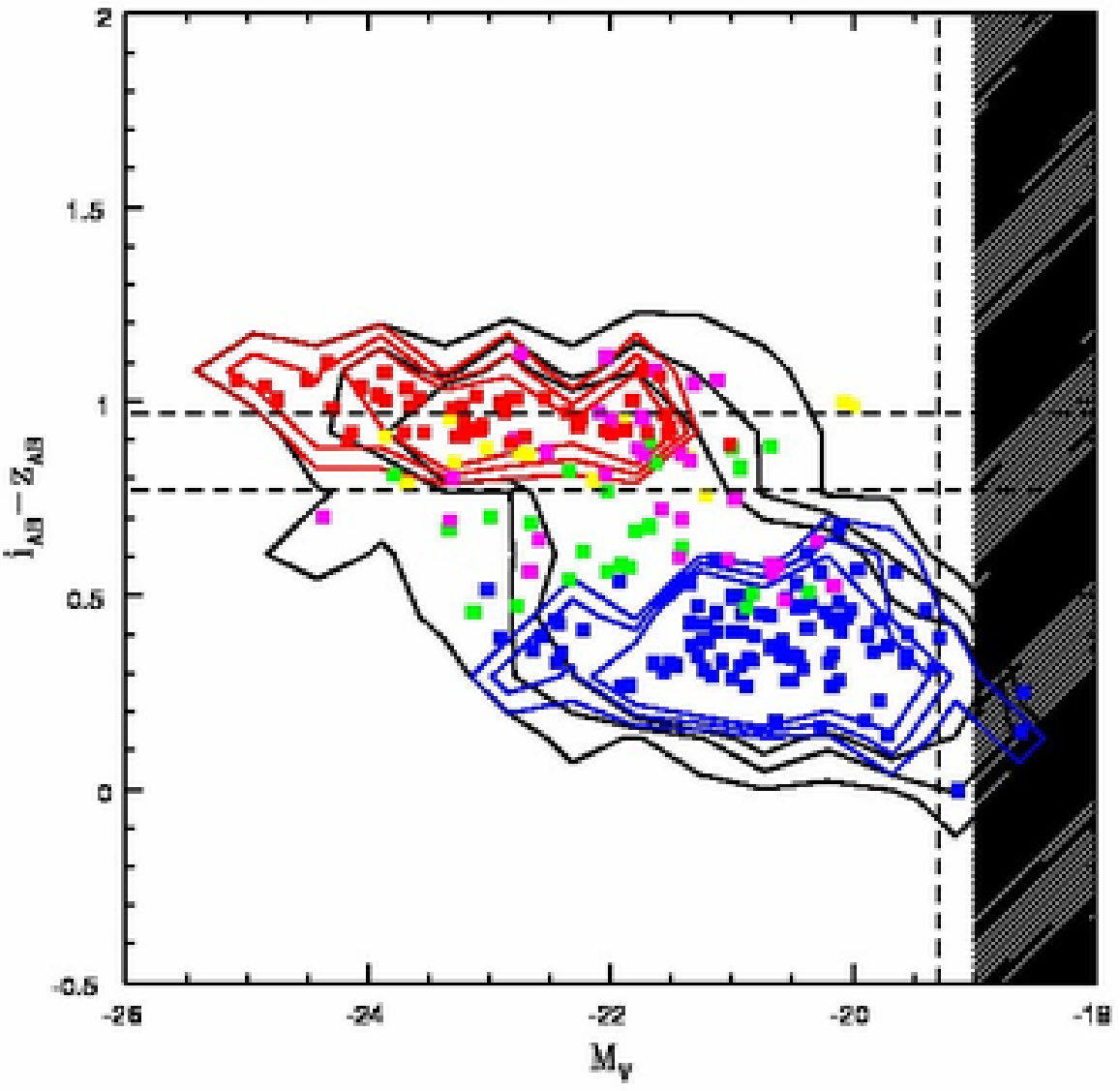}
\caption{The Butcher \& Oemler effect in the colour-magnitude relation, $i_{AB}-z_{\rm ab}$ against $M_{\rm V}$. The dashed lines show the selection box used by Butcher \& Oemler (1984) - see text for details.
We use the same region, redshift selection and colour coding as in Figure 15.
The contours are colour coded: black lines represent all photometric redshift selected cluster members, red and blue contours represent the early and late type cluster population, respectively.
} 
\label{FigXMMUJ1230_BO}
\end{figure}       

The CMR shows a tight red sequence of early type galaxies (ETG). The expected colour for composite stellar population (CSP) models with old age and short e-folding time scales or of old single stellar populations (SSP) is $i_{AB}-z_{AB}\approx$ 1.0 (e.g., see Santos et al.\citealt{santos09}; Demarco et al.\citealt{demarco07}).
We apply a linear fit to the 9 spectroscopic confirmed RS members at $z_{AB}<24$ and obtain $i-z = (-0.078 \pm 0.031)\times (z - 24) + 0.849 \pm 0.065$, a result with poor statistics due to the limited number of spectroscopic data. Therefore we also perform a fit to the 49 photo-z RS members at $z_{AB}<24$ and obtain $i-z = (-0.0273 \pm 0.012)\times (z - 24) + 0.935 \pm 0.015$.
We take the best-fitting RS for the photo-z RS members and plot it in Figure \ref{FigXMMUJ1230_CMR}. The slope is consistent in the errors with that obtained by Santos et al. \cite{santos09} for XMMU J1229+0151, the ``twin brother" cluster at the same redshift observed with ACS onboard the HST.

We can observe spiral and star-forming galaxies (yellow, green, blue data points) until $z_{AB} = 25.5$. The ETG (red data points) can only by detected until $z_{AB} = 24$. Note that the 50\% completeness in z-band is 24.5 mag and in i-band is 25 mag, which is the detection band. 
We observe 7 galaxies with 23.4$\le$ z $\le$ 23.8. If the slope of the luminosity function (LF) is $\alpha \approx$ -1 we expect about 10 galaxies with 23.8$\le$ z $\le$ 24.2. We observe only one in this interval. To explore whether this lack of red galaxies is only apparent, caused by incompleteness  of red the galaxy detection, we have plotted $\le$ 50\% completeness region for the i and z detection as a shaded area in Figure \ref{FigXMMUJ1230_CMR}. We estimate the detection efficiency of galaxies with 23.8$\le$ z $\le$ 24.2 and $i-z \approx$ 1 to be $\approx$ 50\%. Hence we expect to detect about 5 galaxies against the only one found with 23.8$\le$ z $\le$ 24.2.
Furthermore we plot the numbercounts of all cluster galaxies and the ETG ones only as a function of z-band and absolute V-band magnitudes in Figure \ref{FigPhotLumFct}.
We can see that ETG's are less and less starting at $z=22.5$ mag, for which the data are 100\% complete. A similar trend can be seen in the absolute V-band magnitude counts, where ETG's with $M_V<-23.5$ become rare.
Similar results have been reported for the EDisCS clusters in De Lucia et al. (2007) and the cluster RDCSJ0910+54 at $z=1.1$ (Tanaka et al.\citealt{tanaka08}). Demarco et al. (2007) report an indication for truncation of the $i_{775}-z_{850}$ versus $z_{850}$ CMR (see Figure 15 in Demarco et al. 2007) at a magnitude of $z_{850}$=24.5 for RDCSJ1252.9+2927 at $z=1.24$. A similar trend at $z_{850}$=24.0 can be seen in Figure 8 of Santos et al. \cite{santos09} for XMMU J1229+0151 at $z=0.975$.

\subsection{Blue galaxy fraction - Butcher \& Oemler effect}

The fraction of blue galaxy types increases from locally a few percent,  25 \% (z = 0.4) to 40 \% (z $\sim$ 0.8), see e.g. Schade~\cite{schade97}. This is called the Butcher \& Oemler effect.

Butcher \& Oemler \cite{butcher78} defined as blue galaxies those which are by at least $\triangle =0.2$ mag bluer in the $B-V$ rest-frame ($i-z$ observer-frame) than early-type galaxies in the cluster. They then counted red galaxies down to an absolute magnitude $M_V = -19.3$ mag within an arbitrary reference radius and applied a statistical subtraction of the background.
In contrast to Butcher \& Oemler \cite{butcher84} we use a redshift/SED selection as defined in the previous section. Note that it has been shown by De Lucia et al. \cite{delucia07} that a photometric redshift selection does not affect the result. 
As seen in Figure \ref{FigXMMUJ1230_BO} the cluster shows a bimodal galaxy population. The dashed lines in Figure \ref{FigXMMUJ1230_BO} mark the selection boundaries if one would follow exactly the work of Butcher \& Oemler. The colour coding of the galaxy types is shown in Figure \ref{FigXMMUJ1230_BO}. 
We can infer from our photometric analysis a fraction of blue galaxies f$_b$ = 0.514 $\pm$ 0.13 for the cluster and f$_b$ = 0.545 $\pm$ 0.04 for the field at the same redshift, with Poisson errors. Note that for the estimate of the fraction of blue galaxies in the field we take the full 26\arcmin $\times$ 26\arcmin FOV into account and exclude the region around the cluster (6\arcmin $\times$ 6\arcmin). \\
Our result is in good agreement with the prediction from literature.
This can be seen if one extrapolates the blue fraction estimated for the EDisDS sample from $z\approx0.8$ to $z=1$ in De Lucia et al. \cite{delucia07}.
Within the framework of a state-of-the-art semianalytic galaxy formation model Menci et al. \cite{menci08}  derived an equal fraction of blue galaxies for the cluster and the field at z$\sim$1.

\subsection{Colour-Colour Diagram}
We now turn our attention to the colour-colour diagram of the galaxies. In Figure \ref{FigXMMUJ1230_CCD} we show $r-z$ against $U-B$ colours and overlaid single stellar population (SSP) models\footnote{Models computed using \texttt{PEGASE2} (Fioc and Rocca-Volmerange 1997).} for passively evolving galaxies with Salpeter-like IMF and for three different metallicities (1, 2 and 3 times solar).
We expect a $U_{AB}-B_{AB}$ colour of $\approx$ 1.5 from old SSP models with solar metallicity, see tracks in Figure \ref{FigXMMUJ1230_CCD}.
The colour-colour diagram shows $U_{AB}-B_{AB}$ values from -0.5 to 2.5 in the distribution of the elliptical galaxies (red objects in Figure \ref{FigXMMUJ1230_CCD}). Note that the $U_{AB}-B_{AB}$ colours are affected by large U-band measurements errors (see Figure \ref{FigCatErr}).
Nevertheless, we estimate the weighted mean of all ETG in the cluster, which results in a mean colour of $U_{AB}-B_{AB} = 0.81 \pm 0.5$ . This could be a mild indication for a recent star formation (SF) episode. 

The observed U-B color of a galaxy at z$\sim$1 corresponds to a rest-frame far-UV color. 
It may be no surprise the fact that early-type galaxies at this redshift exhibit a rest-frame far-UV color about 0.8 mag bluer than expected from an SSP model. In fact, at least 30\% of the early type galaxies in the local Universe seem to have formed 1\%$-$3\% of their stellar mass within the previous Gyr (Kaviraj et al.\citealt{kaviraj07}). Hence, it is plausible that a larger fraction of the stellar mass of an early type galaxy at z$\sim$1 is formed within a look-back time of 1 Gyr. This would produce significantly bluer far-UV colors in the rest-frame if attenuation by internal dust is negligible. Alternatively, such bluer colors may spot the presence of a significant UV upturn phenomenon (see Yi \citealt{yi08}, for a review). We will investigate the nature of the bluer rest-frame UV colors in early-type galaxies in clusters at z$\le$1 in a future paper.

\section{Weak lensing analysis}
\label{sec:wlanal}
In this section we give an overview on the applied shape measurement technique and on
the lensing mass estimators; we use standard lensing notation. For a broad overview
to the topic, see e.g. Mellier \cite{mellier99}, Bartelmann \& Schneider \cite{bartelmann01} and Hoekstra \& Jain \cite{hoekstra08}.
\subsection{Gravitational shear estimate with the KSB method}
\label{sec:KSB}
The ellipticity of a background galaxy provides an estimate of the trace-free components
of the tidal field $\partial^2\psi/\partial\theta_i\partial\theta_j$, where $\psi(\theta)$ 
is the two-dimensional potential generated by the surface density $\kappa(\theta)$ (Kaiser \& 
Squires \citealt{ks93}) of the mass distribution of the cluster. Any weak lensing measurement relies on the accurate measurement of the average distortion of the shape of background galaxies. Therefore it is crucial to correct for systematic effects, in particular for asymmetric contributions to the point spread function (PSF) from the telescope and from the atmosphere.

The most common technique is the so called KSB method proposed by Kaiser et al. \cite{ksb95}
and Luppino \& Kaiser \cite{luppino97}. In the KSB approach, for each source the following
quantities are computed from the second brightness moments
\begin{equation}
Q_{ij} = \int d^2\theta W_{r_g}(|\theta|) \theta_i \theta_j I(\theta) , \:\: i,j \in \{1,2\},
\end{equation}
where $W_{r_g}$ is a circular Gaussian weight function with filter scale $r_g$
\footnote{We choose $r_g$ to be equal to the \SExtractor parameter \texttt{FLUX\_RADIUS}.}
, and $I(\theta)$ is the surface brightness distribution of an image:\\
(1) the observed ellipticity
\begin{equation}
\epsilon = \epsilon_1 + i\epsilon_2 = \frac{Q_{11} - Q_{22} + 2iQ_{12}}{Q_{11} + Q_{22}},
\end{equation}
(2) the smear polarizability $P^{sm}$ and (3) the shear polarizability $P^{sh}$.\\ 
The KSB method relies on the assumption that the PSF can be described as the sum of an isotropic component which circularises the images (seeing) and an anisotropic part introducing a systematic contribution (for a detailed discussion see e.g. Kaiser et al. \citealt{ksb95} , Luppino \& Kaiser \citealt{luppino97} and Hoekstra et al. \citealt{hoekstra98} - KSB+).

The source ellipticity $\epsilon^s$ of a galaxy after PSF smearing (Bartelmann \& Schneider\citealt{bartelmann01}) is related to the observed one, $\epsilon$, and to the reduced shear g by:
\begin{equation}
\epsilon_i - \epsilon^s_i =  P^{g}_{ij}g_j + P^{sm}_{ij}p^{*}_j,
\end{equation}
where the anisotropy kernel $p$ describes the effect of the PSF anisotropy (an asterisk indicates that parameters are derived from the measurement of stars):
\begin{equation}
p^{*}_i =  (P^{sm\ast})^{-1}_{ij}\epsilon^{\ast}_j.
\end{equation}
The $p$ quantity changes with the position across the image, therefore it is necessary to fit it by,
e.g. a third order polynomial, so that it can be extrapolated to the position of any galaxy.
The term $P^{g}$, introduced by Luppino \& Kaiser \cite{luppino97} as the pre-seeing shear
polarizability, describes the effect of the seeing and is defined as:
\begin{equation}
P^{g}_{ij} = P^{sm}_{ij} \big(\frac{P^{sh}_{ij}}{P^{sm}_{ij}} - \frac{P^{sh\ast}_{im}}{P^{sm\ast}_{mj}}\big),
\end{equation}
with the shear and smear polarisability tensors $P^{sh}$ and $P^{sm}$, calculated from higher-order brightness moments as detailed in Hoekstra et al. \cite{hoekstra98}.
 In Figure \ref{FigKSBrh we plot the distribution of the values of half-light radius $r_h$ and $r_{AB}$ magnitude of the galaxies used later on for our weak lensing analysis.} The computed $P^{sh}/P^{sm}$ values for different bins of $r_h$ and $r_{AB}$ are shown in Figure \ref{FigPSFcorr}.

\begin{figure}
\centering
\includegraphics[width=8.cm]{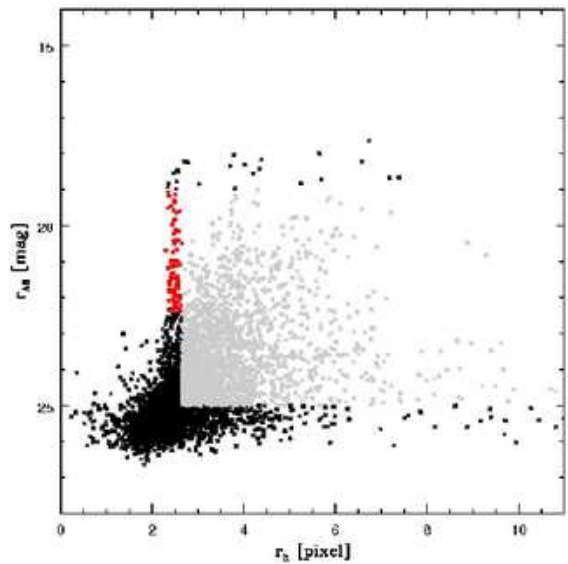}
\caption{The r-band magnitude versus half-light radius. The red zone shows unsaturated stars selected for the PSF anisotropy correction ($19 < r_{\rm AB} < 22.5$ mag, $2.1 < r_{\rm h} < 2.6$ pixel), where one pixel equals 0.224 arcsecond.}
\label{FigStars}
\end{figure}
\begin{figure}
\centering
\includegraphics[width=8.cm]{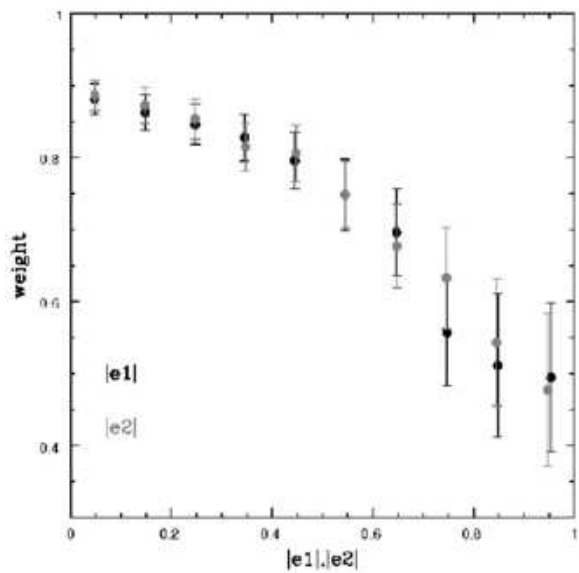}
\caption{Galaxy weight (see Eq. \ref{eq:weight}) as a function of the absolute first and second components of the ellipticity for the weak lensing measurement.}
\label{FigKSBweight}
\end{figure}
\begin{figure*}
\centering
\includegraphics[width=8.cm]{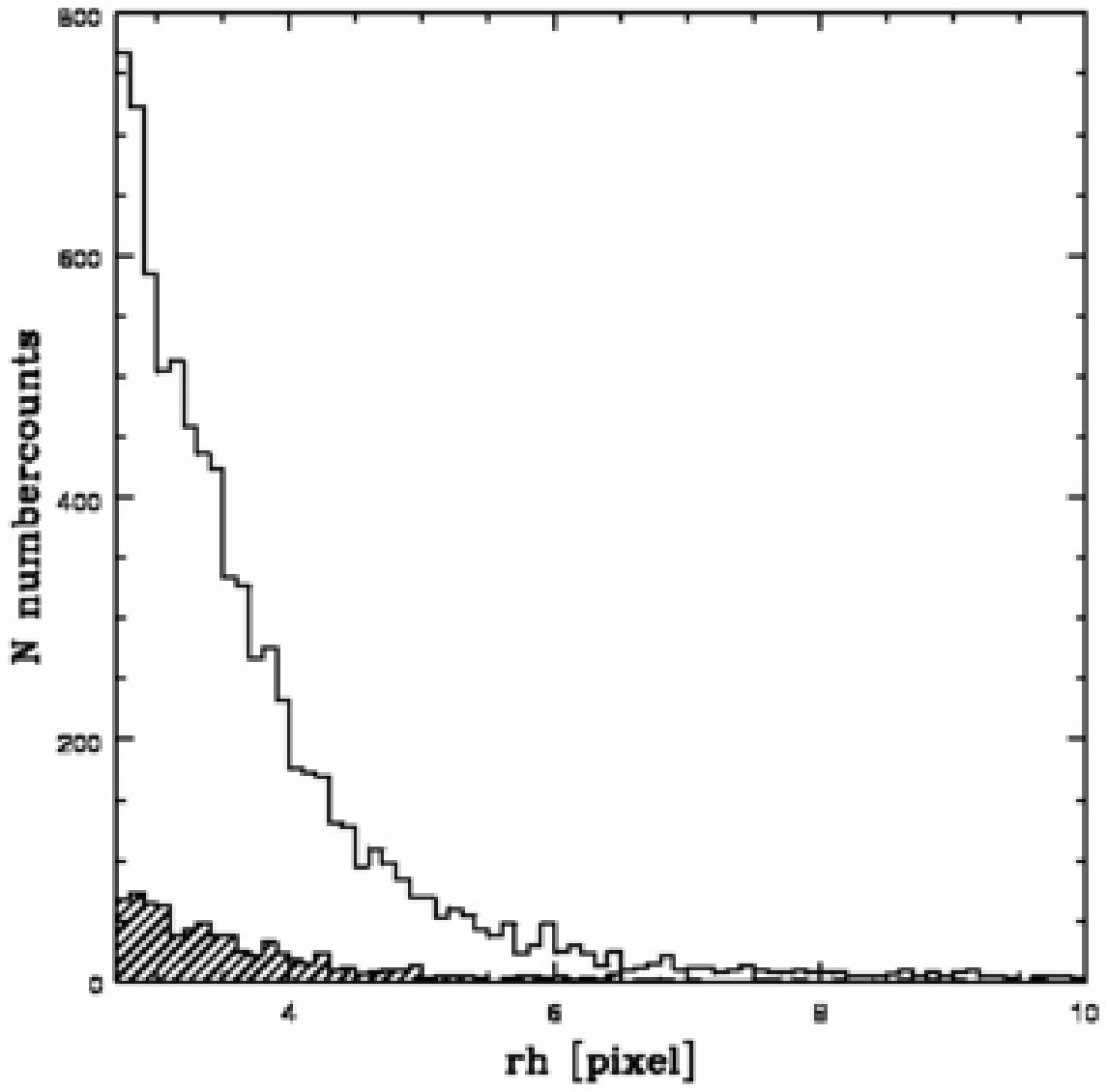}
\hspace{1cm}
\includegraphics[width=8.cm]{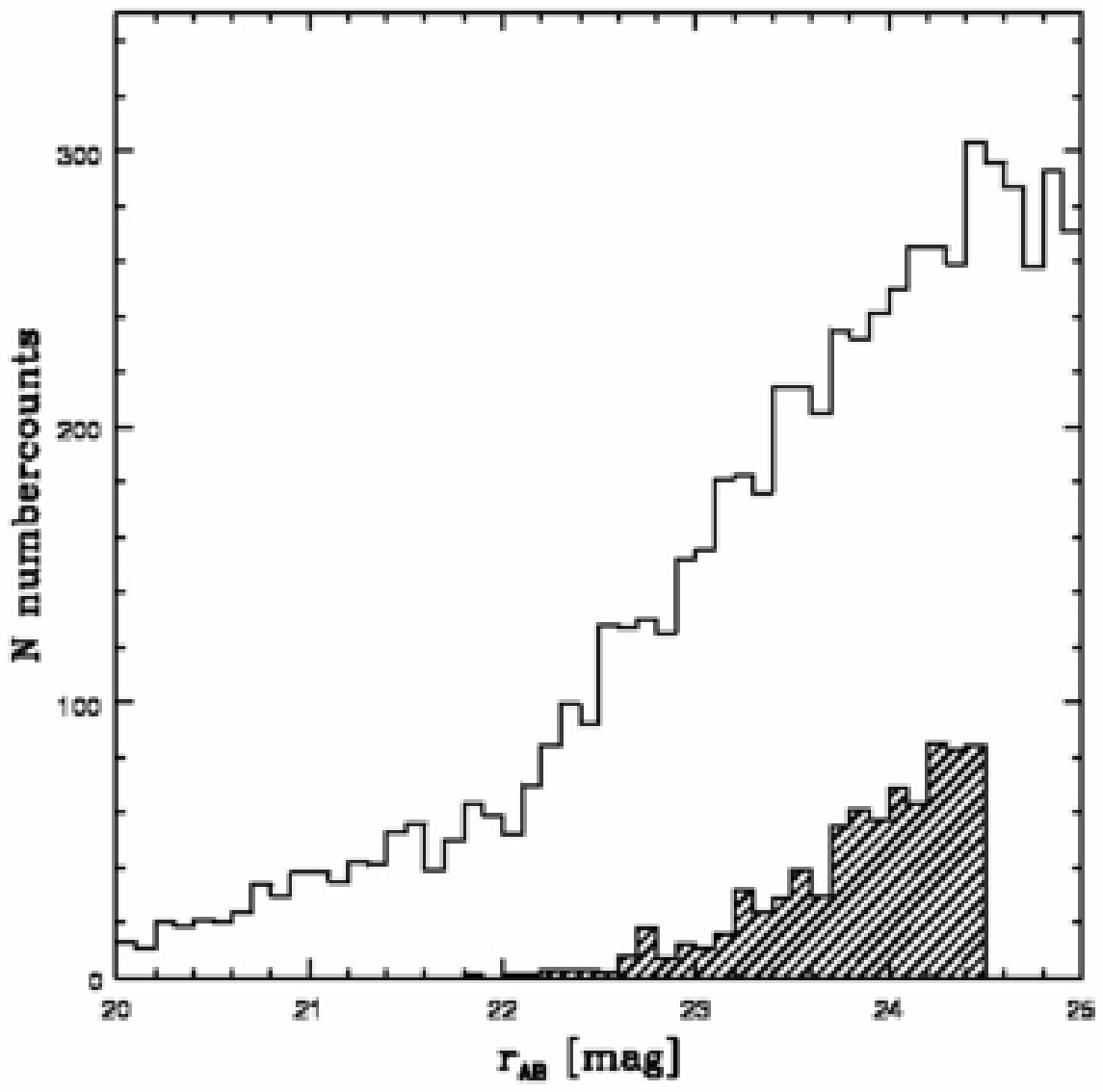}
\caption{Distribution of the $r_h$ and r$_{AB}$ magnitude values of all galaxies used for the weak lensing analysis. The shaded distribution shows the 785 objects used to perform the weak lensing study on XMMU J1230.3+1339.}
\label{FigKSBrh}
\end{figure*}
\begin{figure*}
\centering
\includegraphics[width=8.cm]{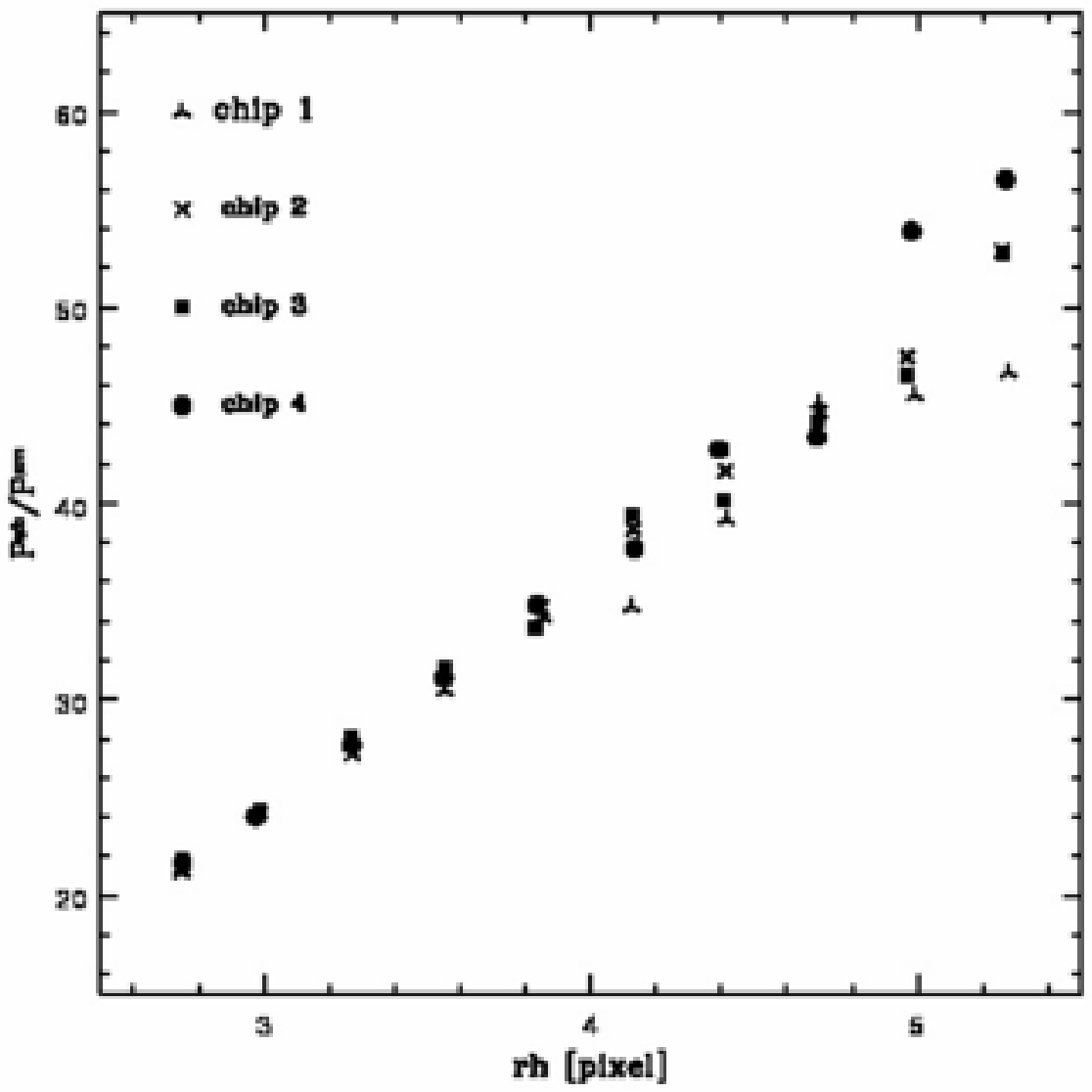}
\hspace{1cm}
\includegraphics[width=8.cm]{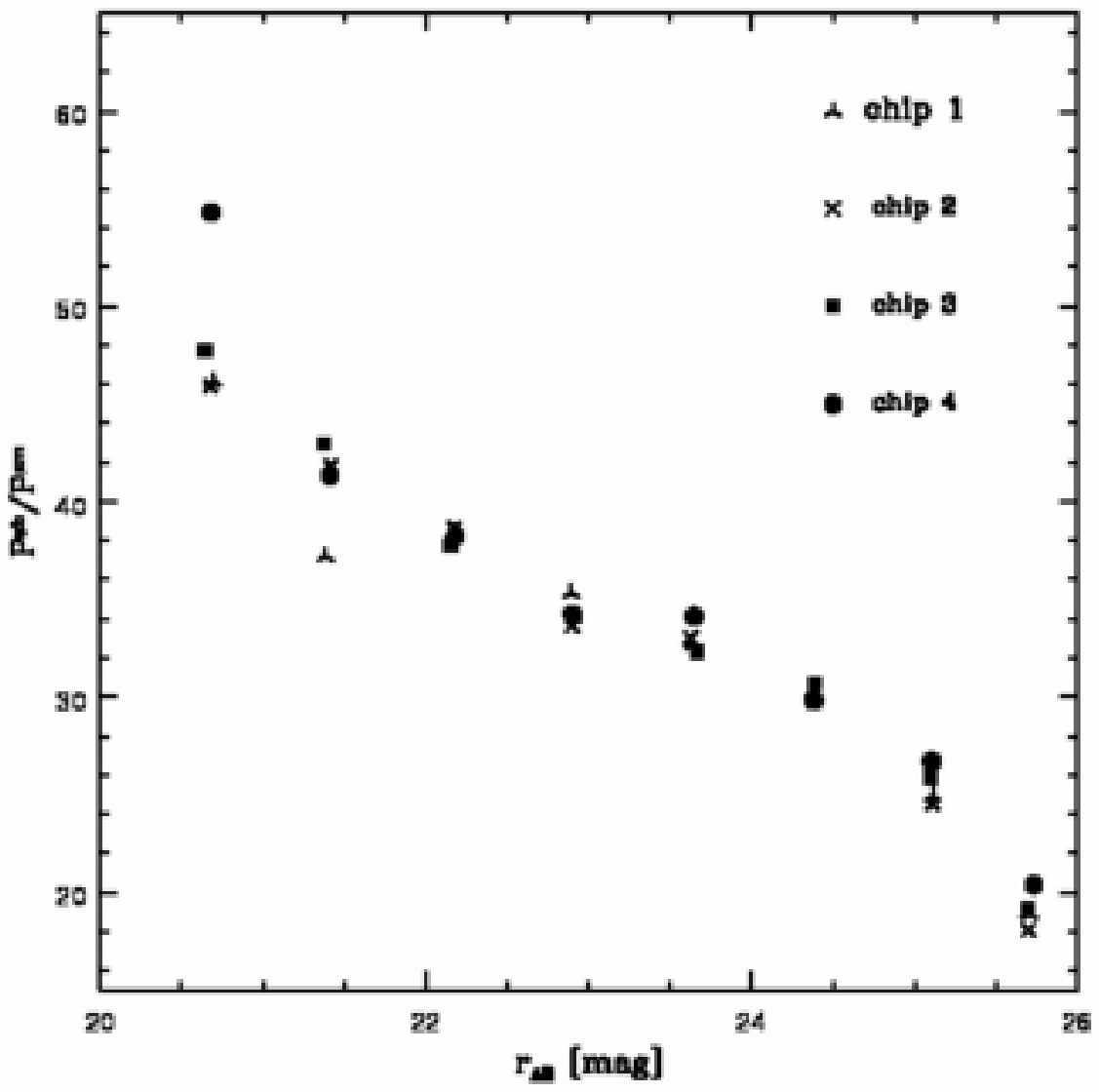}
\caption{PSF correction: $P_{sh}/P_{sm}$ values of all galaxies used for the weak lensing analysis computed in different bins of $r_h$ and r$_{AB}$ magnitude and for each chip.}
\label{FigPSFcorr}
\end{figure*}
\begin{figure*}
\centering
\includegraphics[width=8.cm]{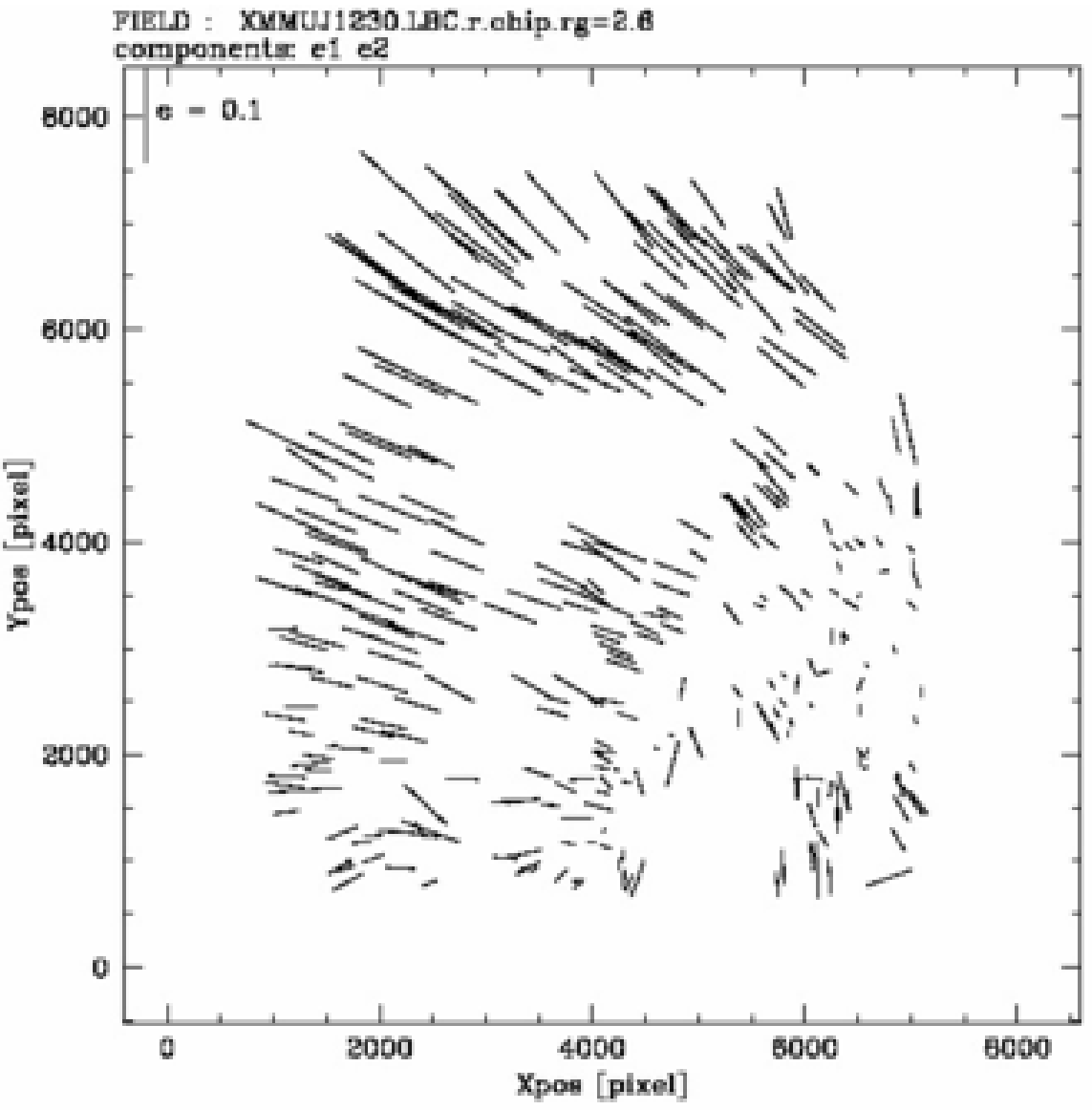}
\hspace{1cm}
\includegraphics[width=8.cm]{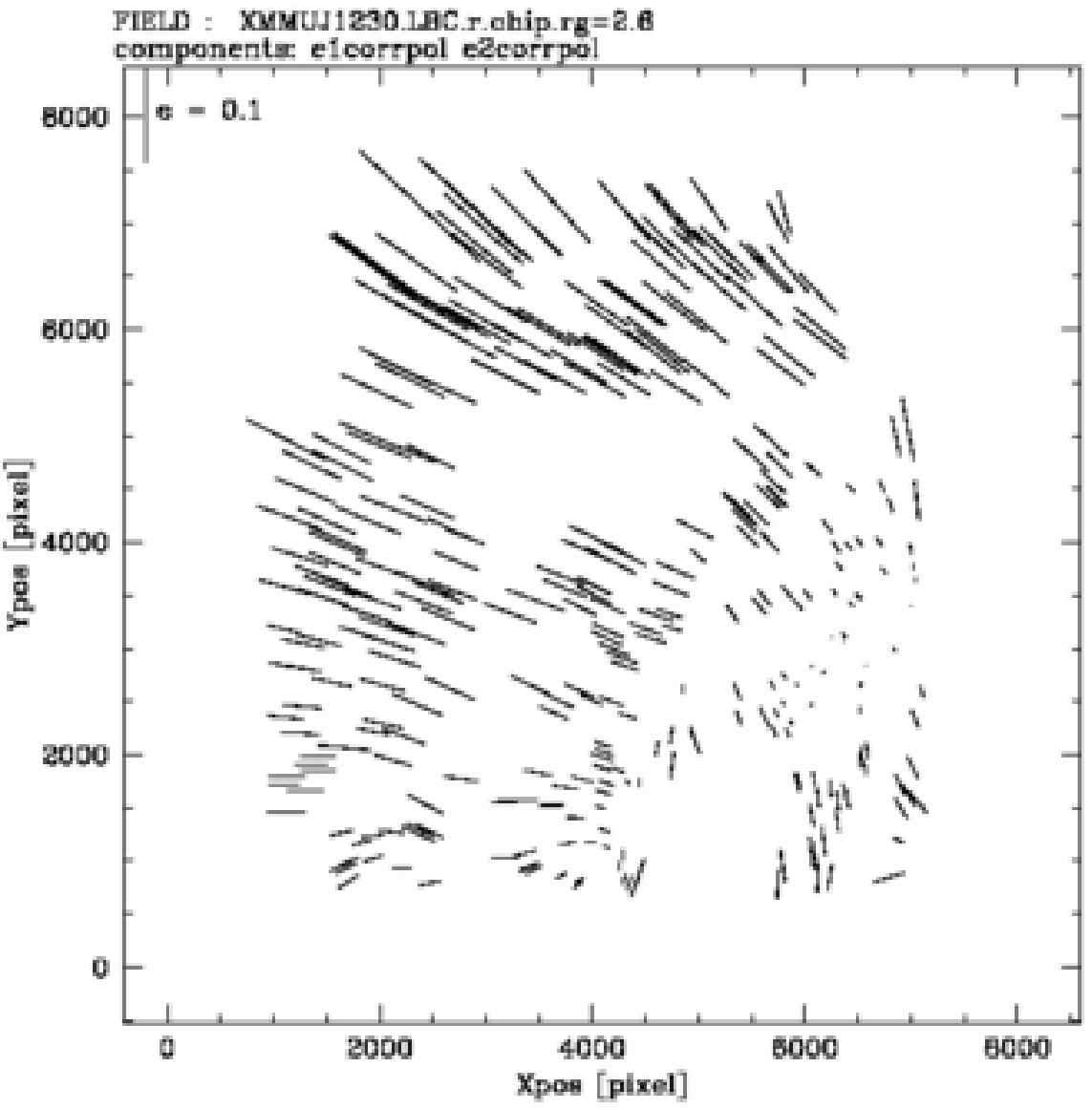}
\includegraphics[width=8.cm]{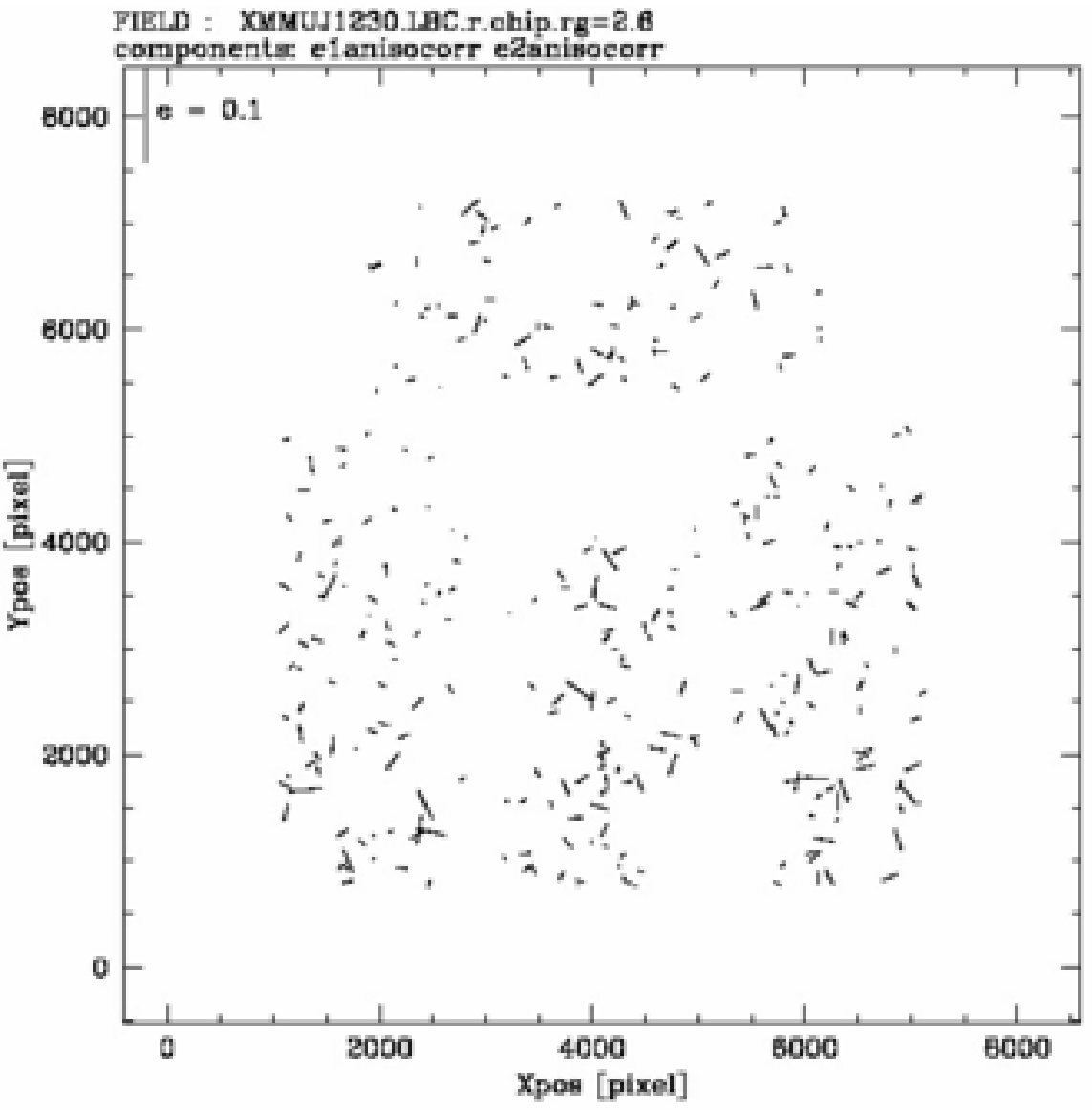}
\hspace{1cm}
\includegraphics[width=8.cm]{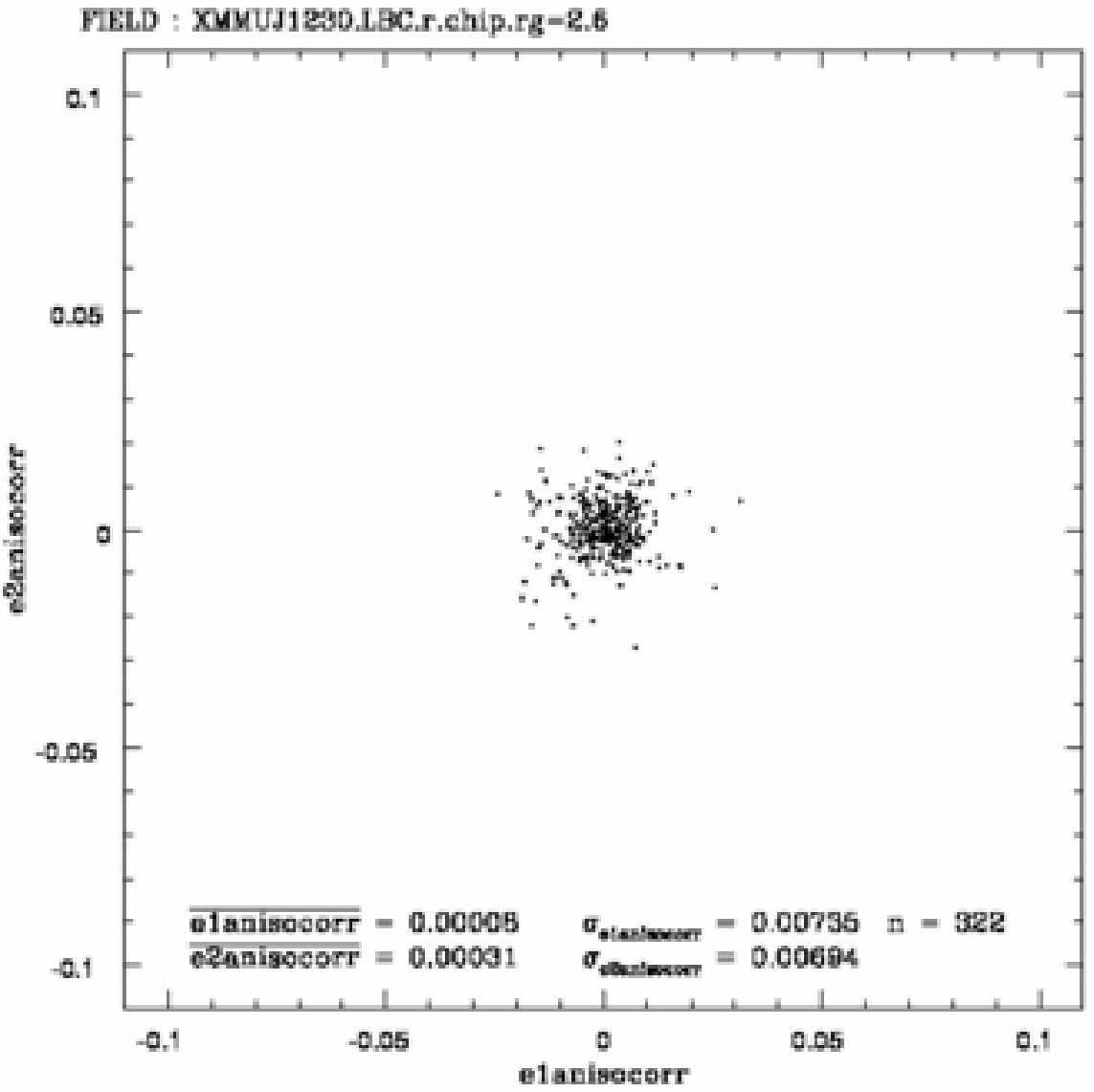} 
\caption{PSF anisotropy correction: ellipticities of the stars in the field of XMMU J1230.3+1339 as observed (top-left panel) and fitted (top-right panel), plus their residuals (bottom-left panel) after correction for PSF anisotropies as a function of their pixel position of the r-band stack. The direction of the sticks coincides with the major axis of the PSF, the lengthscale is indicating the degree of anisotropy. An ellipticity of 10\% is shown in the upper left corner of two upper  and the bottom left panel. The definition of ellipticity is given in Eq. 4.The stars initially had systematic ellipticities up to $\sim 3-8 \%$ in one direction. The PSF correction (bottom-right panel) reduced these effects to typically $< 1.5 \%$.} 
\label{FigXMMUJ1230_PSF}
\end{figure*}       

Following Hoekstra et al. \cite{hoekstra98}, the quantity $P^{sh\ast}/P^{sm\ast}$ should
be computed with the same weight function used for the galaxy to be corrected. The weights on ellipticities are computed according to Hoekstra et al. \cite{hoekstra00},
\begin{equation}
w = \frac{1}{\sigma^2_{g}} = \frac{(P^{g})^2}{(P^{g})^2\sigma^2_{\epsilon_0}+\langle \Delta \epsilon^2 \rangle },
\label{eq:weight}
\end{equation}
where $\sigma_{\epsilon_0} \sim 0.3$ is the scatter due to intrinsic ellipticity of the galaxies, and $\langle \Delta \epsilon^2 \rangle^{1/2}$ is the uncertainty on the measured ellipticity. In Figure \ref{FigKSBweight} we show the galaxy weight as a function of the absolute first and second components of the ellipticity.

We define the anisotropy corrected ellipticity as
\begin{equation}
e^{ani}_{i} = e_{i} - P^{sm}_{ij} p^{\ast}_{j},
\end{equation}
and the fully corrected ellipticity $\epsilon^{iso}$ as
\begin{equation}
e^{iso}_{i} = \frac{1}{P^{g}_{ij}}e^{ani}_{j},
\end{equation}
which is an unbiased estimator for the reduced gravitatonal shear
\begin{equation}
\langle \epsilon_{iso} \rangle = g = g_1 + i g_2 = \frac{\gamma}{1-\kappa},
\end{equation}
assuming a random orientation of the intrinsic ellipticity. For the weak distortions measured, $\kappa \ll 1$, and hence,
\begin{equation}
\langle \epsilon_{iso} \rangle = g  
.
\end{equation}

Our KSB+ implementation uses a modified version of Nick Kaiser's \texttt{IMCAT} tools, adapted from the `TS' pipeline (see Heymans et al.\citealt{step1} and  Schrabback et al.\citealt{schrabback07}), which is based on the Erben et al. \cite{erben01} code.
In particular, our implementation interpolates between pixel positions for the calculation of $Q_{ij}$, 
$P^{sm}$, and $P^{sh}$ and measures all stellar quantities needed for the correction of the galaxy ellipticities as function of the filter scale $r_g$ following Hoekstra et al. \cite{hoekstra98}.
This pipeline has been tested with image simulations of the STEP project
\footnote{\url{http://www.physics.ubc.ca/$\sim$heymans/step.html}}
. In the analysis of the first set of image simulations
(STEP1) a significant bias was identified (Heymans et al.\citealt{step1}), which is usually corrected 
by  the introduction of a shear calibration factor
\begin{equation}
\langle g \rangle = c_{cal} \cdot \langle \epsilon_{iso} \rangle
\end{equation}
with c$_{cal}$ = 1/0.91. In the analysis of the second set of STEP image simulations (STEP2), which takes realistic ground-based PSFs and galaxy morphology into account, it could be seen that with this calibration factor the method is on average accurate to the $\sim 3 \%$ level\footnote{Note: Dependencies of shear on size and magnitude will lead to different uncertainties.} (Massey et al.\citealt{step2}).

Note that for the weak lensing analysis we only consider the r-band stack. We do neither use the i-band stack, because we observe larger residuals of the anisotropy correction, nor we do make use of the z-band stack, which shows remaining fringe patterns after data reduction.
We select stars in the range 19 mag $< r_{AB} <$ 22.5 mag, 0.48\arcsec $< r_h <$ 0.58\arcsec, giving 322
stars (see Figure \ref{FigStars}) to derive the quantities for the PSF correction. Figure \ref{FigXMMUJ1230_PSF} shows the distortion fields of stellar ellipticities before and after PSF anisotropy correction as a function of their pixel position of the r-band stack. The direction of the sticks coincides with the major axis of the PSF, the lengthscale is indicating the degree of anisotropy. The stars initially had systematic ellipticities up to $\sim 3-8 \%$ in one direction. The PSF correction reduced these effects to typically $< 1.5 \%$.  
The achieved PSF correction for LBC is at the same level in terms of the mean and standard deviation of stellar ellipticities as the one for Suprime-Cam as detailed in Table 4 and Figure 2 of Okabe \& Umetsu (2008).

Galaxies used for the shear measurement were selected through the following criteria: $w$ $>$ 3, SNR $> $5, z$_{phot} >$ 1.03, r$_h >$ 0.58\arcsec, r$_{AB} <$ 25.5 mag, and ellipticities,  $e_{iso}$, smaller than one. This yields a total number of 785 background galaxies in the catalogue, and an effective number density of background galaxies ($z_{phot}>1.03$) of $n_{eff} \sim 7$ galaxies per square arcmin.
In Figure \ref{FigXMMUJ1230_CMR} we show the distributions in the in the $i_{\rm AB}-z_{\rm AB}$ against $z_{\rm AB}$ colour-magnitude diagram of the members of the galaxy cluster XMMU J1230.3+1339 and the background galaxies. The area is limited with the dashed-dotted line. Note that without accurate photometric redshifts a separation of background from foreground and cluster member galaxies cannot be achieved. 

The measurement of the shape of an individual galaxy provides only a noisy estimate of the weak gravitational lensing signal. We therefore average the measurements of a large number of galaxies as a function of cluster centre distance. Figure \ref{FigXMMUJ1230_SIS} shows the resulting average tangential distortion $e_t$ of the smear and anisotropy corrected galaxy ellipticity as a function of the distance from the X-ray centroid of the galaxy cluster. The errorsbars on $e_t$ and $e_c$ indicate Poissonian errors.
We detect a significant tangential alignment of galaxies out to a distance of $\sim$ 6\arcmin or $\sim$ 2.8 Mpc from the cluster centre. The black points represent the signal $e_c$, when the phase of the distortion is increased by $\pi/4$. If the signal observed is due to gravitational lensing, $e_c$ should vanish, as observed. 
To obtain a mass estimate of the cluster, we fit a singular isothermal sphere (SIS) density profile to the observed shear. We use photometric redshifts for the computation of angular distances, where $D_s$, $D_d$, and $D_{ds}$ are the distances from the observer to the sources, from the observer to the deflecting lens, and from the lens to the source.
We show our results in Figure \ref{FigXMMUJ1230_SIS}, where the solid line is the best fit singular isothermal sphere model which yields an equivalent line of sight velocity dispersion of  $\sigma$ = 1308 $\pm$ 284 km s$^{-1}$ and $\chi^2/DOF=1.66/5$. 

In addition to the SIS model considered so far from now on we use in addition an NFW model (Navarro et al.\citealt{nfw97}) to describe the density profile of the cluster.
This is because a NFW model automatically provides a virial mass M$_{200}$.
An NFW model at redshift $z$ is defined as
\begin{equation}
\rho(r) = \frac{\delta_c~\rho_{c}}{(r/r_s)(1+r/r_s)^2},
\end{equation}
 where
\begin{equation}
\delta_c = \frac{200}{3} \frac{c^3}{\ln (1+c)-c/(1+c)}
\end{equation}
is the characteristic density, $c = r_{200}/r_s$ is the concentration parameter, $r_s$ is the scale
radius, and $\rho_{c}(z) = 3 H(z)^2/8\pi G$ is the cosmological critical density. The mass within a radius 
$R$ is given by
\begin{equation}
M(r \leq R) = 4~\pi~\delta_c~\rho_{c}(z)~r_s^3~\big[ \ln \big(1+\frac{R}{r_s} \big) - \frac{R/r_s}{1+R/r_s} \big].
\end{equation}
and the corresponding mass $M_{200}$ is given by
\begin{equation}
M_{200} = 200~\rho_{c}(z)~\frac{4~\pi}{3} r_{200}^3.\\
\end{equation}

We perform a likelihood analysis (see Schneider et al. 2000) for details of the method) to fit an NFW model to the measurements. This yields a concentration parameter of c = 4$^{+14}_{-2}$, a scale radius of r$_s$ = 345$^{+50}_{-56}$ kpc and $\chi^2$/DOF=1784/783 within a 50\% significance interval, as shown in Figures \ref{FigXMMUJ1230_NFW}.

\begin{figure}
\centering
\includegraphics[width=8.cm]{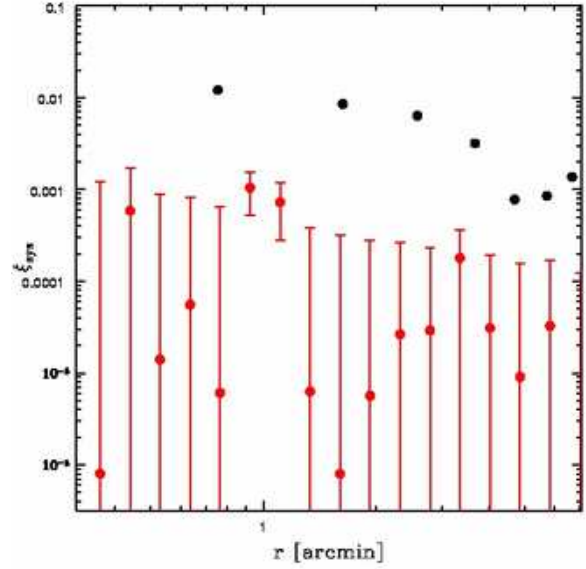}
\caption{PSF systematic residual: the cross-correlation function $\xi_{sys}$ (red points) between galaxies and stars is shown as a function of angular scale.  The amplitude of the cross-correlation is always at least one order of magnitude smaller than the shear signal $\xi_E \sim e_t^2$ (black filled).}
\label{FigPSFsys}
\end{figure}

\begin{figure}
\centering
\includegraphics[width=8cm]{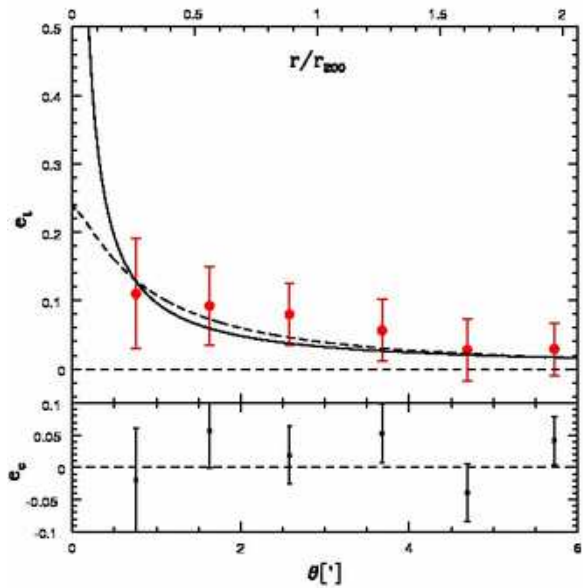} 
\caption{Radial profiles of the tangential shear $\epsilon_t$ (in red, measure of the signal) and the cross shear $\epsilon_c$ (in black, measure of the systematics) around the cluster XMMU J1230.3+1339. The error bars indicate the statistical error in the measurement, due to the intrinsic ellipticity of the individual sources. The black solid line shows the fit of a SIS model to the data, which yields a velocity dispersion of $\sigma=1308 \pm 284$ km/s ($\chi^2/DOF=1.66/5$), while the dashed line shows the NFW model.} 
\label{FigXMMUJ1230_SIS}
\end{figure}

\begin{figure}
\centering
\includegraphics[width=8cm]{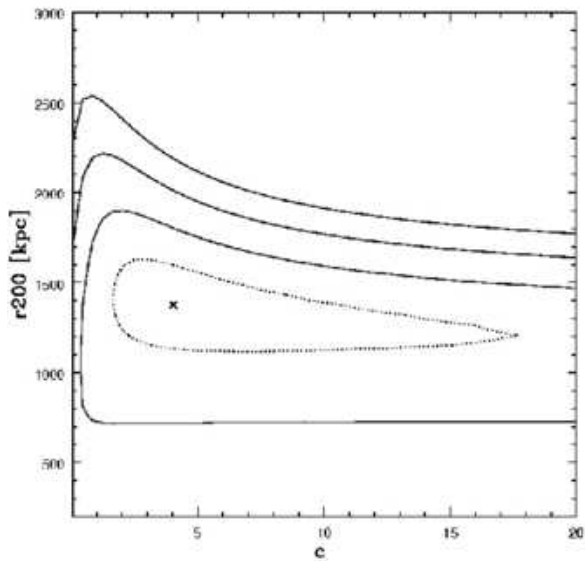}  
\caption{Likelihood confidence contours for the NFW fit to the two shear components on a grid centered on the cluster. The solid line contours are for one, two, and three $\sigma $ confidence levels (68.3\%, 95.4\%, and 99.73\% respectively) with best fit parameters $c = 4.0^{+14}_{-2}$ and $r_{200}=1376^{+200}_{-228}$ kpc ($\chi^2/DOF=1784/783$). The dotted line indicates the  50\% significance level.} 
\label{FigXMMUJ1230_NFW}
\end{figure}

We test our lensing signal for contamination by the imperfect PSF anisotropy correction. Following Bacon et al. \cite{bacon03} we compute the correlation between the corrected galaxy and uncorrected stellar ellipticities. To assess the amplitude we normalise the quantity by the star-star uncorrected ellipticity correlation.
\begin{equation}
\xi_{sys}(r) = \frac{\langle e^*(x)\gamma(x+r) \rangle^2}{\langle e^*(x)e^*(x+r) \rangle},
\end{equation}
where the ``asterisk" indicates a stellar quantity. In Figure \ref{FigPSFsys} we show the cross-correlation compared to the shear signal $\xi_E \sim e_t^2$ up to $10\arcmin$. Note that the amplitude is at least one order of magnitude smaller than the signal. We conclude that the PSF correction is well controlled.

%

\subsection{Aperture mass $M_{ap}$}
%
We make use of the aperture mass statistics, introduced by Schneider \cite{MAP} and worked out further by Schneider \cite{schneider97} to quantify the mass density concentration and the signal of the weak lensing detection. The aperture mass statistic has the advantage that it can be derived from the shear field in a finite region and $M_{ap}$ is not influenced by the mass-sheet degeneracy. It is defined as,
\begin{equation}
M_{ap}(\theta_0^c) = \int d^2\theta~\kappa(\theta)~U(\theta - \theta_0^c) ,
 \end{equation}
where $U(\theta)$ is a radially symmetric weight function, using a compensated filter function
of radius $\theta_0^c$,
\begin{equation}
\int^{\theta_0^c}_{0}~d\theta~\theta~U(\theta) = 0.
 \end{equation}
One can express $M_{ap}$ in terms of the tangential shear $\epsilon_t$
\begin{equation}
M_{ap}(\theta_0^c) = \int d^2\theta~\epsilon_t(\theta)~Q(|\theta-\theta_0^c|). 
 \end{equation}
The filter functions $Q$ and $U$ are related through
\begin{equation}
Q(\theta) = \frac{2}{\theta^2} \int^{\theta}_{0}~d\theta'~\theta'~U(\theta') - U(\theta).
 \end{equation}
It was shown by Schneider \cite{MAP} and Bartelmann \& Schneider \cite{bartelmann01} that the variance $\sigma_{M_{ap}}$ of $M_{ap}$ and $M_{ap}$ itself can be estimated for observational data in a straightforward way. In fact one only has to change the integral into a sum over galaxy ellipticities and the error analysis is simple expressed by
\begin{equation}
M_{ap} = \frac{\pi~\theta_0^2\sum_i~\epsilon_{t,i}~w_i~Q(|\theta_i -  \theta_0^c|)}{\sum_i~w_i},
\end{equation}
\begin{equation}
\sigma^2_{M_{ap}} = \frac{\pi^2~\theta_0^4\sum_i~\epsilon^2_{t,i}~w^2_i~Q^2(|\theta_i -  \theta_0^c|)}{2(\sum_i~w_i)^2},
\end{equation}
where $\epsilon_{t,i}$ are the tangential components of the ellipticities of lensed galaxies relative to each of the grid points investigated, and $w_i$ are the weights defined in Eq.\ref{eq:weight}. 

Therefore the signal to noise ratio (S/N) can by calculated as
\begin{equation}
S/N = \frac{M_{ap}}{\sigma_{M_{ap}}}.
 \end{equation}

We use the polynomial filter function introduced by Schneider et al. \cite{MAP}:
\begin{equation}
Q(\theta) = \frac{6}{\pi}~\frac{(\theta/\theta_0^c)^2}{(\theta^c_0)^2}~\big(1-(\theta/\theta_0^c)^2\big),
\end{equation}
where $\theta$ is the projected angular distance on the sky from the aperture centre, and $\theta_0^c$
is the filter radius. 
We apply this weight function to the data.

We detect a weak lensing signal at a 3.5 $\sigma$ significance. The location of the highest significance coincides with the X-ray measurement of the Chandra analysis (see section \ref{sec:xrayanal}). In Figure~\ref{Fig1aXMMUJ1230} we show an LBT image of XMMU J1230.3+1339 with M$_{ap}$ S/N-contours overlayed.
We observe an elongation in the projected mass distribution from weak lensing in the same direction as the distribution of galaxies associated with the main cluster component and the group bullet (see Paper I), and the X-ray emission.

\section{Kinematic analysis}
\label{sec:kinanal}
For the cluster kinematic analysis the redshifts and positions of galaxies are used to define 
a characteristic velocity dispersion and a characteristic size (scale radius) of the cluster. According to the virial theorem, the RMS velocity dispersion has the property to be completely independent of any
shape and viewing angle (see Carlberg et al. \citealt{carlberg96}). 

\subsection{RMS velocity dispersion}
\label{subsec:rms}
Following Carlberg et al. \cite{carlberg97a} the line-of-sight velocity dispersion is defined as
\begin{equation}
\sigma_{v}^2 = \frac{\sum_i w_i~ v_{pec_i}^2}{\sum_i w_i},
\end{equation}
where the $w_i$ are magnitude dependent geometric weights,
\begin{equation}
v_{pec_i} = \frac{c (z_i - \bar z_{cl})}{1 + \bar z_{cl}},
\end{equation}
the peculiar velocity in the frame of the cluster, and $z_{cl}$ is the mean
redshift of the cluster.

An alternative approach (Carlberg et al.\citealt{carlberg97b}) is the unweighted line-of-sight velocity dispersion, defined as
\begin{equation}
\sigma_{v}^2 = \frac{\sum_i v_{pec_i}^2}{N-1}.
\end{equation}

For a virial mass estimator the projected mean harmonic pointwise
separation can be used
\begin{equation}
\frac{1}{R_H} = {\sum_{i \le j} \frac{w_i~w_j}{|r_i - r_j|}}{\big(\sum_i w_i\big)^{-2}},
\end{equation}
where $ij$ indicate a sum over all pairs.
The three-dimensional virial radius is defined as
\begin{equation}
r_{vir} = \frac{\pi~R_H}{2}.
\end{equation}
The virial mass of the cluster is defined as
\begin{equation}
M_{vir} = \frac{3~\sigma_{v}^2~r_{vir}}{G}.
\end{equation}

We use the model of a singular, isotropic, isothermal sphere with a of velocity dispersion
$\sigma_{SIS} = \sigma_{v}$, and therefore a density profile of 
\begin{equation}
\rho(r) = \frac{\sigma_{v}^2}{2\pi~G~r^{2}}. 
\end{equation}
To compare masses usually an extrapolation to a constant mean interior density 
of $200~\rho_{crit}$ is performed, using
\begin{equation}
M_{200} = M_{vir} \sqrt{\frac{\bar\rho(r_{vir})}{200~\rho_{crit}}},
\end{equation}
where the mean density $\bar\rho(r_{vir})$ inside $r_{vir}$ is defined as
\begin{equation}
\bar\rho(r_{vir}) = \frac{3~M_{vir}}{4~\pi~r_{vir}^3}.\\
\end{equation}

\subsection{Cluster Membership}

Next we define the cluster membership of galaxies, commonly this is determined in two ways:

\subsubsection{$3\sigma$ clipping technique}


The statistical clipping technique (e.g., Zabludoff, Huchra \& Geller\citealt{zabludoff90}) is one method to identify cluster members. It is based on the measured recession velocities of the galaxies. Yahil \& Vidal \cite{yahil77} showed that this is a reasonable assumption, if the velocity distribution of the galaxies follows a Gaussian distribution and clusters are relaxed and isothermal.

\subsubsection{Mass model method}


Another technique to assess the boundaries of a cluster under usage of the spatial and redshift data is the mass model method (Carlberg, Yee and Ellingson\citealt{carlberg97a}). We make use of the model described in Pimbblet et al.~\cite{pimbblet06} for the Las Campanas Observatory and Anglo-Australian Telescope Rich Cluster Survey (LARCS), which is based on Carlberg et al.~\cite{carlberg96}.


For our kinematic mass estimate we use the line-of-sight velocity dispersion $\sigma_{\rm SIS}^{\rm los} = 657 \pm 277$ km s$^{-1}$, compiled in Paper I.

Based on the assumption that a cluster is a singular isothermal sphere (Carlberg et al.\citealt{carlberg97b}), 

\begin{equation}
r_{200} = \frac{\sigma_{v}}{10~ H(z)},
\end{equation}

where $\sigma_{v} \approx \sqrt{3} ~\sigma_{\rm SIS}^{\rm los}$ is the velocity dispersion of the cluster and  $H(z)^2 = {\rm H_0^2} [\Omega_M (1+z)^3 + \Omega_k (1+z)^2 + \Omega_{\Lambda}]$. We estimate $r_{200} = 944 + 397$ kpc and M$_{200} = (2.85 + 3.60) \times 10^{14}$ M$_\odot$; note that due to the poor statistics we only give upper limits. We would need more spectroscopic data of cluster members to improve the estimate together with its uncertainty.

\section{X-ray analysis}
\label{sec:xrayanal}
%

\subsection{Surface brightness profile}
A radial surface brightness profile was computed from a combined, point-source removed image and exposure map over the range 0.29-7.0 keV in $4\arcsec$ annular bins, using the CIAO 4.1 tools DMEXTRACT and DMCALC. 

The surface brightness profile was then fitted with a single $\beta$ model (Cavaliere \& Fusco-Femiano\citealt{cavaliere76}), of the form:

\begin{equation}
I(r) = I_0 \left( 1 + {r^2 \over r_c^2} \right)^{-3\beta+\frac{1}{2}},
\label{sb_eq}
\end{equation} 
where $I_{0}$ is the normalization, $\rm{r}_{\rm{c}}$ is the core radius, and $\beta$ can be determined from the profile at $r \gg r_c$.

\begin{figure}
\centering
\includegraphics[height=8cm]{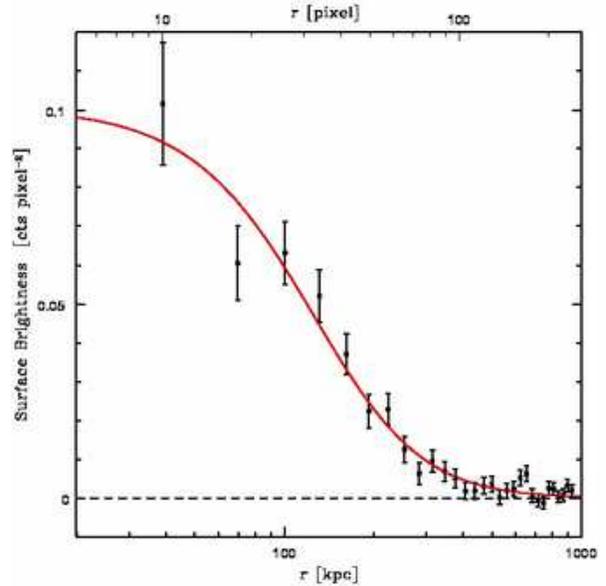}
\caption{Radial surface brightness profile of XMMU J1230.3+1339 for the $0.29-7.0$ keV band sampled in 4 arcsec annular bins (pixel size = 0.492 arcsec/pixel). The red solid line traces the best fitting single $\beta$ model with $r_{c}=39.99$ pixel ($155$ $h_{72}^{-1}$ kpc) and $\beta = 0.67$.
} 
\label{FigXMMUJ1230_X-ray_Beta}
\end{figure}

The best fitting $\beta$ model of this cluster is described by $r_{c}=39.99\pm12.58$ pixel ($155\pm46.5$ $h_{72}^{-1}$ kpc), $\beta = 0.67\pm0.20$ (see Figure \ref{FigXMMUJ1230_X-ray_Beta}).  The reduced $\chi^2$ of the fit was 1.20 for 27 degrees of freedom.

\subsection{Global temperature}
\begin{figure}
\centering
\includegraphics[width=6.0cm,angle=-90]{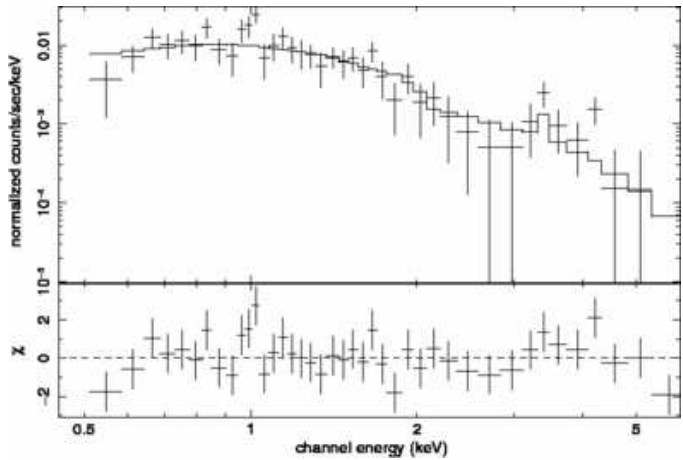}   
\caption{Core spectrum of XMMU J1230.3+1339. The spectrum is fit with a single temperature spectral model, extracted from a circle centered on the emission peak with a radius of 40\arcsec ($\sim$315 kpc) for the $0.5-7$ keV band, grouped to include 20 counts per bin. The fit to the minimally binned data resulted in an ambient cluster temperature of $5.29\pm1.02$ keV. The Fe K-line is visible at 3.3 keV.} 
\label{FigXMMUJ1230_X-ray_Spectra}
\end{figure}       
\begin{figure*}
\centering
\includegraphics[height=5.5cm,angle=-90]{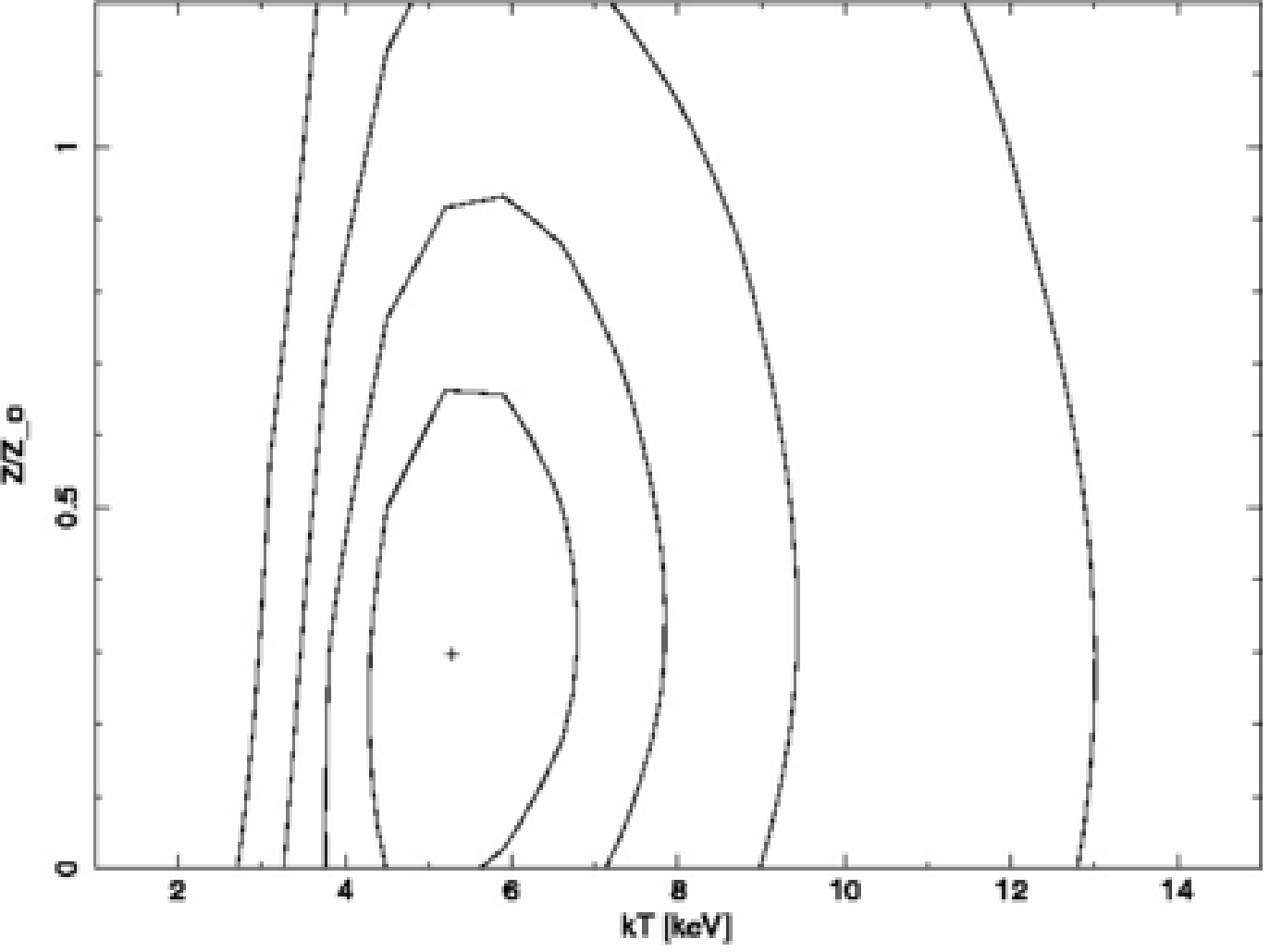}
\includegraphics[height=5.5cm,angle=-90]{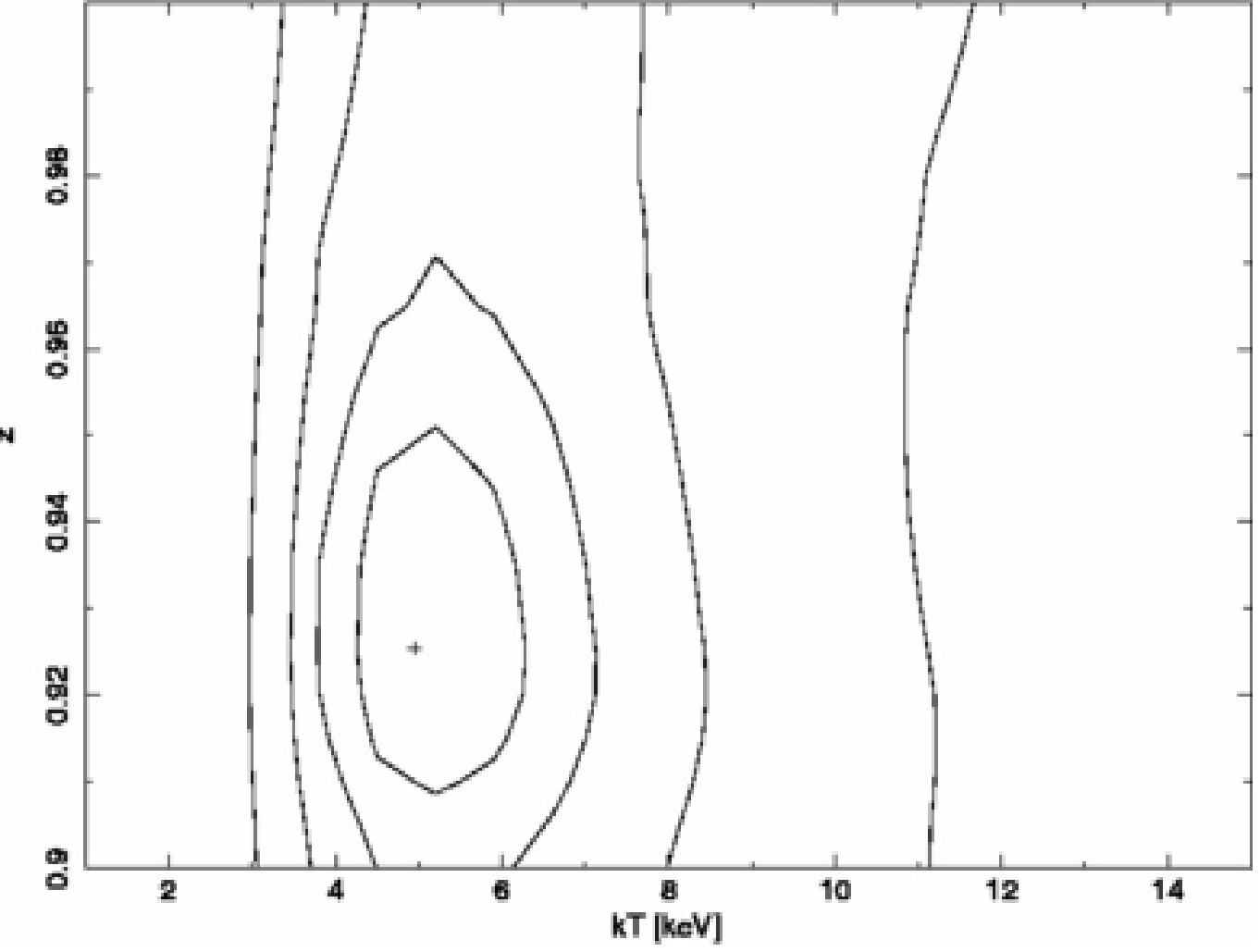}
\includegraphics[height=5.5cm,angle=-90]{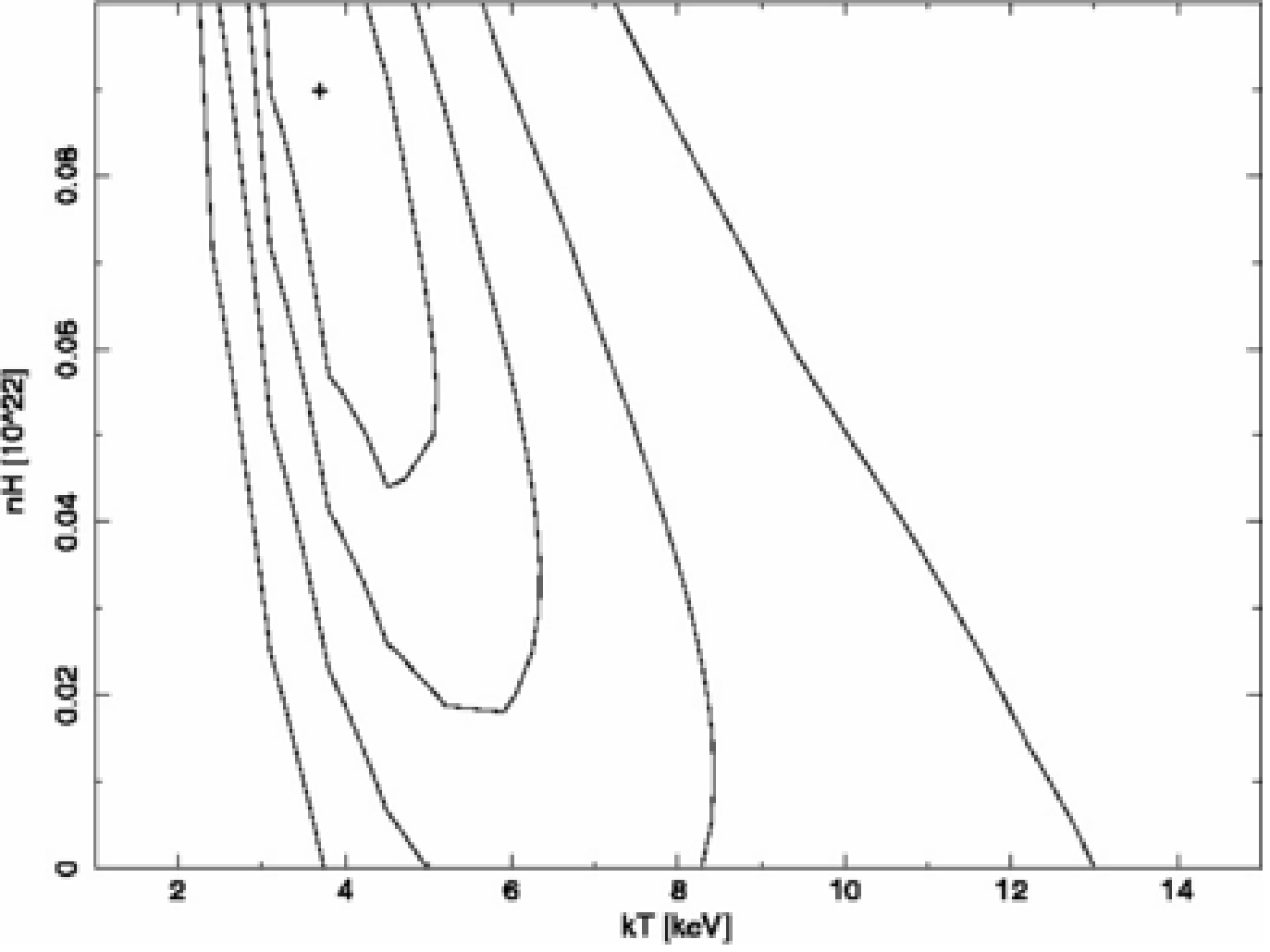}
\caption{Confidence contours obtained from the spectral fit to the X-ray data for XMMU J1230.3+1339. Best fit temperature $T$ and iron abundance $Z/Z_{\odot}$ (left panel), cluster redshift $z$ (middle panel), and Galactic absorption $N_H$ (right panel) are marked as crosses. The solid contours represent 1, 2, 3$\sigma$ confidence levels.} 
\label{FigXMMUJ1230_X-ray_Spectra_Conf}
\end{figure*}      
\begin{table*}
\begin{center}
\caption[]{\rm Comparison of different X-ray statistics.}
\centering
\renewcommand{\footnoterule}{}  
\begin{tabular}{lccccccccccl}
\hline \hline
~ & Cash-stat$^b$ & Cash-stat$^a$ &  $\chi^2$-stat & $\chi^2$-stat & $\chi^2$-stat & $\chi^2$-stat \\
~ & (minimally binned, & (minimally binned, & (20 cts/bin, & (20 cts/bin, & (30 cts/bin, & (30 cts/bin, \\
~ & N$_H$,Z/Z$_{\odot}$,z$_{cl}$ fixed) & N$_H$,z$_{cl}$ fixed) & N$_H$,Z/Z$_{\odot}$,z$_{cl}$ fixed) & N$_H$,z$_{cl}$ fixed) & N$_H$,Z/Z$_{\odot}$,z$_{cl}$ fixed) & N$_H$,z$_{cl}$ fixed)\\
\hline
N$_H$ & 2.64 & 2.64 & 2.64 & 2.64 & 2.64 & 2.64 \\
kT  & 5.29 $\pm$ 1.02 & 5.28 $\pm$ 1.02 & 4.73 $\pm$ 1.12 & 4.85 $\pm$ 1.12 & 4.75 $\pm$ 1.16 & 4.76 $\pm$ 1.17 \\
Z/Z$_{\odot}$ & 0.30 & 0.29 $\pm$ 0.24 & 0.3 & 0.74 $\pm$ 0.43 & 0.30 & 0.54 $\pm$ 0.43 \\
z$_{cluster}$ & 0.975 & 0.975 & 0.975 & 0.975 & 0.975 & 0.975 \\
C/DOF & 486.05/444 &486.05/444 & - & - & - & - \\
$\chi^2_{red}$/prob. & - & - & 0.72/0.965 & 0.71/0.969 & 0.84/0.709 & 0.86/0.679 \\
\hline
\end{tabular}
\end{center}
\label{tab:XraypropRXJ}
Note: $^a$: Confidence contours are shown in Figure \ref{FigXMMUJ1230_X-ray_Spectra_Conf}. $^b$: Used for the X-ray mass estimate. Galactic absorbtion $[N_H]=[10^{20}cm^{-2}]$, global temperature $[kT] =[keV]$, solar abundance $[Z/Z_{\odot}]=[1]$, cluster redshift $[z_{cluster}]=[1]$, likelihood function C/DOF, and min. criterion $\chi^2_{red}$/ Null hypothesis probability.
\end{table*}

Using the X-ray centroid position indicated in Section~\ref{sec:xrayobs}, a point source removed spectrum was extracted from the observation in a circular region with a 40\arcsec ($315~\rm{h}_{72}^{-1}~\rm{kpc}$) radius.  The spectrum was then analysed in XSPEC 11.3.2 (Arnaud\citealt{arnaud96}), using weighted response matrices (RMF) and effective area files (ARF) generated with the CIAO 4.1 tools SPECEXTRACT, MKACISRMF, MKWARF and the Chandra calibration database version 4.1.  The background was extracted from the aimpoint chip in regions free from the cluster emission. The resulting spectrum, after background subtraction, extracted from the observation contains 639 source counts.   
The spectrum is fitted with a single temperature spectral model (MEKAL, see Mewe et al\citealt{mewe85},\citealt{mewe85}; Kaastra et al.\citealt{kaastra92}; Liedahl et al.\citealt{liedahl95}) including Galactic absorption, as shown in Figure \ref{FigXMMUJ1230_X-ray_Model_Spectra}.  The absorbing column was fixed at its measured value of $2.64 \times 10^{20} \rm{cm}^{-2}$ (Dickey \& Lockman\citealt{dickey90}), and the metal abundance was fixed at 0.3 solar (Edge \& Stewart\citealt{edge91}).  Data with energies below 0.5 keV and above 7.0 keV were excluded from the fit.  

The best fit temperature of the spectrum extracted from within this region was ${T}_{\rm{X}}=5.29 \pm 1.02$ keV, with a Cash-statistic $C=486.05$ for 444 degrees of freedom. Using this spectrum, the unabsorbed luminosity of XMMU J1230.3+1339 within 40\arcsec ($315~\rm{h}_{72}^{-1}~\rm{kpc}$) radius is $L_{\rm X} = 1.48 \times 10^{44} \rm{erg}~{\rm{s}}^{-1}$ (0.5-2.0 keV). 
In Figure~\ref{FigXMMUJ1230_X-ray_Spectra} we show a plot of the spectrum with the best fitting model overlaid. The spectrum was grouped to include at least 20 counts per bin.
We performed several tests in order to assess the robustness of our results against systematics, by changing background regions and by using different statistics. A comparison of the Cash-statistic and $\chi^2$-statistic is shown in Table 4; the best fit temperature varies by only $\sim$ 10 \%. The impact of a possible calibration problem of ACIS-S at high-energies has been alleviated by using the latest calibration files.

The resulting temperature, combined with the density profile inferred from the $\beta$ model fit

\begin{equation}
\rho_{\rm gas}(r) = \rho_0 \left[1 + {{r^2}\over{r_c^2}}\right]^{-3\beta/2},
\label{dens_eq}
\end{equation}

\noindent
and the mass equation derived from hydrostatic equilibrium 

\begin{equation}
M_{\rm tot}(<r) = -{{kT(r)r}\over{G \mu m_p}} \left({{\delta~\rm{ln}~\rho}\over{\delta~\rm{ln}~r}} + {{\delta~\rm{ln}~T} \over {\delta~\rm{ln}~r}} \right), 
\label{Mass_eq}
\end{equation}
 
\noindent
(where $\mu m_p$ is the mean mass per particle), were then used to estimate the value of $\rm{r}_{200}$, following Ettori \cite{ettori00}

\begin{equation}
r_{200} = {r_c} \sqrt{{\left[{{3\beta k T}\over{G \mu m_p (4/3) \pi \rho_c(z) r_c^2 200}}\right]}-1},
\label{eq:ettori}
\end{equation}

\noindent
where $r_c$ is the core radius. With $\mu=0.62$, we obtain r$_{200} = 1046 \pm 113~ $kpc. 
The total mass is then calculated out to $\rm{r}_{200}$, using the equation

\begin{equation}
M_{\rm tot}(<r) = {{3\beta}\over{G}}~{{k T r}\over{\mu m_p}}~{{{(r/r_c)}^2}\over{{1+{{(r/r_c)}^2}}}}.
\label{Mass_eq2}
\end{equation}

\section{Cluster mass estimates}
\label{sec:mass-est}
In this section we compare the masses determined with different methods, namely from the kinematics of the cluster galaxies, X-ray emission of the cluster gas, and weak lensing.


%
\begin{table}
\begin{minipage}[t][]{\columnwidth}
\caption{Mass estimates for XMMU J1230.3+1339.}
\centering
\renewcommand{\footnoterule}{}  
\begin{tabular}{lcccl}
\hline \hline
Reference & $r_{200}$     &  $M_{200}$   &  $M_{r\le1{\rm Mpc}}$   & Method \\
 ~                 &       [kpc]        &  [$10^{14}{\rm M}_{\sun}$] & [$10^{14} {\rm M}_{\sun}$]  & (Remark) \\
\hline
This work	 & $~944 +  397$ &  $2.85 + 3.6$  &   $3.01 +  3.8$  &Kinematics\\
This work	 & $1046\pm 113$  &  $3.87 \pm 1.3$ &  $3.72 \pm 1.2$ &  X-ray\\
This work	 & $1376 \pm 218$ &  $8.80 \pm 4.2$  & $6.74 \pm 3.3$  &  WL (NFW)$^a$\\
This work	 & $1125\pm 188$ &  $4.56 \pm 2.3$  &   -  & combination \\
Paper I	 & $1017\pm 68$ &  $4.19 \pm 0.7$  &   -  & combination \\
\hline
\end{tabular}
\end{minipage}
$^a$: Note that we use for the r$_{200}$ the 50\% significance level.
\label{tab:XMMUJ1230_massest}
\end{table}
\begin{figure}
\centering
\includegraphics[height=8cm]{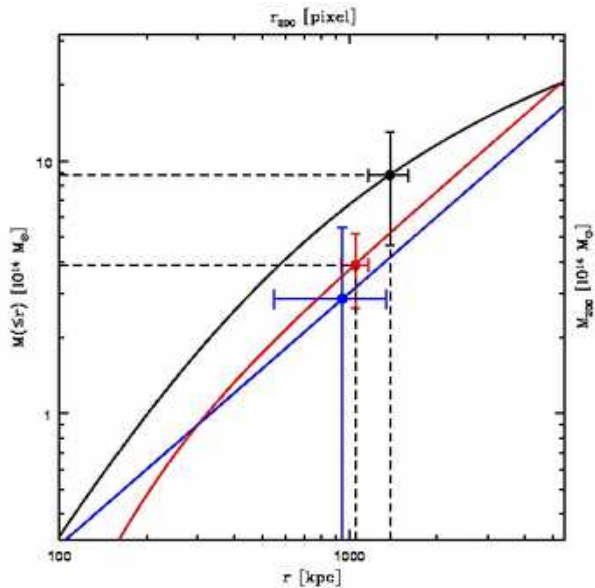} 
\caption{Mass Profile of XMMU J1230.3+1339. In black: best fit NFW profile from weak lensing; in blue: best fit kinematic SIS profile; in red: best fit X-ray $\beta$-model profile. 
We have added the M$_{200}$ and r$_{200}$ values with errorbars for each method.} 
\label{FigXMMUJ1230_massprofile}
\end{figure}       

As seen from Table 5 the mass estimates from the NFW fit, kinematics, and X-ray analyses are in good agreement, within the error bars. In Figure \ref{FigXMMUJ1230_massprofile} we plot the mass estimates against the corresponding radius. The dynamical mass estimate is on the low side but still consistent with the other measurements due to the low statistics.

\section{Discussion}
\label{sec:discon}
Here we discuss the results of our optical, weak lensing, kinematic and X-ray analyses and compare XMMU J1230.3+1339 to other high redshift clusters (see e.g. Appendix Table C1). A more general discussion of the global cluster properties (e.g., mass proxies, scaling relations, large scale structure) can be found in paper I.
\subsection{Colour-Magnitude Relation}
We observe evidence for a deficit of faint red galaxies, down to a magnitude of z$_{AB}$ = 24.5 (50 \% completeness), based on our photo-z/SED classification (see Figures \ref{FigXMMUJ1230_CMR} and \ref{FigPhotLumFct}). Similar results have been reported for several clusters at z $=$ 0.7 - 0.8, for RXJ0152 by Tanaka et al. \cite{tanaka05}, for RXJ1716 by Koyama et al. \cite{koyama07}, and for 18 EDisCS clusters by De Lucia et al. \cite{delucia07}, and for RDCSJ0910+54 at z $=$ 1.1 by Tanaka et al. \cite{tanaka08}. However there are other clusters still under debate, like e.g. MS1054, where Andreon et al. \cite{andreon06} showed the faint end of the CMR and in contrast to that Goto et al. \cite{goto05}, who measured a deficit at 1$\sigma$ level for the same cluster. Some other known clusters seem to have their faint end of the CMR in place, like e.g. the 8 clusters between redshift 0.8 and 1.27 studied in detail by Mei et al. \cite{mei09} and also the ``twin" cluster XMMU J1229+0151 at z $=$ 0.975 (Santos et al.\citealt{santos09}). 

Koyama et al. \cite{koyama07} suggested to collect more observations to study the deficit of faint red galaxies, since it can be seen only in the richest clusters and therefore a larger cluster sample is necessary. 

XMMU J1230.3+1339 is a very rich, massive, high-redshift cluster, and this may have some impact on the build-up of the CMR.
We will explore the build-up of the CMR in clusters at z$\le$1 and do a comparison with semi-analytical modeling of hierachical structure and galaxy formation in an upcoming paper.

\subsection{Blue galaxy fraction - Butcher \& Oemler effect}
About half of the cluster members has a blue galaxy type, as inferred from the photo-z/redshift classification. This is in good agreement with the prediction of De Lucia et al. \cite{delucia07}, Menci et al. \cite{menci08} and Haines et al. \cite{haines09}: an increase in the fraction of blue galaxies with increasing redshift.

Note that the photometric redshift estimates for blue galaxies are usually worse than those for red galaxies, due to the different shape of the SED's (see right panel of Figure 11 for details) which results in larger uncertainties. 
This can not be seen in this data set, because we only have ``red" spectroscopic clusters members (see Ilbert et al. \citealt{ilbert06} and Brimioulle et al. \citealt{brimioulle08} for the dependence of photometric redshift accuracy on different SED types and magnitude intervals).
The larger photometric redshift errors of blue galaxies lead to an under estimation of the blue galaxy fraction due to an increased scattering of object into neighboring redshift bins, see e.g. the small signal bins  z = [0.8 - 0.9] and  z = [1.0 - 1.5] in Figure \ref{FigCorrelGal}.

%
\subsection{Colour-Colour Diagram}
Focussing on the colour coded cluster members, based on the photo-z/SED classification one can easily see that a separation between foreground and background for the blue galaxies is hardly possible without knowing the redshift. The SSP models, computed using \texttt{PEGASE2} (Fioc \& Rocca-Volmerange 1997), nicely reproduce the observed colours of the early type galaxies of the cluster and the field. Note that when observing a cluster at redshift $\sim$ 1, the flux of the cluster members in the blue bands is very low, especially in the U-band, therefore the measurement errors are large.

E.g., additional NIR observations (J and K band down to J$^*+4$ and K$_s^*+5$) can reveal the existence of an additional intermediate age stellar population, and constrain the star formation rate (SFR) of all cluster members together with the available data. With these data we will also be able to the study slope, scatter and zeropoint of the RS in the $J - K$ CMR.
To further understand the evolution of cluster elliptical galaxies around $z=1$ not only the strength of the 4000 Angstrom break is essential, but also the measurement of the fraction of young and intermediate stellar populations. 
%
\subsection{Mass estimates}
\subsubsection{Weak lensing analysis}
Very recently Romano et al. \cite{romano10} performed the first weak lensing study using the LBT. They analysed the galaxy cluster Abell 611 at z $\sim$ 0.3 and demonstrated that the LBT is indeed a suitable instrument for conducting weak lensing studies.
We observe a weak lensing signal for a cluster at a redshift of $z \sim 1$, the most distant cluster so far used in a ground-based weak lensing study. 
We were able to derive a weak lensing mass estimate for the following reasons.
\begin{enumerate}

\item [1.] Deep high-quality multi-band data (U, B, V, r, i, z) have been taken under subarcsecond seeing conditions in order to resolve faint, small and distant objects (cluster members and background galaxies).

\item [2.] We obtained accurate photometric redshifts and SED classifications to separate cluster members, foreground and background galaxies. This avoids a dilution of the shear signal from cluster member and foreground galaxies, which would lead to an additional mass uncertainty of $\sim$ 4 \% (see Hoekstra\citealt{hoekstra07}).

A conventional background separation using the CMR or colour-colour diagram is not possible (see e.g. Medezinski et al.\citealt{medezinski10}) since the background galaxies are not distinguishable from cluster members and foreground galaxies in the colour-magnitude or colour-colour space.

\item [3.] The PSF pattern was stable and isotropic over the entire observation time, and could be modeled with a 5$^{th}$ order polynomial. The systematic error is almost 1/3 of the signal for angular scales considered in the weak lensing analysis.

\end{enumerate}
We had to reject on average one third of the available data (in each band) to produce the stacked images for our scientific exploitation, see Section \ref{sec:optdat} for details. Note that our method (KSB) to measure galaxy shapes and extract the shear signal is similar to the one applied in Romano et al. \cite{romano10}. 

We used spectroscopic data and the two-point angular-redshift correlation analysis to investigate the photometric redshift accuracy. The PSF systematic residual has been tested with the cross-correlation function between galaxies and stars. All tests perform very well and we conclude that the systematics are minimal.
We derive a velocity dispersion of $\sigma_{\rm SIS}^{\rm WL}$ $=$ 1308$\pm$ 284 km/s by fitting a SIS model to the tangential shear profile. From the NFW likelihood analysis we find a best-fit concentration parameter $c = 4.0^{+14}_{-2}$ and a scale radius $r_s= 344 ^{+50}_{-57}$ kpc. The concentration parameter is only weakly constrained. Nevertheless, the best-fit value in good agreement with the relation of Bullock et al. (2001) and Dolag et al. (2004). This results in a total mass of $M_{200}^{\rm NFW}$$=$ (8.80 $\pm$ 4.17) $\times$ 10$^{14}$ M$_{\sun}$.

Hoekstra \cite{hoekstra03} analysed the effect of the large scale structure along the line of sight, which also contributes to the lensing signal, and consequently affects the mass estimates. This introduces a factor $\sim$ 3  for the uncertainties in M$_{200}$ and a factor $\sim$ 2 for the concentration parameter c at the redshift of $z=1$ (extrapolated from Figure 7 in Hoekstra\citealt{hoekstra03}). For deep HST observations of MS 1054-0321 (z=0.89) Jee et al. \cite{jee05b} estimated an additional $\sim$14\% error for the total cluster mass due to cosmic shear. 
Our analysis is limited by the large statistical error of the measurement (shape noise), which is the driving uncertainty of our mass estimate.

XMMU J1230.3+1339 is located behind the local Virgo cluster (z $=$ 0.0038), which covers a sky area of 8 $\times$ 8 square degrees, while XMMU J1230.3+1339 covers 6 $\times$ 6 square arcminutes. 

The Virgo cluster acts like a mass sheet, with a constant surface mass density ($\kappa$ = const.) leading to a constant deflection angle ($\alpha$ = const.) and no shear ($\gamma$ = 0). Note that the 
surface mass density $\kappa$ is indirect proportional to $\Sigma_{crit}$. The critical surface mass density $\Sigma_{crit}$ depends on the ratio $D_s/(D_d D_{ds})$ (Schneider et al.\citealt{MAP}), which leads in the case of the Virgo cluster lens geometry to a high value of $\Sigma_{crit}$. Therefore $\kappa$ becomes negligible small for the Virgo cluster.


We observe an elongation in the projected mass distribution from the M$_{ap}$ statistic similar to the distribution of galaxies associated with the main cluster component and the group bullet (see Paper I) and the X-ray emission. Currently the statistics and the depth are not sufficient to resolve the possible substructure peaks associated with the central group merger event. The WL separation of the main cluster and the group bullet will require high-resolution space-based imaging data.

\begin{figure*}
\centering
\includegraphics[width=8cm,angle=0]{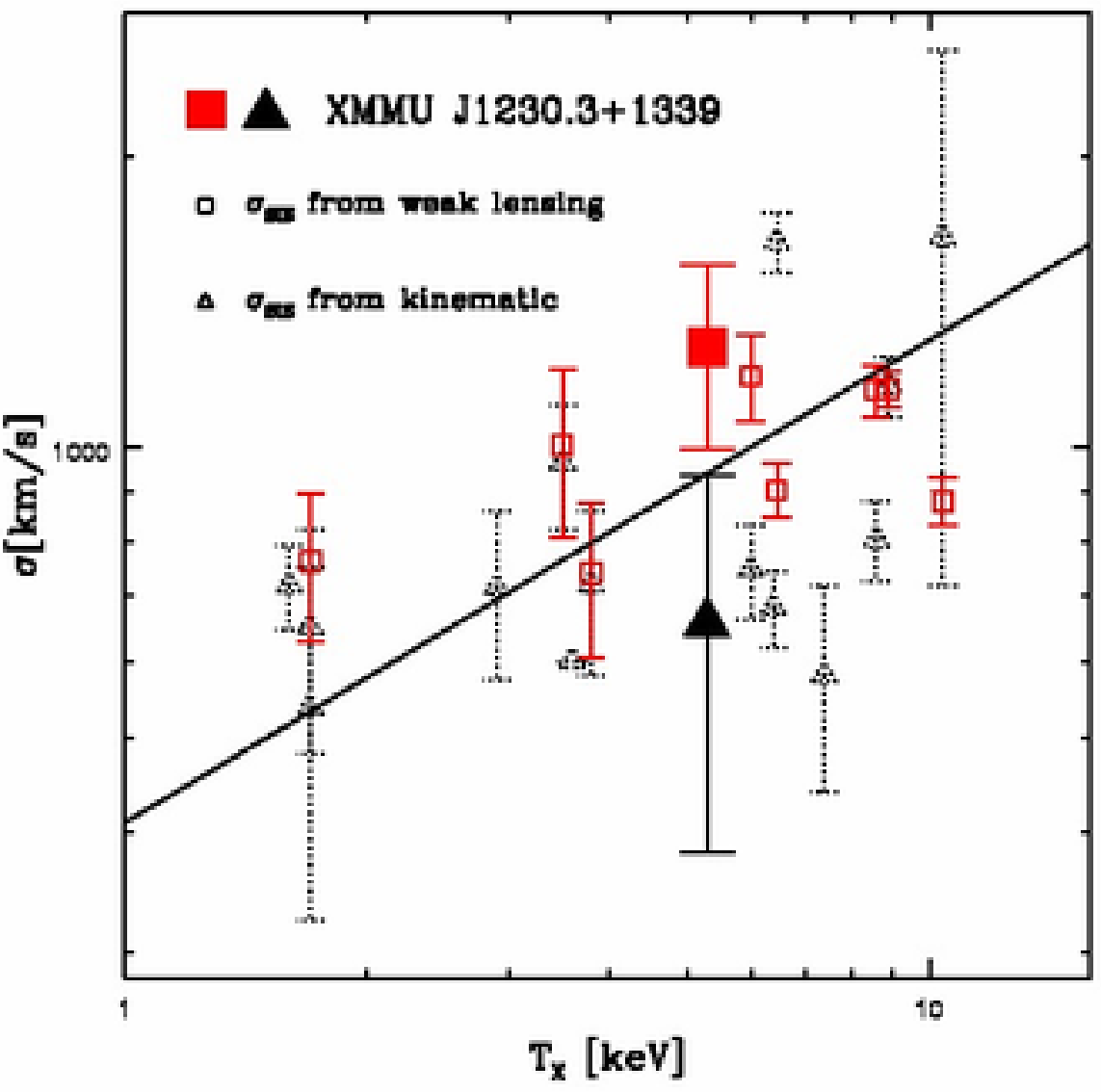} 
\includegraphics[width=8cm,angle=0]{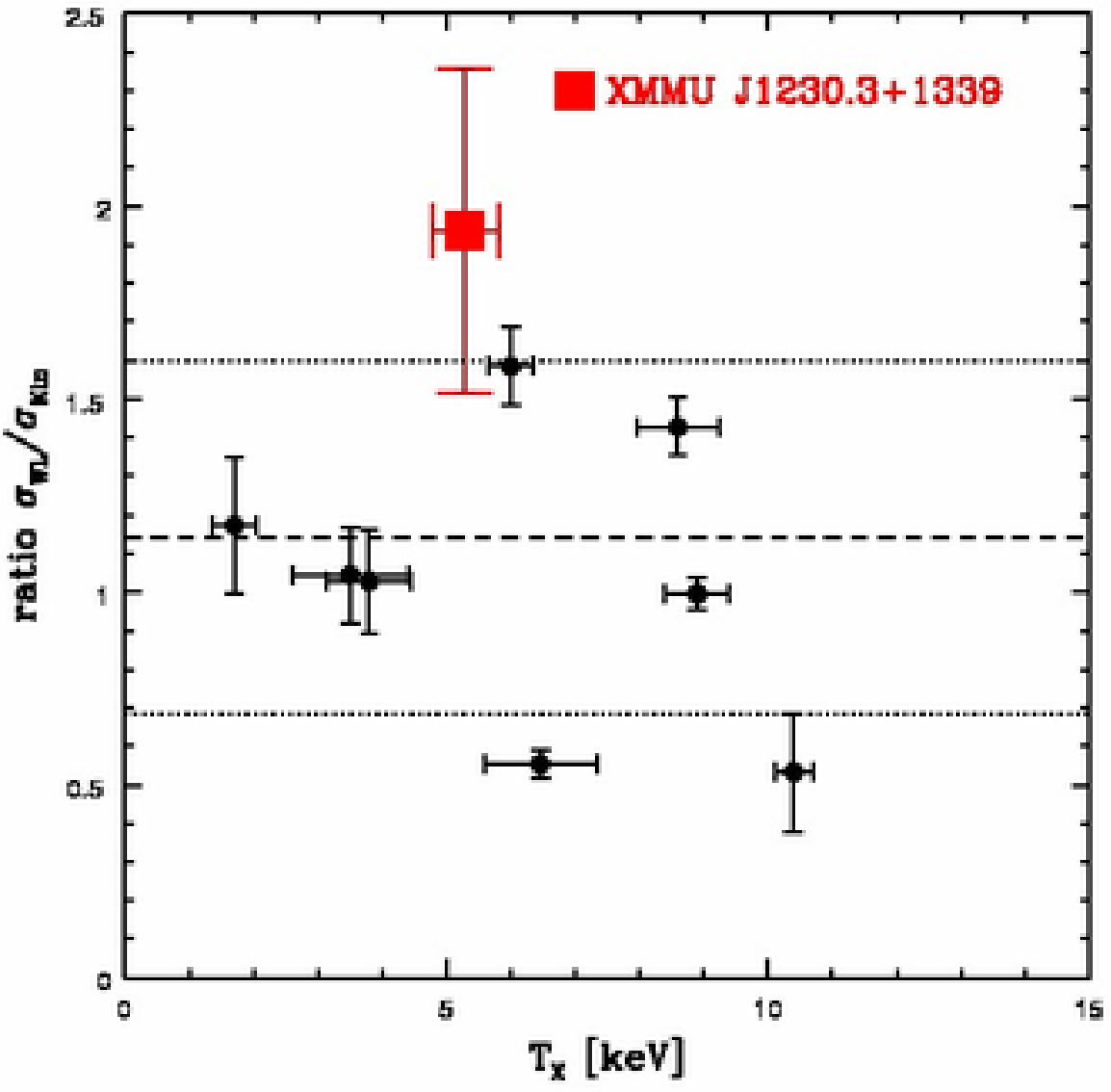}  
\caption{Temperature, $T_{\rm X}$, and galaxy velocity dispersion, $\sigma$, relation for distant ($z>0.8$) galaxy clusters (left panel). Velocity dispersions and temperatures are taken from Table B1. The solid line shows the relation $k_B T_{\rm X} = \mu m_p \sigma^2$ (e.g., see B{\"o}hringer et al. 2010b; Rosati et al. 2002). Red open boxes (black dotted triangles) mark galaxy velocity dispersions taken from weak lensing (kinematic) analyses. XMMU J1230.3+1339 with $T_{\rm X}=5.3 \pm 1$ keV and $\sigma=1308 \pm 284$ km/s is represented as large filled red box. Ratio, $\sigma_{\rm WL}$/$\sigma_{\rm Kin}$, of the velocity dispersion form the weak lensing and the kinematic analysis against temperature $T_{\rm X}$ (right panel) for the nine distant ($z>0.8$) galaxy clusters taken from Table B1. The mean average ratio is $1.14 \pm 0.46$.} 
\label{FigXMMUJ1230_X-ray_Scaling_Relation}
\end{figure*}       

\subsubsection{Kinematic analysis}
The current analysis is based on only 13 cluster members, resulting in a total mass of $M_{200}^{\rm Kin}$$=$ (2.85 $+$ 3.60) $\times$ 10$^{14}$ M$_{\sun}$.  We expect to improve this result substantially by including more spectroscopic data. At the current level the kinematic mass estimate represents only a upper limit of the mass. The analysis can be improved following, e.g., the method described in detail in Biviano et al. \cite{biviano06}. 

\subsubsection{X-ray analysis}
The cluster was recently serendipitously observed with ACIS onboard Chandra with a total integration time of $\sim$ 38 ks. In spite of the surface brightness dimming due to the high cluster redshift, we can see extended X-ray emission out to $\sim$ 0.5 Mpc. This emission shows a regular morphology and is clearly elongated in the same direction as the distribution of red galaxies and the projected matter distribution. 
Our spectral analysis of the Chandra data yields a global temperature of ${T}_{\rm{X}}=5.29 \pm 1.02$ keV, and a total mass of $M_{200}^{\rm X}$$=$ (3.87 $\pm$ 1.26) $\times$ 10$^{14}$ M$_{\sun}$, in good agreement with Paper I. We applied several tests (different background regions, different X-ray statistics) to check the robustness of our findings, which we conclude are robust against several systematics involved in the X-ray spectral analysis.

Any X-ray study relies on the accuracy achieved for the temperature and electron density distribution of the ICM and thus measure the mass distributions with statistical uncertainties below 15\% (e.g., Vikhlinin et al.\citealt{vikhlinin06}). However, the accuracy of X-ray cluster mass estimates based on hydrostatic equilibrium is limited by additional physical processes in the ICM (e.g., merger) and projection effect (e.g., substructure). Recently Zhang et al. \cite{zhang09} showed that for substructure (see B{\"o}hringer et al.\citealt{boehringer10a}) this reflects into $\sim$ 10\% fluctuations in the temperature and electron number density maps. Given the evidence for a cluster bullet and up to six galaxy overdensities in the cluster (see Paper I) these possible will introduce an additional $\sim$ 20\% systematic error in the mass estimate. Nevertheless, our X-ray analysis is still dominated by the poor statistic of the data in hand.

Note that the X-ray global temperature does not follow the $T_{\rm X} - \sigma$ relation shown in Figure \ref{FigXMMUJ1230_X-ray_Scaling_Relation}, which would lead to a global temperature of  ${T}_{\rm{X}}=2.59 $ keV. However together with the other high-z cluster it is still consistent with the relation within its larger scatter. A similar result has been reported for XMMU J1229+0151 (Santos et al.\citealt{santos09}).

We observe evidence for a $\approx$ 14 \% bias in the ratio, $\sigma_{\rm WL}$/$\sigma_{\rm Kin}$, of the velocity dispersion form the weak lensing and the kinematic analysis against temperature $T_{\rm X}$  for the nine distant ($z>0.8$) galaxy clusters (right panel of Figure \ref{FigXMMUJ1230_X-ray_Scaling_Relation}).
For XMMU J1230.3+1339 we can infer a ratio $\sigma_{\rm WL}$/$\sigma_{\rm Kin} =1.93 \pm 0.42$, which is on the upper end of all clusters. We want to stress here again the lack of statistics for the kinematic analysis, which leads to a large uncertainty in the ratio.
Within the large error bars, the ratios of the kinematical and lensing velocity dispersion seem to be the same. The mean average ratio is larger than one ($1.14 \pm 0.46$), but it needs a larger data set to show whether this difference is significant. If the galaxies with measured velocities are still falling into the cluster one would expect a velocity dispersion biased high. 
The velocity dispersion would be biased low if non cluster member galaxies from the surrounding filaments were considered.
Our procedure to eliminate galaxies in the determination of the velocity dispersion is however made to avoid the latter effect. The ratio of weak lensing to dynamical velocity dispersion could also be biased high when matter associated to groups correlated to the cluster and along the line of sight is accounted to the cluster in the weak lensing analysis.
One caveat in estimating the lensing to dynamical velocity dispersion lies in the reliability of the errors. For instance some of the clusters have a weak lensing velocity dispersion of around 900 km/s with a 1 $\sigma$ error of only 50 km/s, or about 5\%. Unless using space data with a high density of objects such an accuracy is very difficult to achieve. Even more so, if one accounts for the fact that clusters at these high redshifts consist of many subclumps, and that the redshift distribution of background objects shows cosmic variance, the weak lensing measurement will be biased.

Given the poor X-ray photon statistics we were able to derive weak constraints on the metal abundance $Z = (0.29 \pm 0.24)~Z_\odot$ of the intracluster medium from the detection of the Fe-K line in the Chandra spectrum.

\subsubsection{Comparison}
We estimate independently and self consistently values of the cluster mass within R$_{200}$, the radius of the sphere within which the cluster overdensity is 200 times the critical density of the Universe at the cluster redshift, from kinematic, X-ray and weak lensing analyses. The X-ray mass profile, based on the assumptions of isothermal gas in hydrostatic equilibrium, is found to be in good agreement with the mass density profile derived from the kinematic analysis. Given the fact that the weak lensing analysis is performed at the feasibility limit of ground based observations and therefore leads to larger errors, the weak lensing mass density profile is consistent with the other data within the errors.
Note that all our individual mass estimates (see Table 5) are in good agreement with each other. The combined weighted mass estimate M$_{200}$$=$ (4.56 $\pm$ 2.3) $\times$ 10$^{14}$ M$_{\sun}$ for XMMU J1230.3+1339 is in very good agreement with the combined mass estimate of Paper I, where a total mass M$_{200}$$=$ (4.19 $\pm$ 0.7) $\times$ 10$^{14}$ M$_{\sun}$ is estimated.

\section{Summary \& Conclusions}
\label{sec:conclu}
This paper describes the results of the first LBT multi-colour and weak lensing analyses of an X-ray selected galaxy cluster from the XMM-Newton Distant Cluster Project (XDCP), the high redshift cluster XMMU J1230.3+1339 at z $\sim$ 1. The high-quality LBT imaging data combined with the FORS2 spectroscopic data allows us to derive accurate galaxy photometry and photometric redshifts.\\

\noindent
Our main conclusions are:

\begin{itemize}
\item{The photometric redshift analysis results in an accuracy of $\triangle z/(1+z)$ = 0.07 (0.04) and an outlier rate $\eta$ = 13 (0) \%, when using all (secure, $\ge$ 95 \% confidence) spectra, respectively.}

\item{We find an $i-z$ against $z$ colour-magnitude relation with a slope of -0.027 and a zeropoint of 0.93. Our $i-z$ against $z$ colour-magnitude relation shows also evidence for a truncated red sequence at $z=24$, while rich low redshift systems exhibit a clear sequence down to fainter magnitudes. Similar evidence has been observed for other clusters (e.g., EDisCS clusters, see De Lucia et al. 2007; RXJ1716.4+6708 at $z=0.81$, see Koyama et al. 2007).}

\item{The cluster clearly follows the Butcher \& Oemler effect, with a blue galaxy fraction of f$_b$ $\approx$ 50 \%.}

\item{We see an indication of recent star formation in the distribution of the early-type galaxies in the $U - B$ against $r - z$ colour-colour diagram.}

\item{The brightest cluster galaxy could be identified spectroscopically and from photometry as an early-type galaxy. A displacement of this galaxy from the X-ray cluster centre of $\sim$ 100 kpc is observed.}

\item{Compared to XMMU J1229+0151 at $z=0.97$ (Santos et al. 2009) we observe a similar $i-z$ against $z$ colour-magnitude relation.}

\item{Compared to RDCS J0910+5422 at $z=1.10$ (Tanaka et al. 2008) we observe also several additional clumps, with similar colours in the colour-magnitude relation.  }

\item{We detect a weak lensing signal at 3.5 $\sigma$ significance using the aperture mass technique. Furthermore we find a significant tangential alignment of galaxies out to a distance of $\sim$ 6\arcmin or $\sim$ 2 r$_{200}$ from the cluster centre. The S/N peak of the weak lensing analysis coincides with the Chandra X-ray centroid.}

\item{We have demonstrated the feasibility of ground based weak lensing studies of massive galaxy clusters out to z $\sim$ 1.}

\item{From our X-ray analysis, we obtain a global temperature of T$_{\rm X}$ $=$ 5.29 keV and a luminosity of L$_{\rm X}$[0.5-2.0keV] $=$ 1.48 $\times$ 10$^{44}$ erg s$^{-1}$, within an aperture of 40\arcsec, in good agreement with the results presented in Paper I.}

\item{Mass estimates inferred from the kinematic, weak lensing and X-ray analyses are consistent within their uncertainties. We derive a total mass of $M_{200}$$=$(3.9 $\pm$ 1.3) $\times$ 10$^{14}$ $M_{\sun}$, $r_{200}$$=$1046 $\pm$ 113 kpc  for the X-ray data, which has the smallest uncertainty of the three mass estimates. From the weak lensing NFW likelihood analysis we find a concentration parameter $c = 4.0^{+14}_{-2}$ and a scale radius of $r_s= 344 ^{+50}_{-57}$ kpc, which is consistent with expectation from simulations (Bullock et al. 2001; Dolag et al. 2004).}

\item{We see evidence for a $\approx$ 14 \% bias in the ratio, $\sigma_{\rm WL}$/$\sigma_{\rm Kin}$, of the velocity dispersion form the weak lensing and the kinematic analysis against temperature $T_{\rm X}$  for distant ($z>0.8$) galaxy clusters.}

\end{itemize}

We have been awarded 75 ks  ACIS-S onboard Chandra and an extensive spectroscopic follow-up on $\sim$ 100 photo-z selected cluster-members, using \texttt{FORS2} to complement our data set. Both will be observed in 2010/11.
With this additional data we will improve the characterization of this remarkable rich, X-ray luminous galaxy cluster. 

%
%
%
\section*{Acknowledgments}

Based on data acquired using the Large Binocular Telescope (LBT). The LBT is an international collaboration among institutions in the United States, Italy, and Germany. LBT Corporation partners are the University of Arizona on behalf of the Arizona university system; Istituto Nazionale di Astrofisica, Italy; LBT Beteiligungsgesellschaft, Germany, representing the Max-Planck Society, the Astrophysical Institute Potsdam, and Heidelberg University; Ohio State University, and the Research Corporation, on behalf of the University of Notre Dame, the University of Minnesota, and the University of Virginia. Also based on observations made with ESO Telescopes at the La Silla or Paranal Observatories program ID 081.A-0312. This research has made use of data obtained from the Chandra Data Archive and software provided by the Chandra X-ray Centre (CXC) in the application packages CIAO, ChIPS, and Sherpa.
This research has made use of the NASA/IPAC Extragalactic Database (NED) which is operated by the Jet Propulsion Laboratory, California Institute of Technology, under contract with the National Aeronautics and Space Administration.\\
We are grateful to our anonymous reviewer for helpful comments and recommendations.
M.L. would like to thank A. Bauer, D. Clowe, T. Erben, C. Heymans, H. Hoekstra, Y. Mellier, P. Schneider, T. Schrabback, P. Simon, R. Suhada, P. Spinelli, A. Taylor and L. van Waerbeke for fruitful discussions and helpful comments.
This work was supported by the DFG Sonderforschungsbereich 375 "Astro-Teilchenphysik", the DFG priority program 1177, Cluster of Excellence "Origin of the Universe" 
and the TRR33 "The Dark Universe".
M.L. thanks the European Community for the Marie Curie research training network "DUEL" doctoral fellowship MRTN-CT-2006-036133. H.Q. thanks partial support from the FONDAP Centre for Astrophysics.\\

\bibliography{Abell}

\appendix

\section {ICM model spectra}

An absorbed ICM model spectrum of the bremsstrahlung continuum and line emission for a plasma, representing the galaxy cluster is shwon in Figure A1.\\

\begin{figure}
\centering
\includegraphics[width=6.2cm,angle=-90]{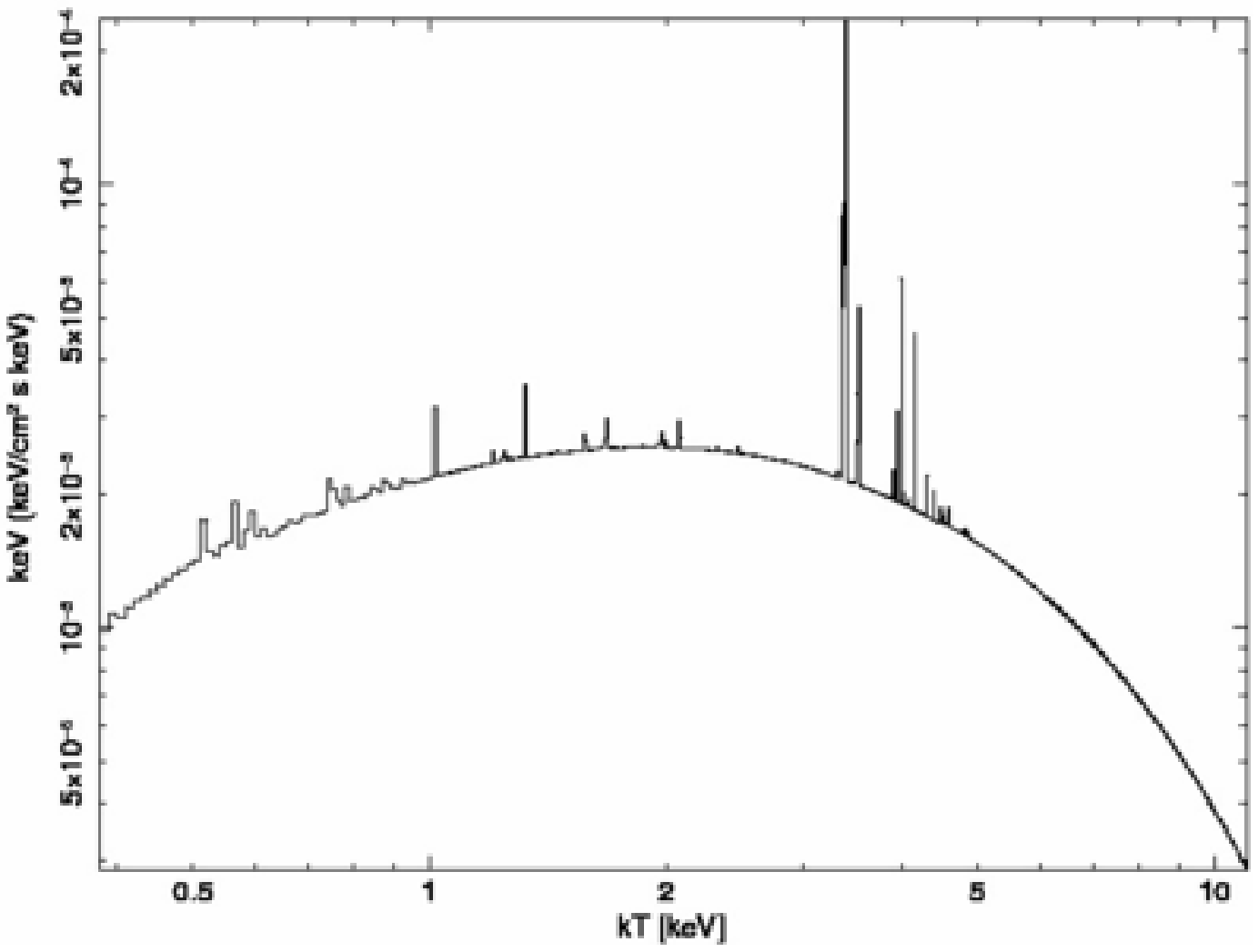}  
\caption{Absorbed ICM model spectrum of the bremsstrahlung continuum and line emission for a plasma at a redshift of $z=0.975$ with $Z/Z_{\odot}=0.3$ and temperature of $T = 5.29$ keV observed through a galactic hydrogen column density of $N_H=2.64\times 10^{22}$ cm$^{-2}$. The redshifted Fe K-line (6.7 keV at rest-frame) is visible at 3.3 keV.} 
\label{FigXMMUJ1230_X-ray_Model_Spectra}
\end{figure}       

\section{PSF anisotropy pattern}

\begin{figure*}
\centering
\includegraphics[width=5.5cm]{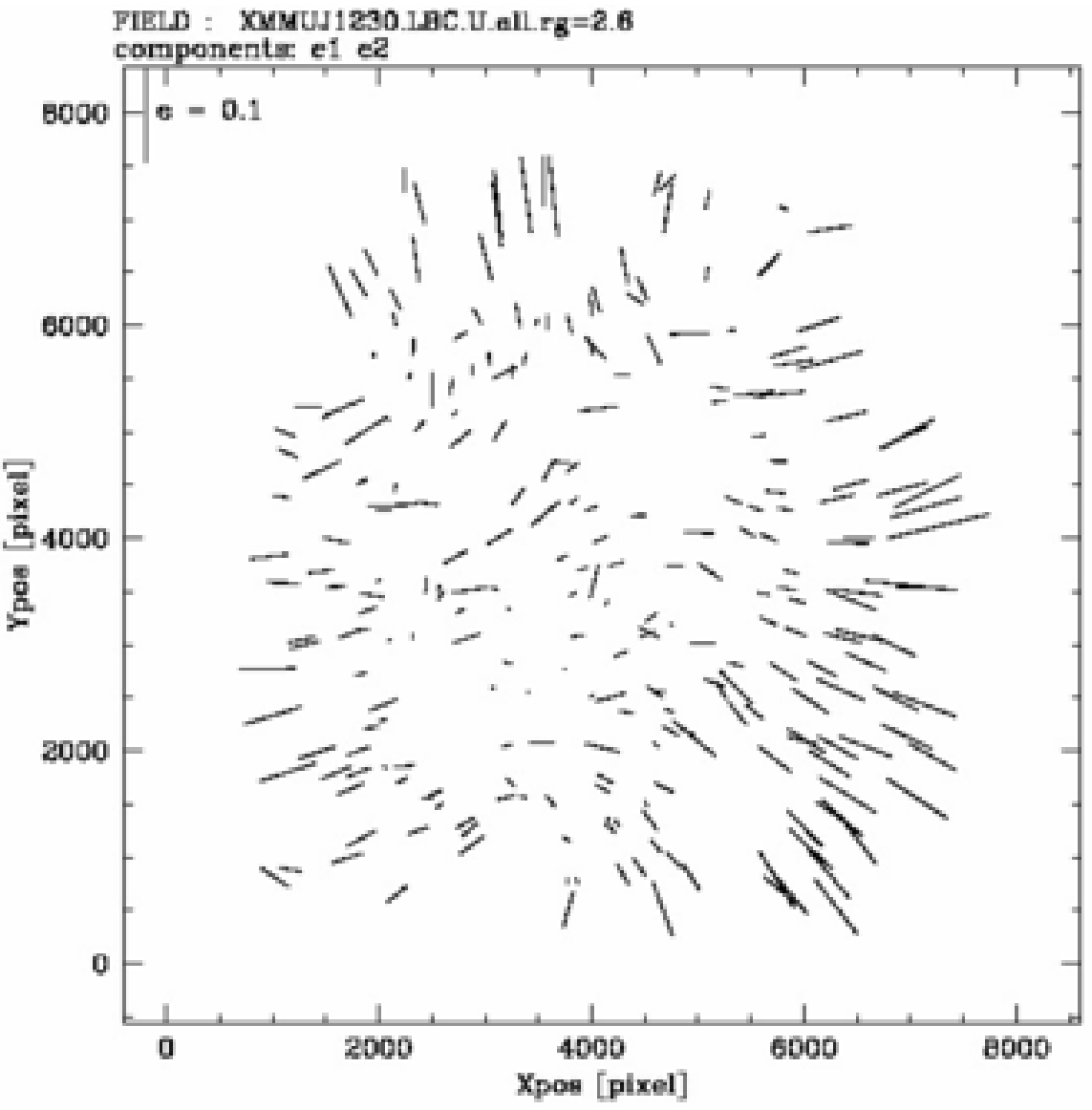}
\includegraphics[width=5.5cm]{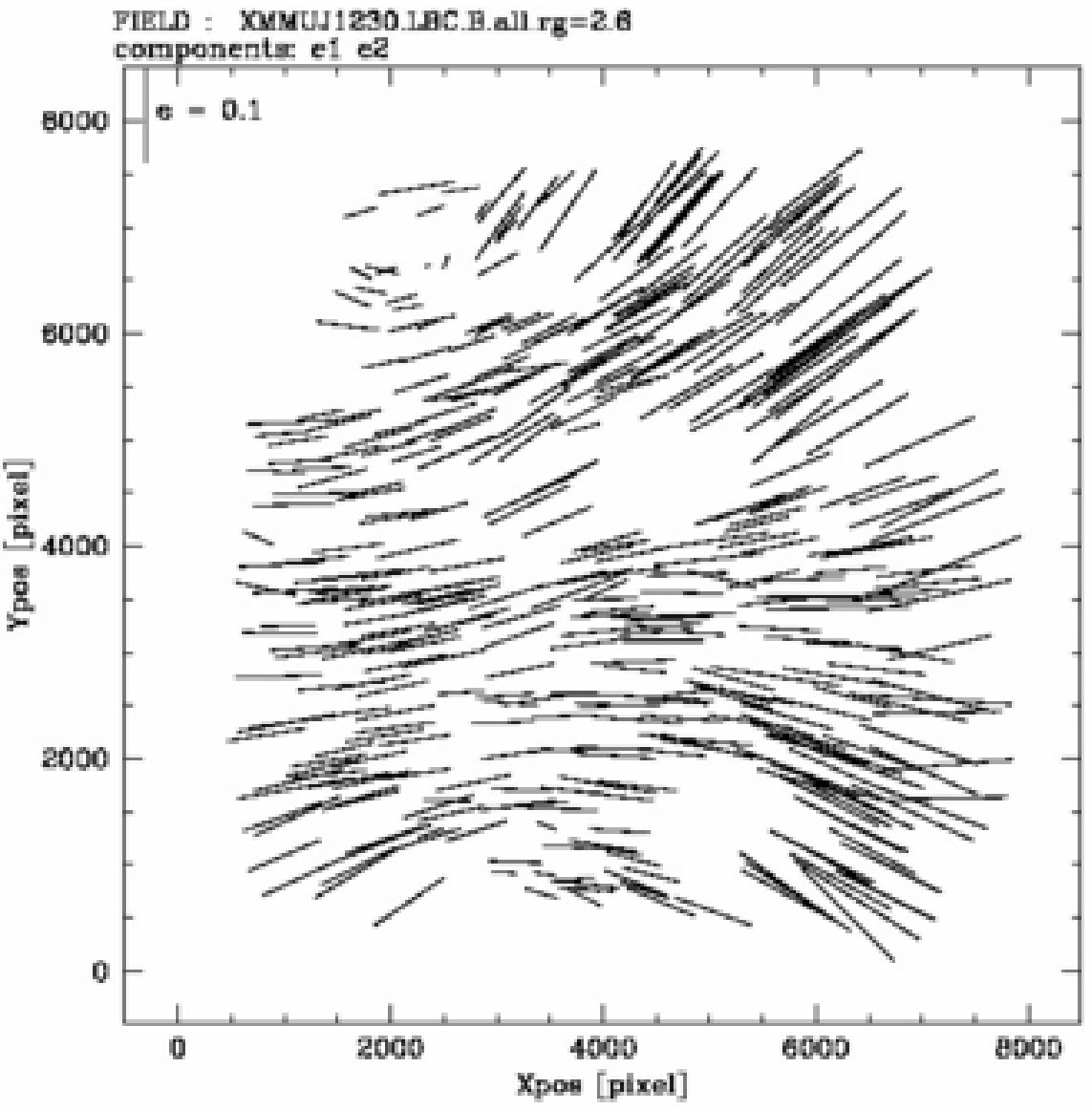}
\includegraphics[width=5.5cm]{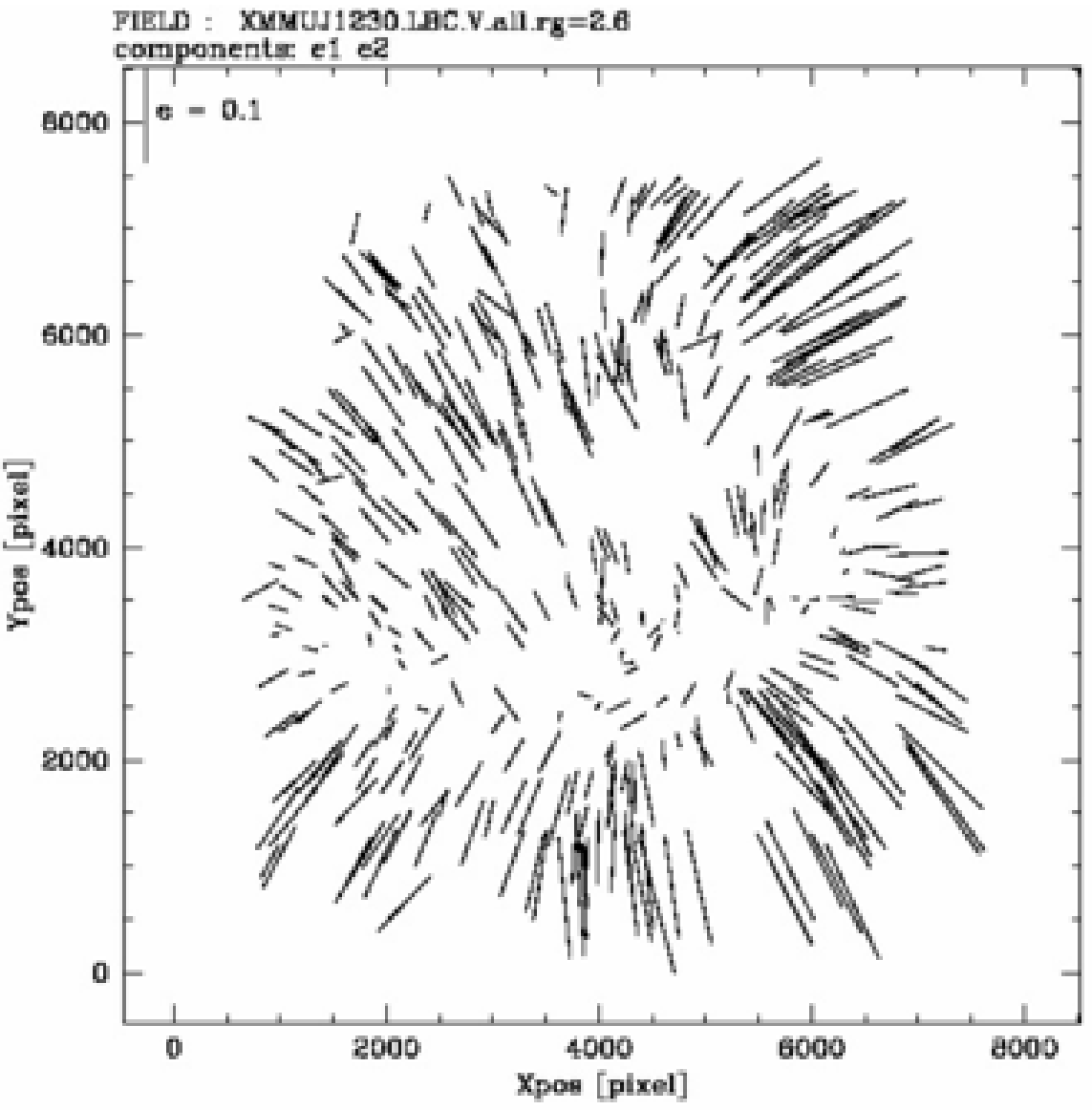}
\includegraphics[width=5.5cm]{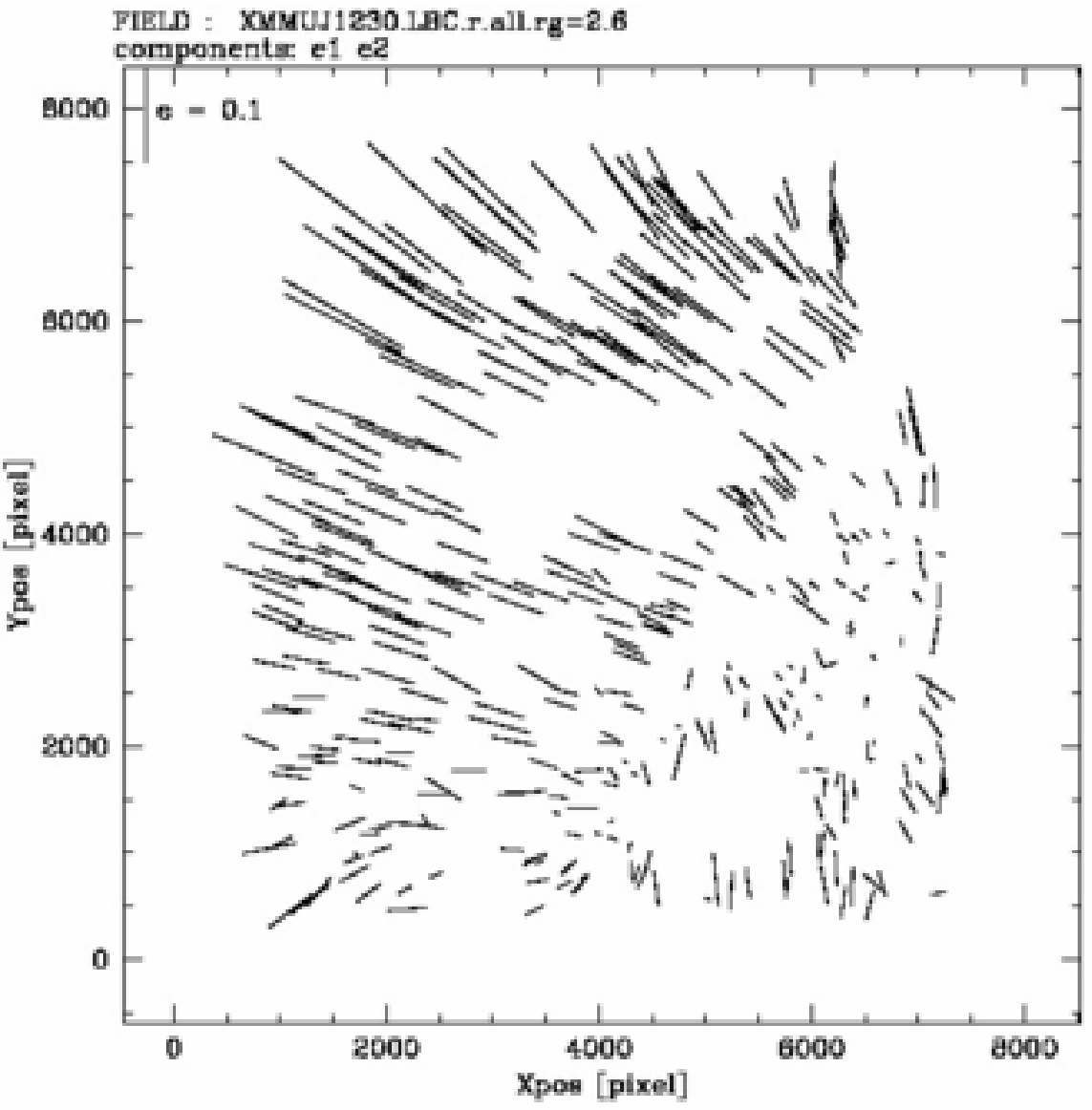}
\includegraphics[width=5.5cm]{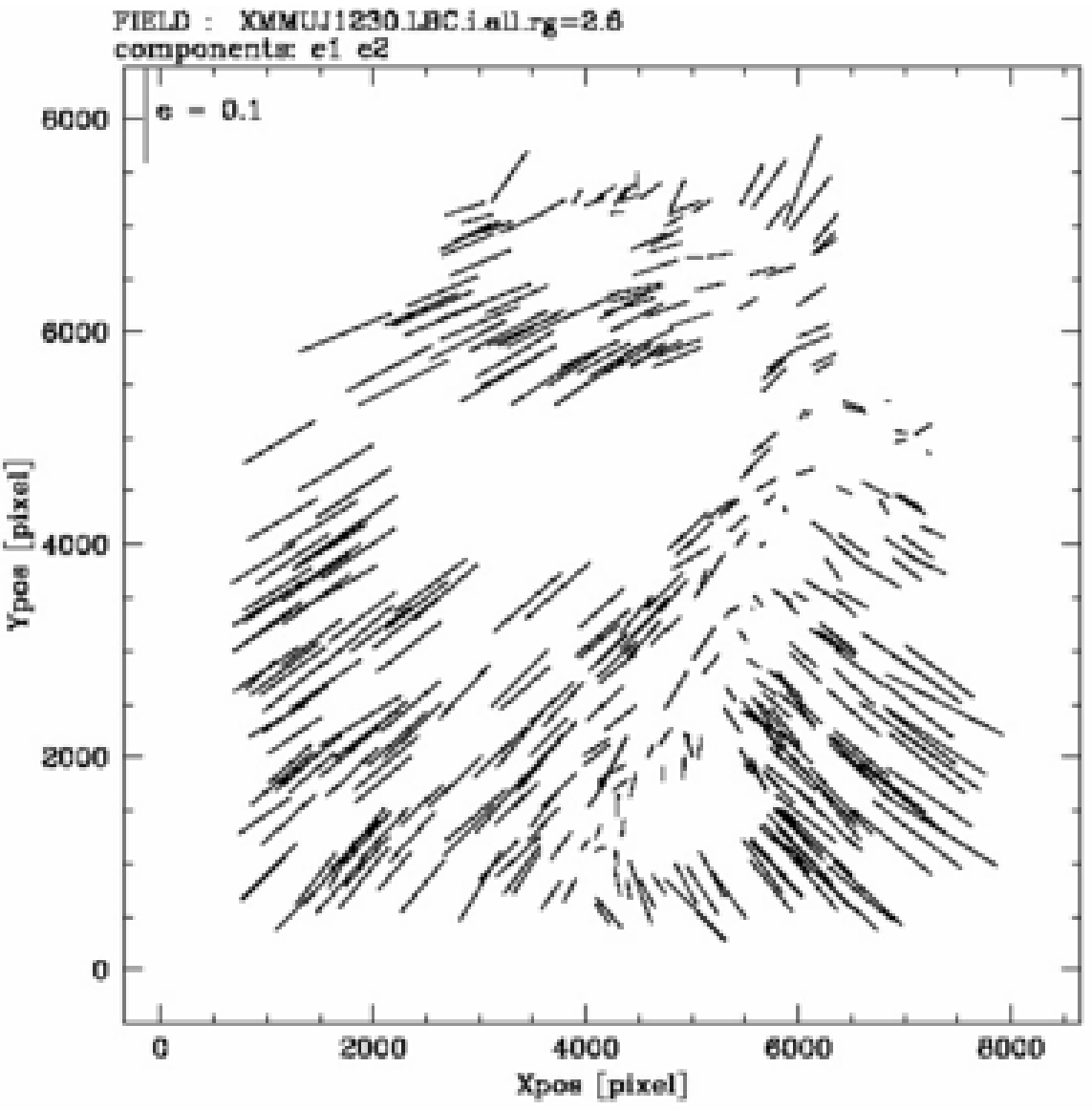}
\includegraphics[width=5.5cm]{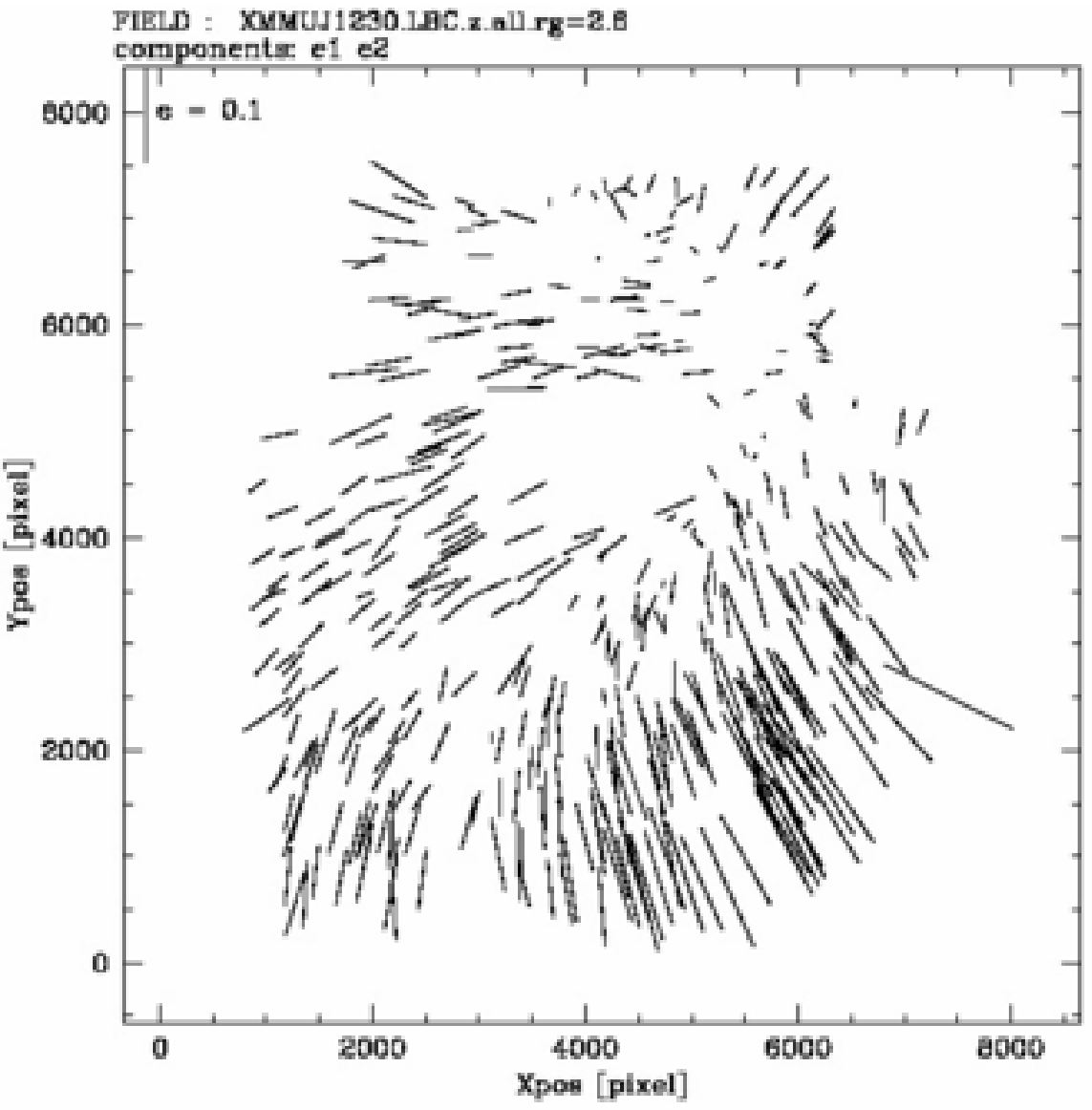}
\caption{PSF anisotropy pattern of the image stacks in the 6 (U, B, V, r, i, z) filters. The sticks mark the observed ellipticities of high signal to noise but unsaturated stars in the image stacks. The direction of the sticks coincides with the major axis of the PSF, the length is indicating the degree of anisotropy. An ellipticity of 10\% is shown in the upper left corner of each panel. The definition of ellipticity is given in Eq. 4. The measured PSF widths of stars on the stacks are summarised in Table 1.}
\label{FigPSF}
\end{figure*}

In Figure \ref{FigPSF} we plot the measured PSF patterns of all 6 bands, represented by the unsaturated stars imaged on the field at high SN. The measured PSF widths of stars on the stacks are summarised in Table 1. The anisotropy pattern of LBC is similar in terms of orientation and size to Suprime-Cam shown in Figure 3 of Okabe \& Umetsu (2008).

\section {High-z cluster}

In Table C1 we give an overview of recently studied high-z galaxy clusters. So far only 10 of them were studied with weak lensing. Note that only XMMU J1230.3+1339 (z$\sim$1), ClG J1604+4304 (z$\sim$0.9), RX J1716.6+6708 and ClG 1137.5+6625 (z$\sim$0.8) have a weak lensing detection based on ground based data.\\

\bsp

\begin{landscape}
\begin{table}
\vspace{0.5cm}
\caption[]{\rm {List of high-z galaxy clusters and their basic optical/X-ray/weak lensing properties: (1) sequence number, (2) right ascension, (3) declination, (4) number of spectroscopic objects used to estimate the cluster redshift (6) and the line-of-sight velocity dispersion (7), (8) bolometric X-ray luminosity, (9) ICM temperature, (10) velocity dispersion from a weak lensing analysis, more information on peculiar individual clusters is provided in the associated reference (11) .}} 
\begin{center}
\hspace{-0.5cm}
\resizebox{24.5cm}{!} {
\begin{minipage}{25cm}
\begin{center}
\renewcommand{\footnoterule}{}  
\begin{tabular}{llccccccccl}
\hline\hline
Id & Cluster name & R.A. & Dec. & N$_{gal}$ & Redhift & $\sigma_l^{Kin}$ & L$_{bol}^x$ & T$_x$ & $\sigma_{sis}^{WL}$ & Ref.\\
~ & ~ & (J2000) & (J2000) & ~ & ~ & [km/s] & [10$^{44}$ erg s$^{-1}$] & [keV] & [km/s] & ~ \\
(1) & (2) & (3) & (4) & (5) & (6) & (7) & (8) & (9) & (10) & (11) \\
\hline\hline
1 &  SXDF-XCLJ0218-0510 & 02$^h$18$^m$21.3$^s$ & -05$^{\circ}$10\arcmin27\arcsec & 9 & 1.6230 $\pm$ 0.001 & 537 $\pm$ 213 & 0.34 $\pm$ 0.16 & 1.7 $\pm$ 0.3 & ... $\pm$ ... & Tanaka et al. (2010); Papovich et al. (2010) \\
\hline
2 &  XMMXCS J2215.9-1738 & 22$^h$15$^m$58.5$^s$ & -17$^{\circ}$38\arcmin03\arcsec & 31 & 1.457 $\pm$ 0.001 & 720 $\pm$ 110 & 2.92 $\pm$ 0.35 & 4.1 $\pm$ 0.9 & ... $\pm$ ... & Hilton et al. (2010); Stanford et al. (2006) \\
\hline
3 &  XMMU J2235.3-2557 & 22$^h$35$^m$20.8$^s$ & -25$^{\circ}$57\arcmin40\arcsec & 30 & 1.390 $\pm$ ... & 802 $\pm$ 77 & 8.5 $\pm$ 0.4 & 8.6 $\pm$ 1.3 & 1145 $\pm$ 70 & Jee et al. (2009a); Rosati et al. (2009);\\
~ & ~ & ~ & ~ & ~ & ~ & ~ & ~ & ~ & ~ & Mullis et al. (2005)\\
\hline
4a &  ClG J0848+4453, Lynx W & 08$^h$48$^m$34.7$^s$ & +44$^{\circ}$53\arcmin42\arcsec & 9 & 1.273 $\pm$ ... & 650 $\pm$ 170 & 1.5 $\pm$ 0.8 & 1.7 $\pm$ 0.7 & 762 $\pm$ 133 & Mei et al. (2009); \\
4b &  ClG J0849+4452, Lynx E & 08$^h$48$^m$56.3$^s$ & +44$^{\circ}$52\arcmin16\arcsec & 11 & 1.261 $\pm$ ... & 720 $\pm$ 140 & 2.1 $\pm$ 0.5 & 3.8 $\pm$ 1.3 & 740 $\pm$ 134 & Jee et al. (2006)  \\
\hline
5 &  RDCS J1252.9-2927 & 12$^h$52$^m$48.0$^s$ & -29$^{\circ}$27\arcmin00\arcsec & 38 & 1.2370 $\pm$ 0.0004 & 747 $\pm$ 84 & 6.6 $\pm$ 1.1 & 6.0 $\pm$ 0.7 & 1185 $\pm$ 120\footnote{For a non-singular isothermal sphere (NSIS) with free coordinates for its center.}  & Demarco et al. (2007); Lombardi et al. (2005); \\
~ & ~ & ~ & ~ & ~ & ~ & ~ & ~ & ~ & ~ & Rosati et al. (2004)\\
\hline
6 &  RDCS J0910+5422 & 09$^h$10$^m$00.0$^s$ & -54$^{\circ}$22\arcmin00\arcsec & 161 & 1.1005 $\pm$ 0.0016 & 716 $\pm$ 141 & 2.9 $\pm$ 0.4 & 2.9 $\pm$ 0.4 & ... $\pm$ ... & Maughan et al. (2008a); Tanaka et al. (2008) \\
\hline
7 & XLSSJ022404.1-041330 & 02$^h$24$^m$04.1$^s$ & -04$^{\circ}$13\arcmin30\arcsec & 5 & 1.050 $\pm$ ... & ... $\pm$ ... & 4.5 $\pm$ 0.1 & 4.3 $\pm$ 1.1 & ... $\pm$ ... & Maughan et al. (2008b); Pacaud et al. (2007)  \\
\hline
8 & RzCS052 & 02$^h$21$^m$42.0$^s$ & -03$^{\circ}$21\arcmin47\arcsec & 5 & 1.016 $\pm$ ... & 710 $\pm$ 150 & ... $\pm$ ... & ... $\pm$ ... & ... $\pm$ ... & Andreon et al. (2008) \\
\hline
9 & XMMU J1229.4+0151 & 12$^h$29$^m$29.2$^s$ & +01$^{\circ}$51\arcmin26\arcsec & 27 & 0.975 $\pm$ 0.001 & 683 $\pm$ 62 & 3.0 $\pm$ ... & 6.4 $\pm$ 0.7 & ... $\pm$ ... & Santos et al. (2009) \\
\hline
10 & XMMU J1230.3+1339 & 12$^h$30$^m$16.0$^s$ & +13$^{\circ}$39\arcmin04\arcsec & 13 & 0.9745 $\pm$ 0.002 & 657 $\pm$ 277 & 6.5 $\pm$ 0.68 & 5.29 $\pm$ 1.02 & 1308 $\pm$ 284 & This work \& Paper I\\
\hline
11a & ClG J1604+4304 & 16$^h$04$^m$23.7$^s$ & +43$^{\circ}$04\arcmin52\arcsec & 67 & 0.9001 $\pm$ ... & 962 $\pm$ 141 & 1.576 $\pm$ ... & 3.5 $\pm$ 1.82 & 1004 $\pm$ 199 & Kocevski et al. (2009); Margoniner al. (2005); \\
11b & ClG J1604+4314 & 16$^h$04$^m$25.8$^s$ & +43$^{\circ}$14\arcmin23\arcsec & 62 & 0.8652 $\pm$ ... & 719 $\pm$ 71 & 1.164 $\pm$ ... & 1.6 $\pm$ 0.65 & ... $\pm$ ... & Gal et al. (2004) \\
\hline
12 & ClG J1226+3332 & 12$^h$26$^m$58.1$^s$ & +33$^{\circ}$32\arcmin47\arcsec & 6 & 0.8877 $\pm$ 0.004 & 1650 $\pm$ 930 & 5.12 $\pm$ 0.12 & 10.4 $\pm$ 0.6 & 880 $\pm$ 50 & Jee et al. (2009b); Maughan et al. (2007); \\
~ & ~ & ~ & ~ & ~ & ~ & ~ & ~ & ~ & ~ & Ebeling et al. (2001)\\
\hline
13 & RX J1257.2+4738 & 12$^h$57$^m$12.2$^s$ & +47$^{\circ}$38\arcmin07\arcsec & 18 & 0.866 $\pm$ ... & 600 $\pm$ ... & ... $\pm$ ... & 3.6 $\pm$ 2.9 & ... $\pm$ ... & Ulmer et al. (2009) \\
\hline
14 & ClG 1054-0321 & 10$^h$57$^m$00.2$^s$ & -03$^{\circ}$37\arcmin27\arcsec & 129 & 0.826 $\pm$ ... & 1156 $\pm$ 82 & 3.1 $\pm$ 0.1 & 8.9 $\pm$ 1.0 & 1150 $\pm$ 51 & Tran et al.(2007); Jee et al. (2005b); \\
~ & ~ & ~ & ~ & ~ & ~ & ~ & ~ & ~ & ~ & Hoekstra et al. (2000)\\
\hline
15 & ClG J0152-1357 & 01$^h$52$^m$41.0$^s$ & -13$^{\circ}$57\arcmin45\arcsec & 102 & 0.837 $\pm$ 0.001 & 1632 $\pm$ 115 & ... $\pm$ ... & 6.46 $\pm$ 1.74 & 903 $\pm$ 57 & Jee et al. (2005a); Demarco et al. (2005); \\
 ~ & ~ & ~ & ~ & ~ & ~ & ~ & ~ & ~ & ~ & Maughan et al. (2006)\\
\hline
 16 & RX J1716.6+6708 & 17$^h$16$^m$49.6$^s$ & +67$^{\circ}$08\arcmin30\arcsec & 37 & 0.809 $\pm$ 0.0051 & 1522 $\pm$ 215 & 1.50 $\pm$ 0.4 & 5.7 $\pm$ 1.4 & 1030 $\pm$ ... & Clowe et al. (2000); Gioia et al. (1999) \\
\hline
 17 & ClG 1137.5+6625 & 11$^h$40$^m$23.3$^s$ & -66$^{\circ}$09\arcmin01\arcsec & 22 & 0.784 $\pm$ 0.0003 & 884 $\pm$ 185 & 1.90 $\pm$ 0.4 & 5.7 $\pm$ 2.1 & 1190 $\pm$ ... & Clowe et al. (1998); Luppino et al. (1995); \\
  ~ & ~ & ~ & ~ & ~ & ~ & ~ & ~ & ~ & ~ & Donahue et al. (1999)\\
\hline
\end{tabular}
\end{center}
\end{minipage}
}
\end{center}
\label{tab:highzclusters}
\end{table}
\end{landscape}

\label{lastpage}

\end{document}